\documentclass[
  aps,
  prd,
  reprint,
  superscriptaddress,
  nofootinbib,
  longbibliography,
  floatfix
]{revtex4-2}

\usepackage{amsmath,amssymb}
\usepackage{graphicx}
\usepackage{bm}
\usepackage{booktabs}
\usepackage{xcolor}
\usepackage[hidelinks]{hyperref}

\newcommand{\cms}{\mathrm{cm^3\,s^{-1}}}
\newcommand{\dd}{\mathrm{d}}

\begin{document}

\title{Right Energy, Wrong Profile:\\Why the 43 GeV Cluster Line Is Unlikely to Be Dark Matter.}

\author{Stefano Profumo}
\affiliation{Department of Physics and Santa Cruz Institute for Particle Physics,
University of California, Santa Cruz, California 95064, USA}

\date{\today}

\begin{abstract}
A narrow gamma-ray line would be a distinctive signature of dark matter, but its interpretation depends on where the emission lies as much as on its energy. We reanalyze the reported 43 GeV feature in Fermi-LAT observations of Virgo, Fornax, and Ophiuchus. The excess is reproduced in the same public data and survives event-type partitions, a test using data withheld from the original selection, and detector-coordinate permutation tests. Hadronic emission and standard inverse-Compton emission from realistic electron populations are far too broad to account for its narrowness. We then perform a spatial--spectral likelihood analysis restricted to Good Time Intervals (GTIs) and including Galactic and isotropic diffuse emission and all catalogued sources. The surviving line-like component is spatially broad. Nearly uniform surface brightness and substructure-enhanced annihilation are preferred, whereas smooth NFW annihilation and a central point source are unlikely explanations. An independent event-level annular analysis agrees: the events associated with the excess extend through much of the virial region. A direct calibration of the maximum over 20--70 GeV and all five morphologies gives a conditional global significance of approximately $2.7\sigma$, with the maximum at 44.5 GeV for uniform brightness. This calibration is conditional on the preselected three-cluster sample. Because smooth-halo annihilation is the scenario a line most directly suggests, this morphology is the central obstacle to a dark-matter interpretation. A dark-matter explanation would require several conditions to hold simultaneously. Photons must be a leading annihilation channel, annihilation must remain efficient in cold subhalos, and subhalos must supply nearly all the cluster annihilation luminosity. However, a substructure boost of the magnitude supported by $\Lambda$CDM drives the required two-photon cross section into the range where Galactic-halo line limits bind most tightly. The most likely explanation is therefore a chance fluctuation amplified by analysis and target-selection effects, possibly compounded by residual Galactic diffuse or unresolved-source mismodeling; if the feature is celestial, its morphology makes a simple dark-matter interpretation unlikely.
\end{abstract}

\maketitle

\section{Introduction}
\label{sec:intro}

The search for monochromatic gamma rays has long occupied a distinctive
place in efforts to identify the particle nature of dark matter. Unlike the
broad continua produced by cosmic-ray interactions and most annihilation
final states, a two-body final state containing a photon can imprint an
energy scale directly related to the dark-matter mass. Early calculations
of photino annihilation established this possibility, and subsequent
one-loop calculations developed it quantitatively for neutralino dark matter
\cite{BergstromSnellman1988,BergstromUllio1997}. For nonrelativistic particles,
$\chi\chi\to\gamma\gamma$ produces photons at $E_\gamma\simeq m_\chi$,
whereas $\chi\chi\to\gamma X$ gives
$E_\gamma\simeq m_\chi[1-m_X^2/(4m_\chi^2)]$, in units with $c=1$;
two-body radiative decay provides a related possibility. Particle motions within a gravitationally bound halo broaden the line
through the Doppler effect, with a characteristic fractional width of order
$v/c$, where $v$ is the relevant line-of-sight velocity scale
\cite{Speckhard2016}. Even for cluster velocities of order
$10^3\,{\rm km\,s^{-1}}$, this corresponds to a width of only a few
parts in $10^3$, appreciably narrower than the energy resolution of the
Fermi Large Area Telescope (LAT) at tens of GeV
\cite{Ackermann2012,FermiLAT2015}. The observed line profile is therefore
governed mainly by the instrumental energy dispersion. For electrically neutral dark matter, direct
annihilation into photons is commonly loop-suppressed, making such signals
faint but potentially highly diagnostic. Sharp spectral features can also
arise from internal bremsstrahlung, although these need not have the shape of
a monochromatic line \cite{Bringmann2008}. Establishing the spectral shape,
as well as its celestial origin, is therefore essential to interpreting a
candidate.

Observational searches have progressed from the Energetic Gamma-Ray
Experiment Telescope (EGRET) aboard the Compton Gamma-Ray Observatory
\cite{Thompson1993EGRET} to the much larger photon samples accumulated
by the LAT. An EGRET search of the Galactic-center region found no line between
0.1 and 10 GeV \cite{Pullen2007}. The first LAT line search used 11 months
of observations to constrain features between 30 and 200 GeV
\cite{FermiLAT2010Lines}; subsequent searches expanded the energy range and
introduced more detailed treatments of energy dispersion and systematic
uncertainties \cite{FermiLAT2013Lines,FermiLAT2015}. Particular attention
followed independent reports in 2012 of a feature near 130 GeV toward the
Galactic center \cite{Weniger2012,SuFinkbeiner2012}. Reprocessed data and an
improved response model reduced its significance, and the 5.8-year Pass-8
search found no significant line over 0.2--500 GeV
\cite{FermiLAT2013Lines,FermiLAT2015}. Studies of Earth-limb photons and
instrument-coordinate selections also raised concerns about possible
instrumental structure \cite{Finkbeiner2013,FermiLAT2013Lines}. This history
illustrates why a locally significant spectral excess must be accompanied
by control samples, a treatment of search trials, and tests in additional
data before it can support a physical interpretation.

Other instruments provide complementary tests with different energy
resolutions and accessible ranges. Five years of DArk Matter Particle Explorer
observations yielded no Galactic line candidate between 10 and 300 GeV
\cite{DAMPE2022Lines}, and imaging atmospheric Cherenkov telescopes have
searched from 0.3 to 100 TeV toward the Galactic center
\cite{HESS2018Lines,MAGIC2023Lines,HESS2026Lines}. None of these searches
directly tests a 43 GeV cluster feature. Comparisons between targets in any
case require a specified emission model, because the relative fluxes depend on
the density distribution, the substructure content, and potentially the
dark-matter velocity distribution.

A dark-matter line is also rarely an isolated prediction. The
interactions that generate a loop-induced photon channel generally also allow
annihilation into charged particles or other Standard-Model states, whose
radiation and decays produce a broad continuum that can be much brighter than
the line. This continuum-to-line ratio was a central model-building issue in
interpretations of the 130 GeV candidate \cite{Abazajian2012,Asano2013}. It is
not universal---heavy charged states inaccessible as final states, different
mediator couplings, and different velocity dependences all alter it---but
explaining a prominent line while respecting continuum limits requires an
explicit particle model rather than a line normalization alone. We return to
this in Sec.~\ref{sec:dm}, where the spectral requirement, the measured
morphology, and the absolute line normalization are combined into a short list
of conditions any dark-matter explanation would have to satisfy at once; they
prove to pull against one another through the cluster substructure boost.

The same interactions must operate in other systems. A cluster
interpretation predicts corresponding emission from the Galactic center and
halo and from dwarf spheroidal galaxies, with a common intrinsic line energy
and relative fluxes set by the emission mechanism and the dark-matter
distribution \cite{FermiLAT2015,Huang2012Consistency,Lacroix2022}. This does
not require a detectable line in every system, since sensitivity, profile
uncertainty, substructure, and velocity dependence determine which counterparts
should be visible. It does require a model that explains any predicted
counterpart exceeding an observational limit: invoking cold subhalos to enhance
cluster emission, for example, must remain consistent with the low-velocity
dark matter in dwarfs.

Galaxy clusters are particularly useful in this respect because their large
dark-matter reservoirs can be studied as spatially extended sources. For
nearby systems, the virial region is resolved by the LAT at tens of GeV,
allowing the radial distribution to test the origin of a spectral feature.
Smooth-halo annihilation weights the density squared and is centrally
concentrated, whereas decay weights the density and produces a shallower
profile. A population of surviving subhalos can redistribute annihilation
luminosity toward larger radii, although the size of the enhancement depends
on uncertain extrapolations of the subhalo population
\cite{Navarro1997,Gao2012,Han2012,Ando2019}. Velocity-dependent annihilation
can further change the relative contributions of the dynamically hot host
halo and its colder subhalos \cite{Lacroix2022}. Clusters thus offer more
than additional targets for a line search: their resolved morphology can
distinguish emission scenarios that would be difficult to separate using
the line energy alone.

Clusters also host conventional mechanisms for producing gamma rays.
Structure-formation and merger shocks, active galactic nuclei, and
galactic activity supply cosmic rays to the intracluster medium.
Inelastic collisions of cosmic-ray protons with the thermal gas
produce neutral pions, whose decays generate a broad gamma-ray
continuum; primary and secondary electrons can contribute through
inverse-Compton scattering and bremsstrahlung. The relative
importance and spatial distribution of these components depend on
particle injection, confinement, transport, and the gas and
radiation fields \cite{BrunettiJones2014}. The competition between
cosmic-ray and dark-matter emission in clusters and groups, and the
use of spectra, morphology, and multiwavelength counterparts to
distinguish them, were investigated explicitly by
Jeltema, Kehayias, and Profumo \cite{JeltemaKehayiasProfumo2009}.
Consequently, spatial extension or association with a cluster is
insufficient by itself to identify a dark-matter signal.

Hadronic emission further links gamma-ray and radio observations,
because the charged pions produced in the same collisions inject secondary
electrons and positrons that radiate synchrotron emission; LAT limits and
radio-halo measurements jointly constrain this scenario and the intracluster
magnetic field \cite{JeltemaProfumo2011,AckermannClusterCR2014}. Standard
hadronic and leptonic processes generally produce broad spectra rather than an
isolated, instrumentally narrow feature at tens of GeV, but that expectation
must be tested quantitatively after convolution with the instrumental response
and in the presence of foreground and source-model uncertainties.

These challenges motivate a sequence of tests before selecting a
particle interpretation. First, one must assess instrumental and other
noncelestial explanations, including energy-response artifacts,
selection-dependent features, and statistical fluctuations. Event
partitions, temporal holdouts, instrument-coordinate tests, and matched-sky
controls address different aspects of this question; agreement among them
cannot exclude a response effect common to all selections. Second, one must
test non-dark-matter celestial explanations, asking whether hadronic
emission, inverse-Compton emission, unresolved sources, or foreground
mismodeling can reproduce the feature's narrow spectrum and spatial extent.
Third, conditional on dark matter, one must weigh the complementary signals
that the same hypothesis predicts: continuum emission, lines in other
halos, target-to-target luminosity scaling, and secondary radiation.
Morphology and redshift coherence connect these stages. Redshift coherence
tests association with the clusters without identifying the emission
mechanism, whereas the radial distribution discriminates among physical
origins without by itself establishing that the feature is celestial.

The feature considered here was first reported at a historical observer-frame
energy near 42.7 GeV in a Pass-8 search of nearby galaxy clusters
\cite{Liang2016}. Fan et al. subsequently
analyzed approximately 15.5 years of P8R3 observations of 13 nearby clusters
and found that Virgo, Fornax, and Ophiuchus dominate a stacked feature near
43 GeV \cite{Fan2026}. Their aperture scan reaches its largest line
preference at approximately the virial radius. This motivates a spatial
analysis, but a cumulative aperture scan is not a radial measurement:
adjacent apertures share photons, and background and target weighting can
alter the likelihood statistic. Throughout, Table~\ref{tab:evidencehierarchy}
summarizes the strength of each class of statement made below, so that
reconstruction, sky-model, and conditional-template results are not read at a
common level of confidence. We independently reconstruct the reported
candidate, examine the control and alternative-origin tests, and compare
observer-frame with cluster-frame energy coherence. Independent annular
fits and a mask-consistent event-level likelihood then determine where the
line support lies and which physical templates can describe it. Finally,
we set out the conditions that a dark-matter interpretation would have to
satisfy, given that morphology.

Throughout, ``43 GeV candidate'' denotes the feature generically. Exact
likelihood results use the present source-frame reference energy
$E_0=43.2$ GeV; 42.7 GeV is reserved for the explicitly historical
observer-frame energy used in the temporal holdout.
The spectral reconstruction uses the same observations as the reported
candidate and is not an independent discovery. Likewise, a preference for
a shallow radial profile neither establishes substructure annihilation nor
measures its boost; the model-building analysis retains these distinctions.
The strict-Good Time Interval (GTI) spatial--spectral likelihood introduced here supersedes the
masked annular ranking as the main assessment of sky-background and
point-source alternatives.

Reading the result forward into model building, the conditions it imposes turn
out to interlock rather than simply accumulate. The mass is spectrally
favorable, since every on-shell electroweak partner channel is closed at
$2m_\chi\simeq86.4$ GeV, so a photon-dominated operator can produce an
essentially isolated line. The morphology is the difficulty: it requires
subhalos to supply nearly all of the cluster annihilation luminosity, which in
turn fixes the sign of any velocity dependence and, because the required
two-photon cross section scales inversely with the substructure boost, pushes
that cross section toward the thermal scale---where Galactic-halo line limits,
rather than the dwarf-spheroidal limits one might expect, are the binding
constraint. 

The paper is organized around a single question: the candidate's
spatial distribution. Section~\ref{sec:framework} describes the data and
defines the likelihood conventions. Section~\ref{sec:nondm} examines
instrumental, statistical, and conventional celestial explanations, including
forward-folded pion-decay spectra and inverse Compton emission in both the
Thomson and Klein--Nishina regimes. Section~\ref{sec:fullsky} presents the
strict-GTI source-plus-diffuse spatial--spectral likelihood, which is the
principal measurement of this work, together with the global calibration of
its energy-and-morphology search; Sec.~\ref{sec:morphology} presents the
independent event-level radial analysis that corroborates it and shows which
photons supply the support. Section~\ref{sec:redshift} tests energy coherence
between clusters, and Sec.~\ref{sec:dm} sets out the conditions a
dark-matter interpretation would have to meet, given the measured morphology. Section~\ref{sec:discussion} collects
the conclusions. The appendices contain the reproduction of the reported
excess, instrumental and bootstrap validation, technical response and spatial
checks, and phenomenological continuum fits. Run identities, seeds, hashes,
optimizer settings, and protocol manifests are collected in a separate
reproducibility supplement.

\begin{table*}[tbp]
\centering
\small
\caption{Hierarchy of evidence used throughout this work. Conditional template
statements assume the selected three-cluster sample, candidate energy,
response and background treatment, and fixed plug-in spatial models. Each row
should be read at its own strength; the rows are not independent factors that
may be combined into a single detection probability.}
\label{tab:evidencehierarchy}
\begin{ruledtabular}
\begin{tabular}{p{0.15\textwidth}p{0.28\textwidth}p{0.24\textwidth}p{0.24\textwidth}}
Level & Result & Establishes & Does not establish\\
\hline
Reconstruction & Exact P8R3 analysis recovers the reported feature & Reproducibility in the public top-three data & Independent detection or celestial origin\\
Data-level checks & Event partitions, time splits, matched skies, masks, and annuli & No identified simple partition-specific pathology; virial-scale support survives catalog masking & Exclusion of common-mode response effects, foreground mismatch, unresolved sources, or fluctuation\\
Complete sky model & Broad templates retain summed ${\rm TS}=13.20$--14.45; the calibrated search gives $p_{\rm global}=3.6\times10^{-3}$ & The candidate survives catalog sources and the nominal diffuse model at a conditional global significance of about $2.7\sigma$ & An unconditional detection, correction for selecting three of 13 clusters, or immunity to diffuse-model systematics\\
Conditional templates & Shallow profiles outrank smooth NFW annihilation; substructure outranks NFW decay & Strong conditional mismatch of smooth annihilation within the tested family & Physical exclusion, a measured boost, or unique identification of substructure\\
Spectral alternatives & Hadronic and standard-field IC spectra do not absorb the narrow residual & Ordinary broad cluster emission is not an economical explanation & Proof that the residual is a line\\
Energy coherence & Modest source-frame preference, dominated by Ophiuchus & A consistency check against cluster association & A measured redshift law or independent evidence of a celestial line\\
\end{tabular}
\end{ruledtabular}
\end{table*}

\section{Data and Likelihood Framework}
\label{sec:framework}
\label{sec:signal}

\subsection{Signal definition and independent reproduction}

Our starting point is the top-three spectral candidate of
Ref.~\cite{Fan2026}, not a newly selected target sample or energy interval.
Using the public LAT photons and the exact P8R3 energy-dispersion response,
we recover 44 events between 40 and 46 GeV after the strict spacecraft-quality
selection, exactly matching the published count. The baseline sliding-window
fit gives ${\rm TS}=30.62$ at the reported source-frame energy 43.2 GeV,
compared with approximately 30.1 in the published analysis. Its maximum is
${\rm TS}=30.81$ at 43.25 GeV. Appendix~\ref{sec:replication} gives the scan
and implementation cross-checks.

For all primary radial tests we freeze the source-frame energy at
$E_0=43.2$ GeV, corresponding to $E_0/(1+z_i)$ in cluster $i$. We do not
rescan the energy in each annulus. This choice isolates the spatial question
conditional on the reported spectral feature; adopting a source-frame line
does not demonstrate that the data prefer source-frame over observer-frame
alignment. The choice of frame is tested separately through energy coherence.

The spectral and spatial TS values answer different questions. The baseline
spectral scan pools the unmasked top-three sample, whereas the primary spatial
comparison masks sources, retains cluster--annulus identity, and imposes
template-dependent signal weights. Its smaller TS is not a revised estimate
of the published discovery significance. The strict-GTI sky-model analysis has
its own global calibration over energy and morphology, presented in
Sec.~\ref{subsec:globalcalibration}; it is not combined with the annular,
instrumental, temporal, or redshift-coherence diagnostics into a post hoc
omnibus significance. Instrumental and temporal validation, including a historically
fixed-energy post-2015 holdout with ${\rm TS}_{\rm hold}=11.02$, is documented
in Appendix~\ref{sec:systematics}.

\subsection{Public LAT data and regions of interest}
\label{subsec:sample}

For the event-level analyses in this work we focus first on the three clusters
that dominate the reported excess: Virgo, Fornax, and Ophiuchus. Photon data
were obtained independently from the public Fermi LAT Data Server for the
same 15.5 yr interval used in Ref.~\cite{Fan2026}, from MET 246823875 to
736304518, corresponding to 2008 October 27 through 2024 May 2. The adopted
ROI centers, redshifts, and radii are listed in Table~\ref{tab:top3data}.
These are the values used in the original analysis and in our replication.

\begin{table}[tbp]
\centering
\caption{Top-three cluster sample used for the event-level reanalysis. The
radius is the circular extraction radius used in the baseline analysis.}
\label{tab:top3data}
\begin{ruledtabular}
\begin{tabular}{lrrrr}
Cluster & RA [deg] & Dec. [deg] & $z$ & $\theta_{\rm ROI}$ [deg]\\
\hline
Virgo     & 187.704 &  12.391 & 0.0038 & 3.47\\
Fornax    &  54.669 & -35.310 & 0.0046 & 3.02\\
Ophiuchus & 258.111 & -23.363 & 0.0280 & 1.56\\
\end{tabular}
\end{ruledtabular}
\end{table}

Two energy selections were downloaded. A 20--70 GeV sample is used for
the narrow-line reconstruction and the redshift-coherence analysis, while a
broader 2.5--450 GeV sample is used to test continuum and cosmic-ray
alternatives. The latter range matches the systematic background-model study of
Ref.~\cite{Fan2026}. The remaining ten clusters of the published sample are not
refit here; a common event-level likelihood over all thirteen systems is left
to future work.

\subsection{Event selection}
\label{subsec:events}

The downloaded photon files were processed with the Fermitools
\texttt{gtselect} task. We require the P8R3 ULTRACLEAN event class
($\texttt{evclass}=512$), EDISP1+EDISP2+EDISP3
($\texttt{evtype}=896$), and zenith angle $<90^\circ$. The same MET interval
and ROI radii are imposed again locally. We then apply the spacecraft-quality
selection with \texttt{gtmktime}. Ref.~\cite{Fan2026} explicitly uses
\begin{equation}
 (\texttt{DATA\_QUAL}=1)\land(\texttt{LAT\_CONFIG}=1),
 \label{eq:strictgti}
\end{equation}
and this strict selection is therefore adopted for the final reproduction.
For comparison we also evaluate the commonly used
$\texttt{DATA\_QUAL}>0$ selection; it changes the broad-band sample by only
one event and leaves the 40--46 GeV count unchanged.

After the strict good-time cut, the merged top-three 2.5--450 GeV sample
contains 6390 photons and exactly 44 photons between 40 and 46 GeV, matching
the published event count. We construct a mission-long livetime cube with
\texttt{gtltcube} and ROI-averaged exposures with \texttt{gtexpcube2}, using
P8R3\_ULTRACLEAN\_V3 and EDISP1+2+3. The same response products are used for
the broad-band continuum fits and for the exact line-response calculation.

\subsection{Response treatment}

The final line fits use the non-Gaussian P8R3\_ULTRACLEAN\_V3 energy
response, separately evaluated for each selected EDISP type and weighted by
the mission-long incidence-angle distribution and exposure. Cluster redshifts
enter the line energy before the responses are combined. The response
construction is given in Appendix~\ref{subsec:response}; the approximate
central 68\% reconstructed-energy interval for a 43.2 GeV source-frame line
is 40.42--44.50 GeV. Spatial blurring is treated separately through the
PSF-convolved template fractions and the containment-radius variations of
Sec.~\ref{subsec:radialpsf}. An exact energy-dispersion kernel does not remove
uncertainty in the spatial response.

\subsection{Statistical and response conventions}
\label{subsec:conventions}

The baseline spectral reconstruction, hadronic test, event partitions,
historical holdout statistic, and radial analyses use strict quality selection
and exact P8R3 line responses. The phenomenological broad-band
fits retain Gaussian line kernels and are labeled as secondary diagnostics.
These analyses are not interchangeable.

Observed statistics and their calibrations are also distinct. We do not use
the legacy Gaussian-response holdout simulations to assign a p-value to the
exact-response holdout, nor an asymptotic EDISP-energy estimate as a formal
consistency probability. Non-nested radial likelihood gaps are calibrated
with the matched conditional bootstrap of Sec.~\ref{subsec:radialcalibration}
and Appendix~\ref{subsec:trials}, not a Gaussian-significance mapping.
Robustness checks explored after inspecting the data are not combined into a
new global discovery significance.

\subsection{Conditional spectral likelihoods}
\label{subsec:likelihood}

Within a given energy interval, the basic background shape is a normalized
power law,
\begin{equation}
 b(E\mid\gamma)=\frac{E^\gamma}
 {\int_{E_{\min}}^{E_{\max}}E'^\gamma\,\dd E'}.
 \label{eq:bkgPL}
\end{equation}
In the exposure-weighted fits, $E^\gamma$ is multiplied by the appropriate
energy-dependent exposure before normalization. This describes a locally
smooth count spectrum, not a full spatial model of the diffuse sky.
For a line fraction $f\ge0$, the conditional unbinned probability density is
\begin{equation}
 p(E)=(1-f)b(E\mid\gamma)+f s(E\mid E_0).
 \label{eq:mixtureline}
\end{equation}
Because the overall normalization is free under both hypotheses, conditioning
on the observed event count removes a nuisance parameter without changing the
shape likelihood ratio. The local test statistic is
\begin{equation}
 {\rm TS}=2\left[\ln{\cal L}_{\rm line}-\ln{\cal L}_{0}\right],
 \label{eq:TSdefours}
\end{equation}
with the line amplitude constrained to be non-negative.

For the redshift test we retain cluster identity and replace
Eq.~(\ref{eq:mixtureline}) by a product of three cluster-specific likelihoods,
with independent line fractions and background slopes. For the broad-band
continuum tests we again merge the photons, matching the logic of the
published stacked-spectrum analysis.

The narrow-window background absorbs locally smooth foreground, point-source,
and cluster-continuum spectra; it does not assign separate catalog-source
components. A few-photon source fluctuation can nevertheless bias a radial
profile. We therefore excise catalogued-source regions and reprofile
each annular background. This masking test is distinct
from a complete spatial source-and-diffuse likelihood.

\section{Alternative Origins and Control Tests}
\label{sec:nondm}

A cluster-scale radial pattern does not uniquely identify dark matter.
We first test detector, temporal, and background nulls without assuming a
celestial line, and then examine whether conventional celestial mechanisms
can account for the narrow spectrum. The spatial tests in
Sec.~\ref{sec:morphology} provide additional
constraints on these explanations. Table~\ref{tab:nondmsummary} separates the
tests already performed from those still required.

\begin{table*}[tbp]
\centering
\caption{Status of the principal non-dark-matter explanations for the 43 GeV
candidate. ``Disfavored'' refers to the tests performed here and should not be
read as a universal theorem excluding arbitrarily contrived source or detector
models. Instrument-coordinate permutation tests, a 36-field matched-sky
Gaussian screening, and the strict-GTI source-plus-diffuse likelihood are
complete; a fully response-matched instrumental control likelihood remains
outside the present analysis.}
\label{tab:nondmsummary}
\begin{ruledtabular}
\begin{tabular}{p{0.19\textwidth}p{0.24\textwidth}p{0.42\textwidth}}
Alternative & Generic expectation & Result/status in this work\\
\hline
Hadronic CR $pp\to\pi^0\to\gamma\gamma$ & Broad continuum from pion-production kinematics & Physical forward-folded proton models leave ${\rm TS}_{\rm line}=21.49$--$26.38$ at 43.2 GeV; ordinary hadronic CR emission does not replace the line.\\
CMB/IR/optical IC & Broad continuum even for monoenergetic electrons & Only 7--8\% of 20--70 GeV IC photons fall in the exact-line central-68\% energy interval; no narrow bump is produced.\\
Deep-KN IC on EUV/X-rays & Can be line-like for an extremely cold electron population & Requires $\epsilon\gtrsim46$ eV even for a 30\% line fraction and enormous target energy density; unavoidable CMB IC dominates by factors of $10^5$--$10^6$ for a generous virial-scale X-ray field.\\
Unresolved compact sources & A special cold pair wind can in principle make a narrow IC line & The common energy in several clusters, source masking, and virial-scale extension make such a population highly contrived.\\
LAT response/systematic & A common energy-reconstruction or selection artifact can mimic a line & Exact P8R3 response and independent EDISP/FRONT/BACK partitions disfavor a simple class-specific pathology, but a common-mode instrument-coordinate effect is not yet excluded.\\
Statistical fluctuation & Localized excess with no physical counterpart & Not excluded by the present work; the historically frozen post-2015 holdout is nontrivial evidence against a transient fluctuation, but no new combined global significance is claimed.\\
\end{tabular}
\end{ruledtabular}
\end{table*}

\subsection{Instrumental and statistical nulls}
\label{subsec:nondminstrument}

Exact-response reproduction and the presence of the feature in multiple
EDISP and FRONT/BACK partitions disfavor a simple single-class pathology.
The historically fixed post-2015 holdout also provides useful temporal
evidence, but its exact-response statistic has not received a matching
background-only Monte-Carlo calibration. These checks are not independent
factors that can be multiplied into a discovery probability
(Appendix~\ref{sec:systematics}).

The calibrated detector-coordinate screening test described in
Appendix~\ref{sec:systematics} finds no significant amplitude or centroid
variation: the four-axis omnibus probabilities are 0.257, 0.119, and 0.454
for amplitude, centroid, and their combined statistic. The principal remaining
nulls are a common-mode response effect and a statistical or source-background
fluctuation. The completed 36-field matched-sky analysis applies identical
masked, fixed-window spectral screening to targets and controls. Its nominal
Cartesian tail fractions are 0.023 for an observer-frame scan and
0.014--0.016 for source-frame tests, but the stacks reuse fields and their
upper tail is sensitive to one Ophiuchus-matched field
(Appendix~\ref{subsec:blankcompleted}). This is a control-field comparison,
not a new global probability or an upper bound on instrumental contamination.
Complementary Earth-limb
or other control photons matched in incidence angle, instrument azimuth,
conversion type, rocking angle, and epoch can test detector phase-space
structure \cite{Ackermann2012,AckermannEnergyScale,FermiLAT2013Lines,FermiLAT2015}.
The cluster-model bootstrap does not substitute for these controls: it
simulates specified line-plus-background models, not unknown LAT or sky
systematics. Even after these controls, the morphology constrains the
interpretation of a candidate whose celestial origin remains unsettled.

\subsection{Physical hadronic cosmic-ray spectra}
\label{subsec:hadronic}
\label{subsec:nondmcr}

Relativistic protons accelerated by shocks and injected by galaxies and
active galactic nuclei can accumulate in the intracluster medium and collide
with its thermal gas. Neutral-pion decay then produces gamma rays, while
charged-pion decay injects electrons and positrons that contribute radio
synchrotron and inverse-Compton emission. The gas density, cosmic-ray
transport, and magnetic field jointly determine these signals
\cite{BrunettiJones2014,JeltemaKehayiasProfumo2009,JeltemaProfumo2011}.
The issue here is whether this emission can reproduce the narrow spectral
feature, rather than merely supply an extended continuum.

A phenomenological photon broken power law can create sharper curvature than
a realistic cosmic-ray spectrum because hadronic production convolves the
parent proton distribution with broad pion-production kinematics. We
therefore test the cosmic-ray alternative with an explicit physical
calculation of
\begin{equation}
 \phi_\gamma(E_\gamma)\propto
 \int \dd E_p\,N_p(E_p)v_p
 \frac{\dd\sigma_{pp\rightarrow\gamma}(E_p,E_\gamma)}{\dd E_\gamma}.
 \label{eq:hadronicconv}
\end{equation}
We use the \textsc{naima} implementation \cite{Zabalza2015} of the
Kafexhiu et al. $pp$ parameterization \cite{Kafexhiu2014}, adopting the
Pythia8 high-energy interaction model and including the standard nuclear
enhancement. We consider parent proton power-law (PL$_p$), exponentially
cutoff power-law (ECPL$_p$), and broken-power-law (BPL$_p$) spectra.

The final calculation uses the strict-GTI 2.5--450 GeV top-three sample, containing
6390 photons, including 44 between 40 and 46 GeV, and forward-folds the hadronic spectrum through
the energy-dependent exposure of each cluster. For a common cluster-frame
hadronic spectral shape $h(E)$, the merged count spectrum is
\begin{equation}
 \lambda_{\rm CR}(E_{\rm obs})\propto
 \sum_i W_i\,\epsilon_i(E_{\rm obs})(1+z_i)
 h[(1+z_i)E_{\rm obs}],
 \label{eq:crforward}
\end{equation}
where $\epsilon_i$ is the ROI-averaged exposure and the baseline weights are
$W_i=\Omega_i$, corresponding to a common surface-brightness normalization.
The overall normalization of the CR component is profiled jointly with a
smooth PL component. The narrow feature is modeled with the exact stacked
P8R3 ULTRACLEAN EDISP1+2+3 kernel derived in
Sec.~\ref{subsec:response}; no Gaussian approximation is used for the
line in this test. The likelihood is the same conditional unbinned
energy-shape likelihood over 2.5--450 GeV used for the other broad-band
comparisons.

As a baseline, a PL continuum plus the exact line gives
\begin{equation}
 {\rm TS}_{\rm line}=21.81,\qquad
 \widehat E_{\rm line}=43.2~{\rm GeV}.
 \label{eq:hadronicbaseline}
\end{equation}
Profiling a physical hadronic component does not remove the narrow residual.
For the three proton spectra we find
\begin{align}
 {\rm PL+CR}({\rm PL}_p):&\quad
 {\rm TS}_{\rm line}=21.49,\\
 {\rm PL+CR}({\rm ECPL}_p):&\quad
 {\rm TS}_{\rm line}=26.17,\\
 {\rm PL+CR}({\rm BPL}_p):&\quad
 {\rm TS}_{\rm line}=26.38.
 \label{eq:hadronicTS}
\end{align}
Thus a proton power law leaves the line likelihood improvement essentially
unchanged, while the more flexible cutoff and broken proton spectra increase
the conditional line TS after broad curvature is profiled. The fitted line
energy is consistent with 43.2 GeV across all four continuum choices; differences below the displayed precision are not physically meaningful at the LAT energy resolution.

\begin{table*}[tbp]
\centering
\caption{Final exposure-weighted hadronic fits using the exact P8R3
EDISP1+2+3 line response. Information-criterion differences are evaluated
over the complete set of line and no-line models in this subsection. The AIC
marginally prefers the ECPL$_p$+line model, whereas the BIC prefers the
simpler PL+line model; neither criterion provides evidence that the additional
hadronic degrees of freedom are required.}
\label{tab:hadronicfinal}
\begin{ruledtabular}
\begin{tabular}{lrrrr}
Model with line & ${\rm TS}_{\rm line}$ & $\widehat E_{\rm line}$ [GeV]
& $\Delta{\rm AIC}$ & $\Delta{\rm BIC}$\\
\hline
PL+line                 & 21.81 & 43.2 & 0.58 & 0.00\\
PL+CR(PL$_p$)+line      & 21.49 & 43.2 & 4.17 & 17.11\\
PL+CR(ECPL$_p$)+line    & 26.17 & 43.2 & 0.00 & 19.71\\
PL+CR(BPL$_p$)+line     & 26.38 & 43.2 & 1.48 & 27.96\\
\end{tabular}
\end{ruledtabular}
\end{table*}

\begin{figure*}[tbp]
\centering
\includegraphics[width=0.485\textwidth]{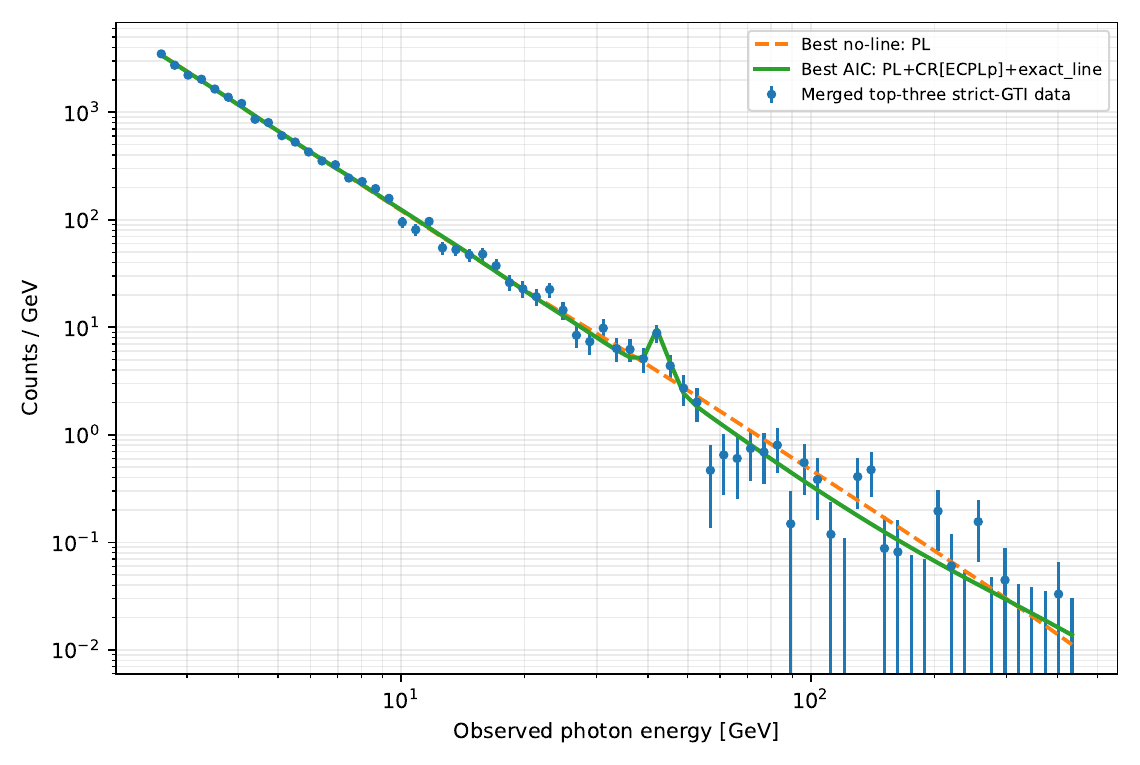}
\hfill
\includegraphics[width=0.485\textwidth]{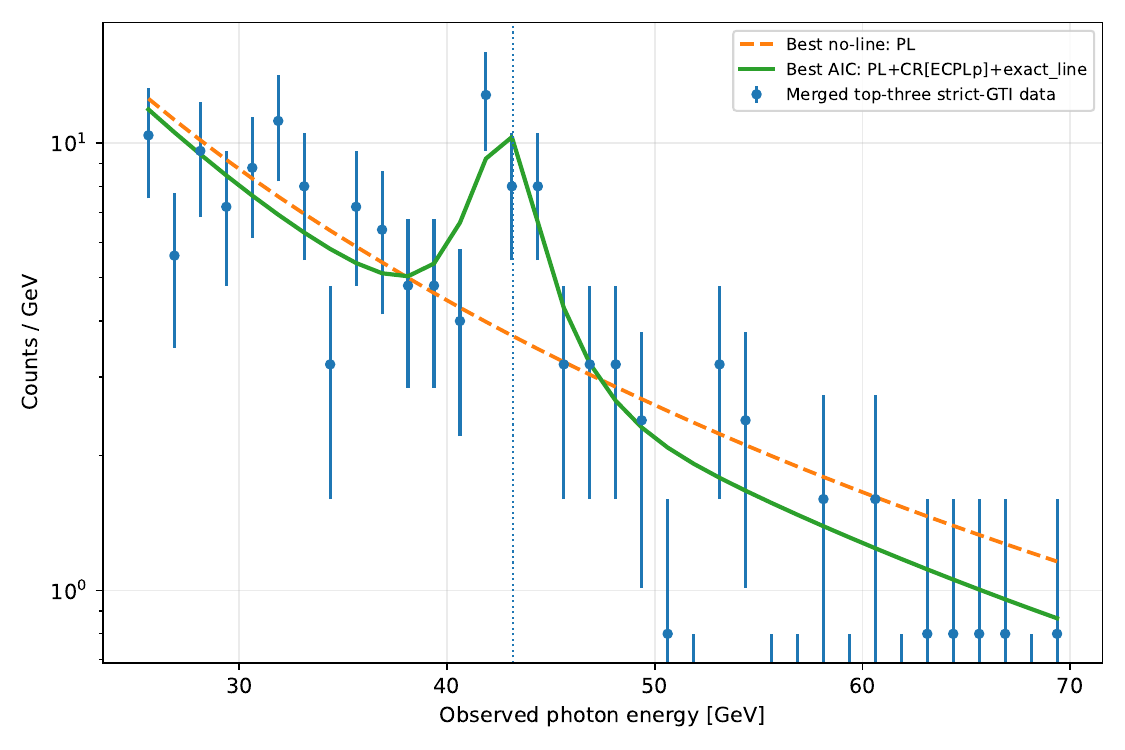}
\caption{Final physical hadronic test with energy-dependent exposure and the
exact P8R3 line kernel. Left: the full 2.5--450 GeV merged spectrum. Right:
zoom on the 25--70 GeV region containing the candidate feature. The orange
curve is the best no-line model, while the green curve is the best-AIC model
including a physical pion-decay component and the exact line. The statistical
analysis is unbinned; the binned points are shown only for visualization.}
\label{fig:hadronic_exact_spectra}
\end{figure*}

\begin{figure}[tbp]
\includegraphics[width=\columnwidth]{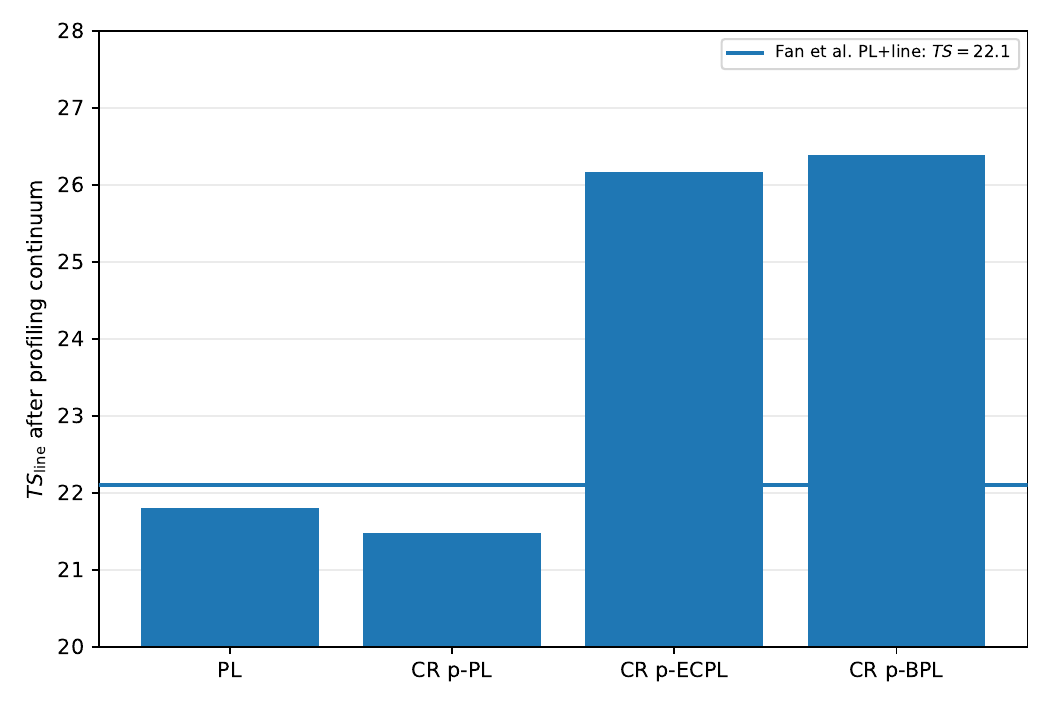}
\caption{Line likelihood improvement after profiling the broad continuum.
The PL baseline and a physical proton power law give nearly identical
${\rm TS}_{\rm line}$, while proton cutoff and broken-power-law spectra leave
an even larger narrow residual. The horizontal line shows the PL+line value reported by
Ref.~\cite{Fan2026}. The ordinate is truncated and does not begin at zero.}
\label{fig:hadronic_TS_final}
\end{figure}

The information criteria clarify the role of the extra continuum freedom.
The ECPL$_p$+line model improves the raw likelihood enough to be marginally
favored by AIC over the simpler PL+line model, but only by
$\Delta{\rm AIC}=0.58$. The stronger BIC penalty instead selects PL+line and
penalizes all CR-augmented line models by $\Delta{\rm BIC}\gtrsim17$.
Moreover, the no-line ECPL$_p$ and BPL$_p$ fits drive proton indices to the
imposed parameter boundaries, indicating that they attempt to emulate the
localized curvature with extreme broad-band shapes. Once the line is
included, the narrow component persists near 43.2 GeV
(Figs.~\ref{fig:hadronic_exact_spectra} and \ref{fig:hadronic_TS_final};
Table~\ref{tab:hadronicfinal}).

A realistic $pp\to\pi^0\to\gamma\gamma$
continuum, forward-folded with the final exposure treatment, does not provide
a satisfactory replacement for the narrow 43 GeV component. The additional
hadronic freedom can alter the broad curvature, but it does not move or
eliminate the localized residual.

\subsection{Inverse Compton emission and the deep-Klein--Nishina loophole}
\label{subsec:icline}

For an isotropic target photon field $n(\epsilon)$ and a monoenergetic electron
of Lorentz factor $\gamma$, the exact inverse-Compton photon-production kernel
can be written \cite{BlumenthalGould1970}
\begin{equation}
 \frac{\dd N_\gamma}{\dd t\,\dd E_\gamma}=
 \frac{3\sigma_T c}{4\gamma^2}
 \int \dd\epsilon\,\frac{n(\epsilon)}{\epsilon}
 F(q,\Gamma),
 \label{eq:ickernel}
\end{equation}
where $\Gamma=4\gamma\epsilon/(m_ec^2)$,
$q=E_\gamma/[\Gamma(\gamma m_ec^2-E_\gamma)]$, and
\begin{equation}
 F(q,\Gamma)=2q\ln q+(1+2q)(1-q)
 +\frac{(\Gamma q)^2(1-q)}{2(1+\Gamma q)}.
 \label{eq:icF}
\end{equation}
We use this full isotropic Klein--Nishina kernel to construct an intentionally
optimistic IC null. The electron distribution is a delta function in energy,
which is narrower than any realistic shock-accelerated or cooled population,
and for each target field the electron energy is tuned so that
$E_\gamma^2\dd N_\gamma/\dd E_\gamma$ peaks at 43.2 GeV. We compare the intrinsic IC spectra with the reconstructed line interval
before applying LAT energy dispersion; this isolates the broadening already
present in the emission kernel.

\begin{figure}[tbp]
\centering
\includegraphics[width=\columnwidth]{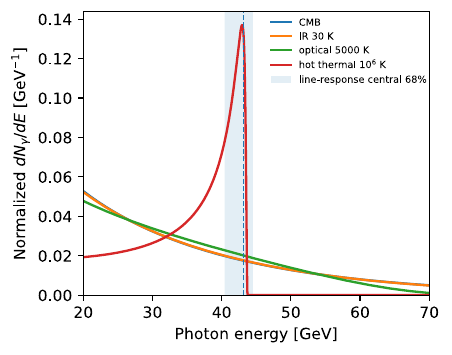}
\caption{Maximally favorable IC spectra for delta-function electron
populations, with the electron energy separately tuned for each target field
so that the energy-flux spectrum peaks near 43.2 GeV. Each curve is normalized
to unit photon number over 20--70 GeV. The shaded band is the central 68\%
reconstructed-energy interval, 40.42--44.50 GeV, for a true 43.2 GeV line in
our exact stacked P8R3 response. CMB, infrared, and optical IC remain broad and
monotonically decreasing in photon counts across the line region; omitting LAT
energy smearing makes this a conservative test.}
\label{fig:icstandard}
\end{figure}

For CMB, 30 K infrared, and 5000 K optical blackbody target fields the tuned
electron energies are approximately 2.45 TeV, 767 GeV, and 92.5 GeV. Yet only
7.24\%, 7.30\%, and 8.32\% of the IC photons between 20 and 70 GeV fall within
40.42--44.50 GeV, the central 68\% reconstructed-energy interval of the exact
43.2 GeV LAT line response used in this work. In all three cases the photon
count spectrum is monotonically decreasing through 20--70 GeV. Even an
artificial $10^6$ K thermal field reaches only a 34\% fraction. Figure
\ref{fig:icstandard} illustrates the result. Ordinary cluster IC emission on
CMB/IR/optical photons therefore cannot mimic the observed narrow feature even
when the electron distribution is made unrealistically cold.

A narrow astrophysical IC line is nevertheless possible in principle. The
130 GeV episode prompted Aharonian, Khangulyan, and Malyshev to show that a
cold ultrarelativistic electron--positron wind scattering a sufficiently hard,
quasi-monoenergetic photon field in the deep Klein--Nishina limit can transfer
most of the electron energy to a single gamma ray and generate a line-like
feature \cite{Aharonian2012ColdWind}. We therefore test an even more favorable
version of that loophole: both the electrons and target photons are taken to
be monoenergetic, and the target-photon energy is scanned from the infrared to
hard X-rays.

\begin{figure*}[tbp]
\centering
\includegraphics[width=0.485\textwidth]{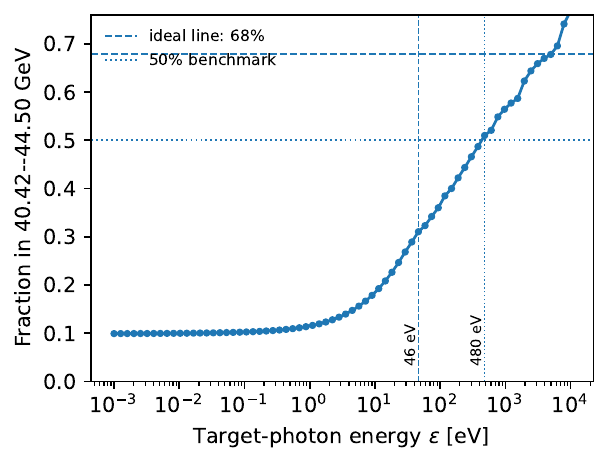}
\hfill
\includegraphics[width=0.485\textwidth]{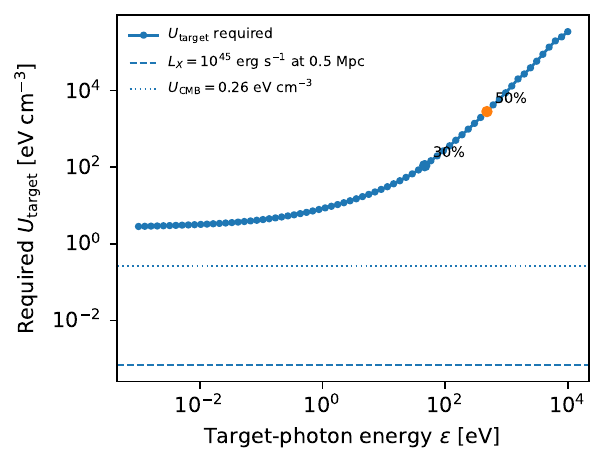}
\caption{Extreme deep-Klein--Nishina IC loophole using monoenergetic electrons
and an artificially monochromatic target photon field. Left: fraction of
20--70 GeV IC photons that fall in the exact-line central-68\% interval. The
30\% and 50\% benchmarks are first reached for target photons of approximately
46 eV and 480 eV, respectively. Right: target radiation energy density required
for the power in that narrow 43 GeV interval to equal the total CMB-IC power
of the same electrons. The dashed horizontal comparison assumes the deliberately
extreme virial-scale field obtained from $L_X=10^{45}$ erg s$^{-1}$ at
0.5 Mpc; the dotted line is the CMB energy density. The line-like KN solutions
require radiation densities many orders of magnitude above plausible
virial-scale EUV/X-ray fields.}
\label{fig:ickn}
\end{figure*}

The left panel of Fig.~\ref{fig:ickn} shows that obtaining even 30\% of the
20--70 GeV IC photons inside the exact-line central-68\% interval requires
$\epsilon\gtrsim46$ eV, with $E_e\simeq45.8$ GeV and
$\Gamma\simeq32.6$. Reaching a 50\% line fraction requires
$\epsilon\gtrsim480$ eV, $E_e\simeq43.5$ GeV, and
$\Gamma\simeq320$: the target photons must be EUV/soft X-rays and the
scattering must be deep in the KN regime.

The energetics pose a severe difficulty for a virial-scale cluster source. The same
$\sim44$--46 GeV electrons inevitably scatter the CMB. For each
monoenergetic target field we therefore compute the target photon energy
density required for the power deposited in the 40.42--44.50 GeV band to equal
the \emph{total} CMB-IC power from the same electrons. At the 30\% and 50\%
line-fraction thresholds the required densities are
\begin{equation}
 U_{\rm target}\simeq 1.09\times10^2
 \quad\hbox{and}\quad
 2.80\times10^3~{\rm eV\,cm^{-3}},
 \label{eq:icUreq}
\end{equation}
respectively. For comparison, even assigning an extreme cluster X-ray
luminosity $L_X=10^{45}$ erg s$^{-1}$ to a radius of only 0.5 Mpc gives
$U_X\simeq6.96\times10^{-4}$ eV cm$^{-3}$. The required radiation field is
therefore larger by factors of $1.6\times10^5$ and $4.0\times10^6$ in the two
benchmark cases. Equivalently, if such an X-ray field powered the observed
43 GeV IC component, the unavoidable CMB scattering from the same electrons
would carry those factors more power, peaking at roughly 6--7 MeV. The right
panel of Fig.~\ref{fig:ickn} displays this energetic mismatch.

The measured extension adds a spatial constraint to this energetic argument. EUV and X-ray
radiation fields capable of driving the KN process are concentrated around the
hot intracluster gas and luminous galaxies, whereas the line-like support in
this work extends through a large fraction of $R_{200}$ after catalogued gamma
sources are masked. An IC explanation would thus have to arrange simultaneously
(i) an extraordinarily narrow $\sim44$ GeV electron population in multiple
clusters, (ii) an intense EUV/soft-X-ray photon bath extending to virial scales,
(iii) suppression of the vastly larger CMB-IC component, and (iv) nearly the
same output energy in Virgo, Fornax, and Ophiuchus. This combination makes the tested conventional leptonic explanation
implausible on virial scales. The
cold-wind mechanism remains an existence proof that astrophysics can make a
narrow gamma-ray line, but not a credible model for the extended cluster
feature observed here.

\subsection{Unresolved compact sources}
\label{subsec:nondmsources}

A population of unresolved AGN, pulsars, or cluster-member galaxies is a more
general astrophysical alternative. Such sources can certainly generate hard
gamma-ray continua, and the cold-pair-wind mechanism just discussed shows that
special compact objects can even generate narrow IC features in principle
\cite{Aharonian2012ColdWind}. The requirements here are substantially more
restrictive, however. The same characteristic energy must recur in several
clusters; the signal must survive 4FGL-DR4 masking; and the aggregate radial
profile must remain extended over $0.2$--$1.0R_{200}$ rather than tracing a
small number of bright central objects. The disappearance of the small central
Virgo contribution after masking M87 is consistent with ordinary point-source
contamination being removed, while the extended component persists. We cannot
mathematically exclude a finely tuned population of unresolved line emitters,
but no known cluster source class naturally predicts such a common narrow
energy and virial-scale distribution.

Catalog masking rules out the excised regions as the sole origin of the broad
pattern. It does not rule out unresolved-source shot noise, a transient absent
from the 14-year catalog during the 15.5-year photon interval, or diffuse
foreground mismodeling. In particular, narrow-window power laws can absorb
smooth spectral curvature but do not test spatial errors in the Galactic
diffuse model. The complete likelihood of
Sec.~\ref{subsec:fullskylike} therefore treats Galactic and isotropic diffuse
emission, point sources, and catalogued extended sources simultaneously. Its
result retains a modest broad component and shows no conspicuous
Ophiuchus-specific foreground residual, although unresolved-source
fluctuations and diffuse-model systematics cannot be eliminated.

\section{The Candidate in a Complete Sky Model}
\label{sec:fullsky}
\label{subsec:fullskylike}

The tests of the previous section treat the background locally: a
smooth spectral continuum inside an aperture, with catalogued sources removed
by masking. That treatment is transparent about which photons drive a
preference, but it cannot say whether a structured Galactic foreground, or the
fitted wings and spectra of nearby catalog sources, can account for the same
photons. This section replaces it with an explicit model of each cluster field.
It is the principal quantitative result of this work, and the analysis to which
we assign the greatest interpretive weight.

The principal sky-model test replaces source masking and local spectral
backgrounds by an explicit model of each cluster field. We analyze a
$10$--$100$ GeV, $8^\circ$-radius region around Virgo, Fornax, and Ophiuchus
with separate EDISP1, EDISP2, and EDISP3 likelihood components. Each component
uses its matching P8R3 ULTRACLEAN isotropic template; the sky model also
contains \texttt{gll\_iem\_v07}, all 4FGL-DR4 sources in the model region, and
the catalogued extended-source templates. Nearby catalog-source
normalizations and the Galactic and isotropic diffuse normalizations are
profiled. The candidate is fixed at 43.2 GeV in the source frame and folded
through energy dispersion. Point-source, uniform-disk, NFW-annihilation,
NFW-decay, and substructure templates are each compared with the same
optimized no-line baseline.

The event preparation is deliberately stricter than relying on Fermipy's
internal file bookkeeping. We first apply \texttt{gtselect} with
\texttt{evclass=512}, the individual EDISP event-type bit, the frozen MET
range, and $z_{\max}=90^\circ$, followed by \texttt{gtmktime} with
Eq.~(\ref{eq:strictgti}). Those already filtered files are then supplied as
the component inputs. Before fitting, the pipeline verifies by
\texttt{RUN\_ID}/\texttt{EVENT\_ID} that every binned component is a subset of
its strict-GTI seed and checks that its GTI ontime does not exceed the seed.
This audit prevents a pre-GTI intermediate file from entering either the
counts or exposure calculation.

\begin{figure*}[tbp]
\centering
\includegraphics[width=0.96\textwidth]{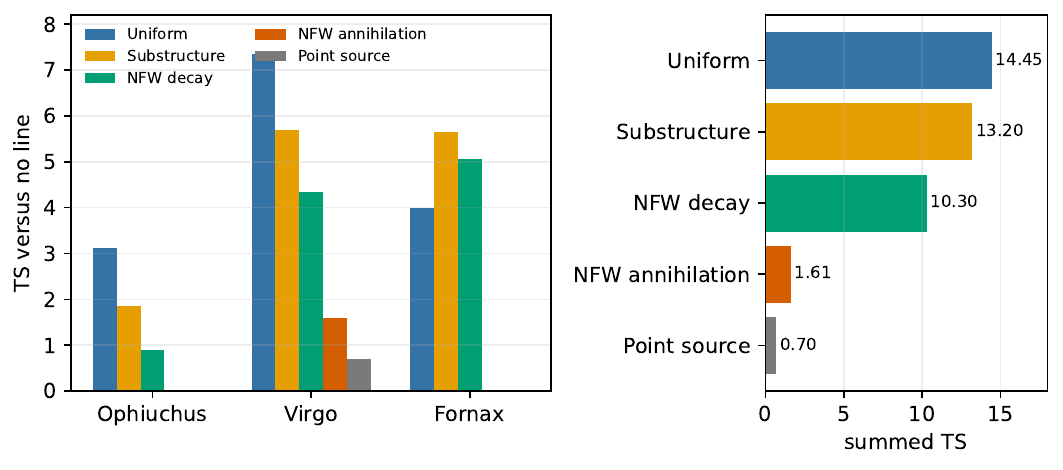}
\caption{Strict-GTI full-sky likelihood comparison. Left: per-cluster TS
relative to the common no-line baseline. Right: sum of the three independently
profiled TS values. Uniform brightness and substructure are nearly tied in the
sum, whereas smooth NFW annihilation and a central point source receive little
support. The bars compare fixed non-nested morphologies and therefore do not
carry pairwise Gaussian significances.}
\label{fig:fullskylikelihood}
\end{figure*}

All nine strict-GTI EDISP components pass the event-identity and ontime audit:
no component contains an event outside its filtered seed. Every null and line
fit converges with MINUIT quality 3 and status 0. Table~\ref{tab:fullskyresults}
and Fig.~\ref{fig:fullskylikelihood} give the likelihood results. Virgo and
Ophiuchus individually prefer uniform brightness, whereas Fornax prefers the
substructure template. Summing the independently profiled cluster likelihoods
gives ${\rm TS}=14.45$ for uniform brightness, 13.20 for substructure, 10.30
for NFW decay, 1.61 for smooth NFW annihilation, and 0.70 for central point
sources. Thus the explicit sky model retains a modest extended component but
does not reproduce the much larger TS of the conditional stacked-spectrum
reconstruction. It independently confirms the important spatial statement:
the support is broad, not centrally concentrated.

This conclusion is not driven by the low-latitude Ophiuchus field. Omitting
Ophiuchus leaves summed Virgo-plus-Fornax values of 11.33 for uniform
brightness, 11.34 for substructure, 9.40 for NFW decay, and 1.61 for smooth
NFW annihilation. The two broad templates are then indistinguishable, but the
separation from smooth annihilation remains.

For the uniform template, the local fixed-energy probability is computed from
the three-parameter boundary distribution
$2^{-3}\sum_{k=0}^{3}{3\choose k}\chi_k^2$, with a point mass for $k=0$.
This is the chi-bar-square form for likelihood-ratio tests with parameters on
the boundary of their allowed domain \cite{SelfLiang1987}.
The summed ${\rm TS}=14.45$ gives $p_{\rm local}=6.21\times10^{-4}$, or
$3.23\sigma$ one-sided. Section~\ref{subsec:globalcalibration} calibrates the
distinct maximum over the energy-and-morphology search. The non-nested
morphology gaps nevertheless remain conditional likelihood rankings rather
than Gaussian significances. In particular,
uniform brightness exceeds substructure by only
$2\Delta\ln\mathcal L=1.25$ in the three-cluster sum, so the data select a
broad family rather than a unique flat profile. The more consequential gap is
between that broad family and smooth NFW annihilation,
$2\Delta\ln\mathcal L=12.84$ for uniform brightness.

\begin{table*}[tbp]
\centering
\caption{Strict-GTI source-plus-diffuse spatial--spectral likelihood at the
fixed source-frame energy 43.2 GeV. Entries are TS values relative to the same
optimized no-line sky model; parentheses give the fitted line flux in units of
$10^{-12}\,{\rm ph\,cm^{-2}\,s^{-1}}$. Cluster amplitudes are independently
constrained to be non-negative. ``Sum'' adds the three profile-likelihood TS
values. Non-nested differences between rows are not Gaussian significances.}
\label{tab:fullskyresults}
\begin{ruledtabular}
\begin{tabular}{lrrrr}
Template & Ophiuchus & Virgo & Fornax & Sum\\
\hline
Uniform & 3.12 (9.74) & 7.35 (14.51) & 3.98 (9.00) & 14.45\\
Substructure & 1.86 (6.80) & 5.68 (14.69) & 5.66 (8.62) & 13.20\\
NFW decay & 0.90 (4.30) & 4.35 (12.75) & 5.05 (7.61) & 10.30\\
NFW annihilation & 0.00 (0.00) & 1.59 (3.94) & 0.03 (0.41) & 1.61\\
Point source & 0.00 (0.00) & 0.70 (2.72) & 0.00 (0.00) & 0.70\\
\end{tabular}
\end{ruledtabular}
\end{table*}

As a secondary population check, we combine the saved one-dimensional line
profiles after imposing common amplitude relations. A common observed flux
gives ${\rm TS}=14.09$ for uniform brightness and 12.55 for substructure.
Scaling the smooth-annihilation fluxes with the adopted $J$-factor ratios gives
only ${\rm TS}=1.04$, while the corresponding $D$-scaled NFW-decay combination
gives ${\rm TS}=9.97$. These combinations use interpolated saved profiles,
rather than a new simultaneous fit in which every nuisance parameter is
reprofiled at each shared amplitude, and are therefore supporting diagnostics.
Their direction is nevertheless consistent with the direct independent-fit
result: tying the clusters according to smooth annihilation does not recover a
line preference.

\begin{figure*}[tbp]
\centering
\includegraphics[width=0.98\textwidth]{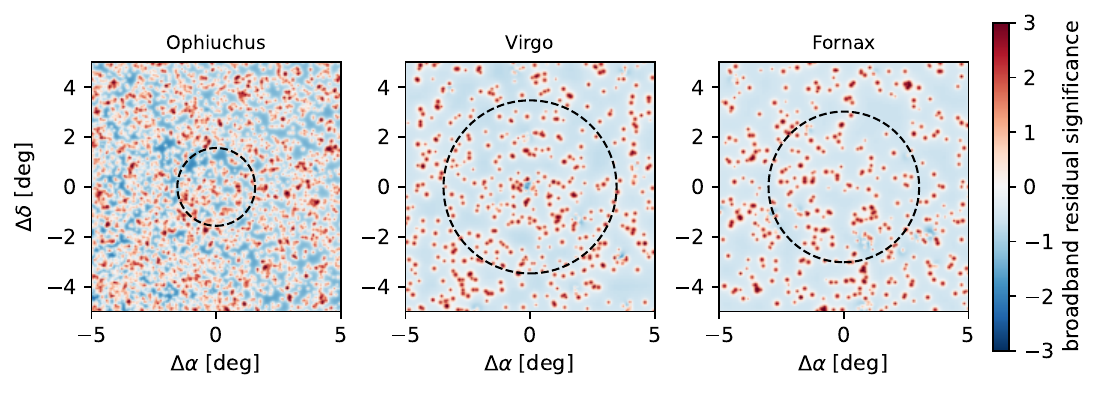}
\caption{Baseline 10--100 GeV residual-significance maps after profiling the
4FGL-DR4 sources and Galactic and isotropic diffuse components. Dashed circles
mark $R_{200}$. The broad-band residual fields show no coherent
cluster-centered structure. Because the fitted line components contain only a
few photons, the corresponding best-fit maps are visually indistinguishable
on this scale and are not shown.}
\label{fig:fullskyresiduals}
\end{figure*}

The nuisance and image diagnostics reveal no compensating fit pathology. At
the preferred morphology, the Galactic/isotropic normalizations are
0.844/1.811 for Ophiuchus, 1.338/0.802 for Virgo, and 0.962/0.587 for Fornax.
The elevated Ophiuchus isotropic normalization is not evidence for an
intrinsically enhanced isotropic sky. At this low Galactic latitude it can
also absorb locally mismodeled Galactic emission, so the Ophiuchus result is
the cluster most exposed to foreground-model systematics.
We therefore repeat the Ophiuchus fit with two progressively more flexible
foreground prescriptions: a free spectral tilt of the Galactic diffuse
template, and that tilt together with a cluster-centered $5^\circ$ disk whose
power-law continuum normalization and slope are profiled. At 43.2 GeV the
uniform-template TS is 3.121, 3.118, and 3.122 for the nominal, tilted, and
tilted-plus-local models, respectively; the corresponding substructure TS is
1.857, 1.853, and 1.855, and the NFW-decay TS is 0.897, 0.894, and 0.897.
At the calibrated 44.5 GeV maximum the Ophiuchus uniform-template TS is
similarly stable, changing from 7.116 to 7.122 and 7.125. The isotropic
normalization remains 1.831--1.832 in the three null fits, while the added
local component is driven to zero. Thus neither a global Galactic spectral
tilt nor this broad local continuum absorber accounts for the line preference
or changes its morphology ordering. This check does not span the full
uncertainty of alternative interstellar-emission models, but it directly tests
the smooth foreground freedom most likely to be hidden by the elevated
isotropic normalization.
None lies on an allowed boundary. Across the five line morphologies, the
diffuse normalizations change by less than 1.3\% in Ophiuchus, 10\% in Virgo,
and 3.5\% in Fornax; well-constrained nearby-source normalizations are likewise
stable. Figure~\ref{fig:fullskyresiduals} shows the baseline 10--100 GeV
residual-significance maps. They contain no coherent cluster-centered
broadband excess or conspicuous large-scale failure. Adding the preferred line
template changes these decade-wide maps by less than 0.011 in residual
significance within $R_{200}$, as expected for fitted components containing
only 4.15, 5.88, and 3.61 predicted photons in Ophiuchus, Virgo, and Fornax.
The maps are therefore a background-model diagnostic, not an image of the
narrow feature.

\subsection{Global calibration of the energy-and-morphology search}
\label{subsec:globalcalibration}

The fixed-energy values above are local in both energy and spatial morphology.
We therefore calibrate the maximum obtained when the line energy and all five
spatial hypotheses are searched. For cluster $c$, source-frame energy $E$, and
morphology $m$, we define
\begin{equation}
 \begin{aligned}
 {\rm TS}_{c}(E,m)={}&2\left[\ln\mathcal L_c(\widehat\mu_c;E,m)
 -\ln\mathcal L_c(\mu_c=0)\right],\\
 &\widehat\mu_c\geq0,
 \end{aligned}
 \label{eq:globalclusterts}
\end{equation}
where the line amplitude is independent among clusters but shared by the three
EDISP components within each cluster. The statistic calibrated below is
\begin{equation}
 T_{\max}=\max_{E,m}\sum_{c\in\{\mathrm{Oph,Vir,Forn}\}}
 {\rm TS}_{c}(E,m).
 \label{eq:globalmaxstat}
\end{equation}
We scan 20--70 GeV in the source frame at 0.5 GeV spacing, include 43.2 GeV
explicitly, and test the point-source, uniform, substructure, NFW-decay, and
NFW-annihilation templates. Evaluating the maximum directly in every
simulation retains the correlations among neighboring energies and among the
five morphologies; it does not treat the 510 scan cells as independent trials.

Repeating the full Fermitools response calculation for every realization is
unnecessary because the exposure, point-spread function, energy dispersion,
and source responses do not change under the null. Starting from the fitted
no-line model, we cache the predicted background counts $b_i$ and the line
response $s_i(E,m)$ in every spatial, energy, and EDISP bin $i$. For each null
realization $r$ we draw a single complete count cube,
\begin{equation}
 n_i^{(r)}\sim {\rm Poisson}(b_i),
 \label{eq:globalpoissontoy}
\end{equation}
and use that same cube throughout its scan. At each scan point the
non-negative line amplitude maximizes
\begin{equation}
 \ln\mathcal L(\mu)=\sum_i\left[n_i^{(r)}
 \ln\!\left(b_i+\mu s_i\right)-\left(b_i+\mu s_i\right)\right],
 \qquad \mu\geq0.
 \label{eq:cachedpoissonlike}
\end{equation}
Response pixels below $10^{-9}$ of the peak are omitted, with at least
99.999\% of the predicted line counts retained in every cached response.

This accelerated likelihood fixes the background at its no-line optimum
rather than re-profiling the diffuse and catalog-source normalizations in every
toy. We validate that approximation against the complete profiled likelihood
at 43.2 GeV for every cluster and morphology. Across all 15 comparisons,
\begin{equation}
 \max_{c,m}\left|{\rm TS}^{\rm cache}_{c}
 -{\rm TS}^{\rm profile}_{c}\right|=0.105,
 \label{eq:globalcachevalidation}
\end{equation}
well below one TS unit; the cached values are uniformly slightly smaller.
Thus nuisance re-optimization has negligible effect on the fixed-energy line
statistic in the tested region. We additionally evaluate the actual scan
maximum with the complete profiled likelihood. At 44.5 GeV with uniform
surface brightness, the per-cluster TS values are 7.116 for Ophiuchus, 11.237
for Virgo, and 1.585 for Fornax, giving
\begin{equation}
 T_{\max}^{\rm profile}=19.938.
 \label{eq:globalmaximumprofilecheck}
\end{equation}
This differs from the cached statistic by only 0.133 TS. All three null and
line fits have fit quality 3. We retain the cached value below when evaluating
the Monte Carlo tail because the toy ensemble was generated and ranked with
that statistic; the full-profile calculation is its direct validation.

\begin{figure}[tbp]
\centering
\includegraphics[width=\columnwidth]{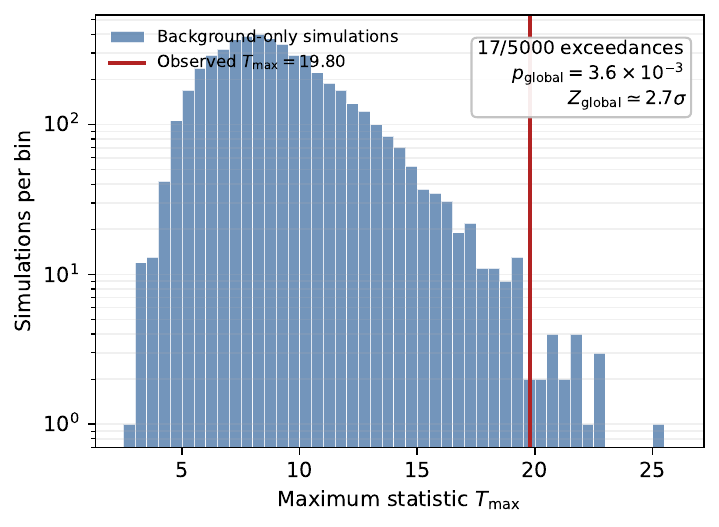}
\caption{Distribution of the maximum statistic in 5000 background-only
simulations. Each realization is scanned over the common 20--70 GeV
source-frame grid and all five morphologies. The vertical line marks the
observed maximum, $T_{\max}=19.805$; 17 simulations reach or exceed it.}
\label{fig:globalcalibration}
\end{figure}

The observed scan reaches
\begin{equation}
 \boxed{T_{\max}^{\rm obs}=19.805}
 \label{eq:observedglobalmax}
\end{equation}
at 44.5 GeV for uniform surface brightness. Of 5000 independent
background-only realizations, 17 reach or exceed this value. With the
finite-simulation estimator
\begin{equation}
 \widehat p_{\rm global}=\frac{N_{\rm exc}+1}{N_{\rm toy}+1},
 \label{eq:addoneglobalp}
\end{equation}
we obtain
\begin{equation}
 \boxed{p_{\rm global}=3.60\times10^{-3},\qquad
 Z_{\rm global}=2.69\sigma}.
 \label{eq:globalresult}
\end{equation}
The exact 95\% binomial interval is
$p_{\rm global}=(1.98$--$5.44)\times10^{-3}$, corresponding to a one-sided
Gaussian-equivalent range of $2.55$--$2.88\sigma$. The appropriate numerical
summary is therefore a conditional global significance of approximately
$2.7\sigma$. Figure~\ref{fig:globalcalibration} shows the simulated maximum
distribution and the observed threshold.

The calibration is global with respect to the stated energy and morphology
search, but remains conditional on the preselected Ophiuchus--Virgo--Fornax
sample and on the fitted null sky model. It does not reproduce the original
selection of these targets from 13 clusters or marginalize over alternative
Galactic diffuse models. The Ophiuchus foreground variations described above
test the stability of the observed statistic, but the 5000-trial global
calibration is not rerun under each variation. It also answers a different question from the
pairwise morphology calibrations below: Eq.~(\ref{eq:globalresult}) quantifies
how often the null produces a line-like maximum this prominent somewhere in
the defined search, whereas the non-nested template comparisons assess the
spatial distribution conditional on the candidate.

Although the historical target selection cannot be represented by a unique
trials factor, its likely scale is informative. If selecting the three clusters
introduced approximately 10--20 effectively independent opportunities---not
the 286 formally possible triplets---the nominal $2.7\sigma$ result would fall
to roughly $1.5$--$1.8\sigma$. This is an illustration rather than a calibrated
correction, but it shows that a conventional statistical fluctuation remains a
natural explanation of the feature.

This analysis is designed to carry greater interpretive weight than the
masked annular likelihood because it tests the candidate in the presence of
structured diffuse emission and individually modeled catalog sources.
Conversely, it is less transparent about which individual photons generate a
preference. We therefore retain the event-level analysis below as a diagnostic
of radial support, not as a substitute for the complete sky model.

\section{Event-Level Radial Support}
\label{sec:morphology}
\label{sec:physicalmorph}

The complete sky model establishes that the surviving component is
broad, but it is opaque about which photons produce that preference and over
what range of radius. We therefore retain an independent event-level analysis
with local backgrounds and explicit source masking. It is a diagnostic of
radial support rather than a competing measurement, and where the two
disagree numerically the sky-model result of Sec.~\ref{sec:fullsky} governs.

We first examine where the line likelihood resides without imposing an
annihilation or decay profile. The enlarged samples extend to $1.4R_{200}$
for Virgo, Fornax, and Ophiuchus. All annular fits use the fixed source-frame
energy defined in Sec.~\ref{sec:signal}, the strict event selection, and the
exact cluster-redshifted line response. Cumulative apertures reproduce the
published trend; disjoint annuli provide a transparent empirical diagnostic.
The primary quantitative assessment will be the complete sky likelihood of
Sec.~\ref{subsec:fullskylike}, which profiles catalog and diffuse components
rather than removing selected sky regions.

\subsection{Cumulative aperture and independent-annulus results}
\label{subsec:radialdata}

We first reproduce the cumulative aperture behavior using
$f_{\rm ROI}=0.4,0.5,\ldots,1.4$, with the same strict ULTRACLEAN
EDISP1+2+3 event selection used in the spectral analysis. The exact P8R3 line
response is used at each cluster redshift and the smooth local power-law
background is forward-weighted by the energy-dependent exposure. For the
top-three stack, the fixed-energy TS reaches
\begin{equation}
 {\rm TS}_{\rm max}=32.00
 \qquad {\rm at}\qquad
 f_{\rm ROI}=1.1 ,
 \label{eq:radialcumtop3}
\end{equation}
while Virgo alone peaks at
\begin{equation}
 {\rm TS}_{\rm max}=14.22
 \qquad {\rm at}\qquad
 f_{\rm ROI}=0.9 .
 \label{eq:radialcumvirgo}
\end{equation}
Thus the public data reproduce the qualitative behavior emphasized by
Ref.~\cite{Fan2026}: adding photons out to approximately the virial radius
strengthens the feature, whereas extending substantially beyond the virial
radius does not.

\begin{figure*}[tbp]
\centering
\includegraphics[width=0.485\textwidth]{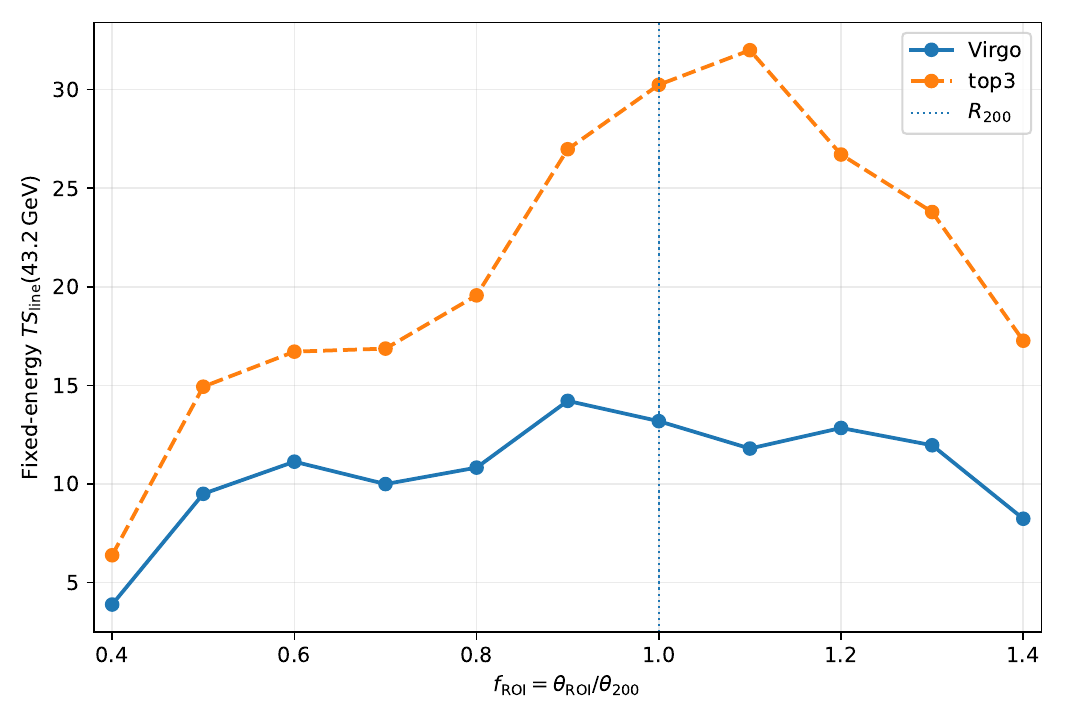}
\hfill
\includegraphics[width=0.485\textwidth]{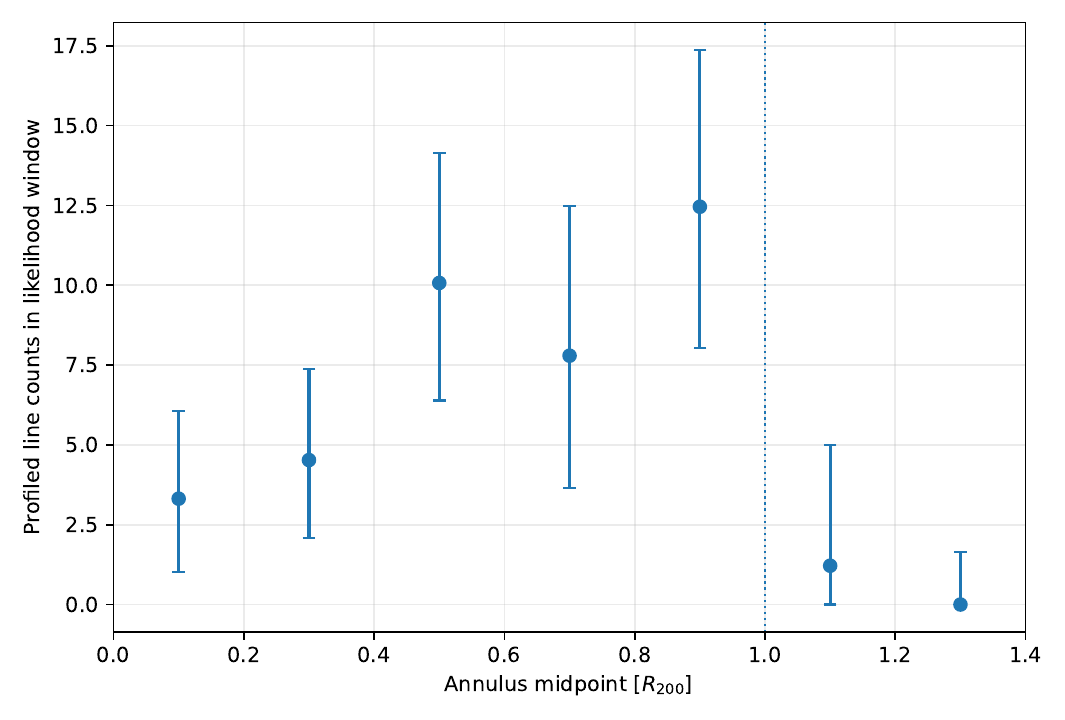}
\caption{Two complementary views of the radial behavior. Left: fixed-energy
cumulative-aperture line TS for Virgo and the top-three stack; neighboring
points are strongly correlated because the apertures are nested. Right:
profiled line counts in disjoint top-three annuli, with profile-likelihood
uncertainties. The independent-annulus representation, rather than the
cumulative TS curve, is the primary empirical morphology diagnostic.}
\label{fig:radial_data_summary}
\end{figure*}

Cumulative TS alone is not a spatial profile. We therefore divide each
cluster into the seven disjoint annuli
\[
[0,0.2],\ [0.2,0.4],\ldots,[1.2,1.4]\,R_{200}
\]
and fit the fixed-energy line amplitude independently in every annulus.
The resulting top-three profile is given in
Table~\ref{tab:annularline}. The largest independent contributions occur in
$0.4$--$0.6R_{200}$ and $0.8$--$1.0R_{200}$, with
${\rm TS}=10.64$ and 10.99, respectively. By contrast, the innermost
$0.2R_{200}$ contributes only ${\rm TS}=2.39$. No significant line excess is
present beyond $R_{200}$. Figure~\ref{fig:radial_data_summary} contrasts the
cumulative-aperture view with the independent-annulus measurement.

\begin{table}[tbp]
\centering
\caption{Independent top-three annular fits at fixed source-frame line energy
43.2 GeV. Quoted line-count intervals correspond to
$2\Delta\ln{\cal L}=1$ profile-likelihood intervals with the local
power-law slope profiled.}
\label{tab:annularline}
\begin{ruledtabular}
\begin{tabular}{lrrr}
Annulus [$R_{200}$] & ${\rm TS}_{\rm line}$ &
$\widehat N_{\rm line}$ & 68\% interval\\
\hline
0.0--0.2 &  2.39 &  3.31 & 1.02--6.07\\
0.2--0.4 &  4.39 &  4.52 & 2.09--7.37\\
0.4--0.6 & 10.64 & 10.07 & 6.39--14.14\\
0.6--0.8 &  4.15 &  7.79 & 3.65--12.48\\
0.8--1.0 & 10.99 & 12.46 & 8.03--17.36\\
1.0--1.2 &  0.14 &  1.22 & 0.00--5.01\\
1.2--1.4 &  0.00 &  0.00 & 0.00--1.64\\
\end{tabular}
\end{ruledtabular}
\end{table}

Normalizing the independently fitted line counts inside $R_{200}$ gives the
empirical cumulative fractions
\begin{equation}
 \begin{aligned}
 &F_{\rm data}(<0.2,0.4,0.6,0.8R_{200})\\
 &\qquad\simeq(0.087,\ 0.205,\ 0.469,\ 0.674).
 \end{aligned}
 \label{eq:empirical_radial_cdf}
\end{equation}
This is far more extended than a smooth NFW-squared annihilation profile.

\subsection{Survival under catalogued-source masking}
\label{subsec:pointsources}

An extended annular profile need not be diffuse cluster emission: catalogued
sources projected into those annuli can supply individual line-band photons.
We therefore cross-match the enlarged fields against 4FGL-DR4
\cite{Ballet2023DR4} and repeat the fits after circular source masks of
$0.15^\circ$, $0.25^\circ$, and $0.35^\circ$. For every cluster--annulus cell
we recompute the retained solid angle, update the exposure-weighted stacked
line response, and reprofile the local background slope. The source inventory,
sky maps, and source-centered diagnostics are retained in
Appendix~\ref{app:sourcechecks}.

\begin{figure*}[tbp]
\centering
\includegraphics[width=0.485\textwidth]{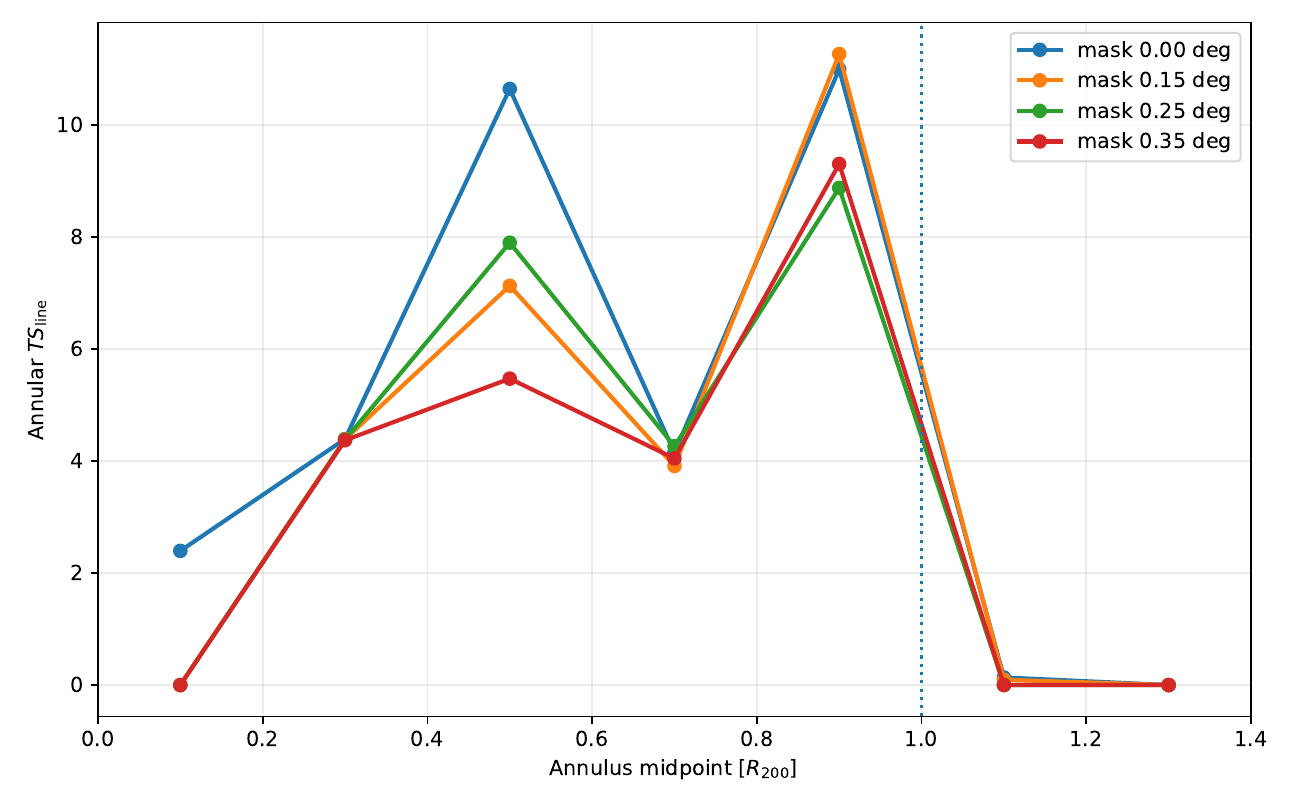}
\hfill
\includegraphics[width=0.485\textwidth]{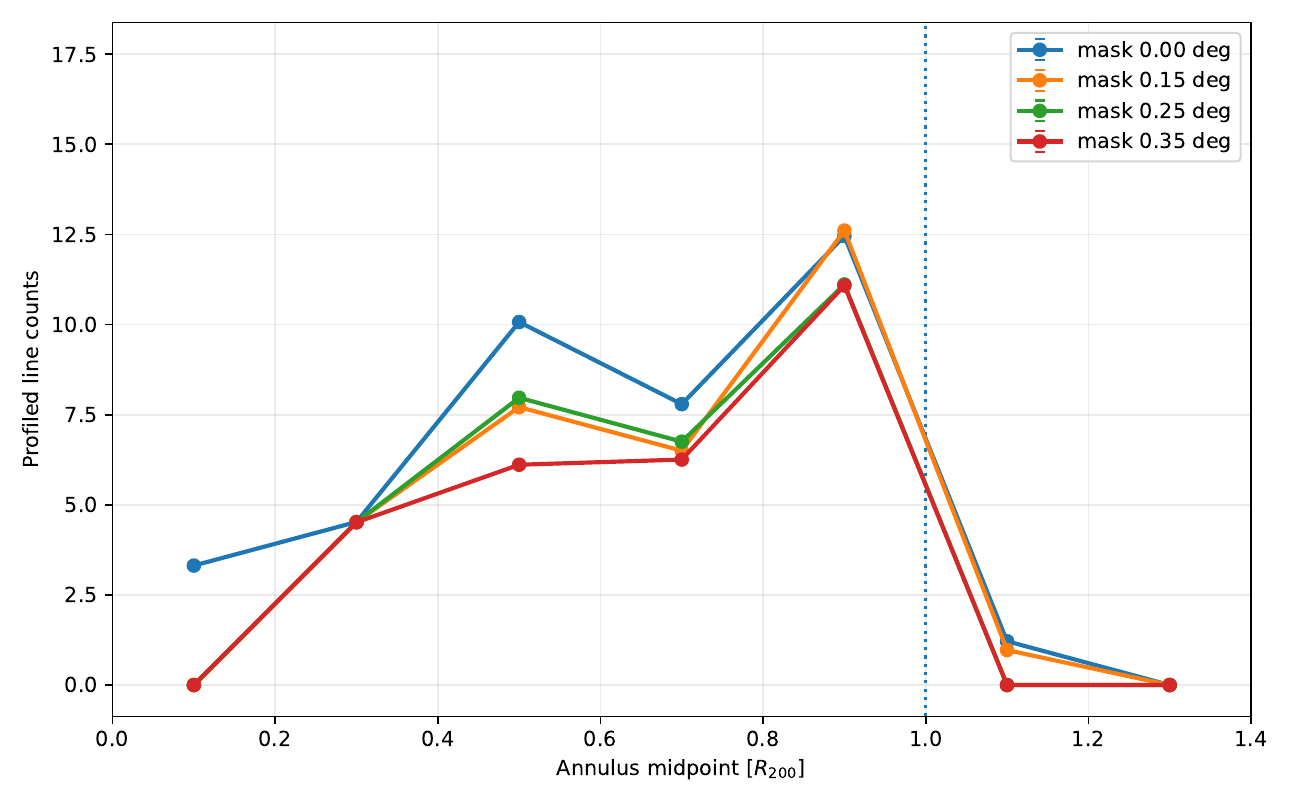}
\caption{Robustness of the independent radial excess to 4FGL-DR4 masking.
Left: annular line TS as a function of radius for four point-source mask
radii. Right: the corresponding profiled line counts. Both panels show
point estimates without error bars. The
strongest $0.8$--$1.0R_{200}$ contribution is stable under aggressive masking,
while the $0.4$--$0.6R_{200}$ contribution is reduced but not removed.}
\label{fig:psmasking}
\end{figure*}

The main result is shown in Table~\ref{tab:psmaskTS} and
Fig.~\ref{fig:psmasking}. The innermost annulus loses its line support under
even the smallest mask; source-centered fits associate much of that small
central Virgo contribution with M87. Masking also reduces, but does not
remove, the $0.4$--$0.6R_{200}$ excess. The remaining annuli between
$0.2R_{200}$ and $R_{200}$ retain line support, including
${\rm TS}=9.30$ in the outermost interior annulus under the conservative
$0.35^\circ$ mask. Thus the broad radial pattern is not supplied solely by
the excised catalogued-source regions.

\begin{table*}[tbp]
\centering
\caption{Fixed-energy annular line TS after masking every 4FGL-DR4 source.
The final column gives the fitted line counts and 68\% profile-likelihood
interval for the most conservative $0.35^\circ$ mask.}
\label{tab:psmaskTS}
\begin{ruledtabular}
\begin{tabular}{lrrrrr}
Annulus [$R_{200}$] & No mask & $0.15^\circ$ & $0.25^\circ$ &
$0.35^\circ$ & $\widehat N_{\rm line}(0.35^\circ)$\\
\hline
0.0--0.2 & 2.39 & 0.00 & 0.00 & 0.00 & $0.00^{+0.77}_{-0.00}$\\
0.2--0.4 & 4.39 & 4.39 & 4.38 & 4.37 & $4.51^{+2.84}_{-2.43}$\\
0.4--0.6 & 10.64 & 7.12 & 7.89 & 5.47 & $6.11^{+3.45}_{-2.99}$\\
0.6--0.8 & 4.15 & 3.91 & 4.26 & 4.04 & $6.25^{+3.91}_{-3.42}$\\
0.8--1.0 & 10.99 & 11.26 & 8.87 & 9.30 & $11.08^{+4.71}_{-4.23}$\\
1.0--1.2 & 0.14 & 0.09 & 0.00 & 0.00 & $0.00^{+3.13}_{-0.00}$\\
1.2--1.4 & 0.00 & 0.00 & 0.00 & 0.00 & $0.00^{+1.20}_{-0.00}$\\
\end{tabular}
\end{ruledtabular}
\end{table*}

The apparent local minimum of the unmasked TS near $0.7R_{200}$ should not
be interpreted as evidence for a physical radial gap. The
$0.6$--$0.8R_{200}$ annulus contains 12 photons between 40 and 46 GeV, the
same number as the stronger $0.8$--$1.0R_{200}$ annulus, but has a larger
total event count in the fitted window and a less favorable arrangement of
individual photon energies relative to the exact line kernel. Its fitted
line count, $7.79^{+4.69}_{-4.14}$, overlaps that of the neighboring outer
bin, $12.46^{+4.90}_{-4.43}$, and its TS remains approximately four for
every source mask. The jagged TS profile is therefore consistent with
ordinary few-photon spectral fluctuations superposed on a smooth extended
source.

\subsection{The virial-scale extent and its limitations}

Both independent annuli outside $R_{200}$ are consistent with no line, before
and after masking. Together with the surviving interior signal, this locates
the observed support on approximately virial scales. It does not measure an
infinitely sharp physical edge: the bin width is $0.2R_{200}$, the counts are
small, and the assumed centers and virial radii have uncertainties. Likewise,
fitted annular line counts are conditional spectral components, not a catalog
of photons identified individually as line emission.

Masking supplies a specific robustness test, not a complete source-background
model. It does not exclude sub-threshold or late-time transient sources,
extended-source leakage, or spatially structured Galactic foregrounds. The
small catalog-source count around low-latitude Ophiuchus is not evidence of
an intrinsically clean field. These remaining backgrounds are discussed in
Sec.~\ref{sec:nondm}. We apply the same masks to the physical templates
as to the data, so that excision is included in the spatial likelihood itself.

The empirical result motivates a comparison of centrally concentrated and
shallow templates. We distinguish the radial shape of each cluster from the
relative signal normalization across clusters. The numerical comparisons
use both; they are conditional tests of the specified joint models.

\subsection{Smooth annihilation and decay templates}

We adopt an NFW density profile
\cite{Navarro1997},
\begin{equation}
 \rho(r)=
 \frac{\rho_s}
 {(r/r_s)(1+r/r_s)^2},
 \qquad
 r_s=\frac{R_{200}}{c_{200}},
 \label{eq:NFW}
\end{equation}
with the halo parameters used in Ref.~\cite{Fan2026}.  For
velocity-independent annihilation the projected surface brightness is
\begin{equation}
 T_s(\theta)
 \propto
 \int_{\rm los}\rho^2[r(s,\theta)]\,\dd s,
 \label{eq:Ts}
\end{equation}
whereas for decay
\begin{equation}
 T_d(\theta)
 \propto
 \int_{\rm los}\rho[r(s,\theta)]\,\dd s.
 \label{eq:Td}
\end{equation}
The difference is substantial: smooth annihilation is strongly concentrated
toward the cluster center, while decay remains sensitive to mass at large
radii.

The templates used in the likelihood are normalized over the adopted
analysis region,
\begin{equation}
 \int_{\Delta\Omega_i} T_{H,i}(\bm{\theta})\,\dd\Omega=1,
 \label{eq:templatenorm}
\end{equation}
and are convolved with an adopted LAT point-spread-function model
before comparison with the data.  The PSF convolution is particularly
important in the innermost radial bins, but the virial angular radii of the
three leading targets are degree-scale, so the distinction between a
centrally concentrated profile and emission extending to $R_{200}$ is not
erased by the instrumental response.

\subsection{Substructure-enhanced annihilation}

Subhalos produce a qualitatively more extended annihilation morphology.
Following the radial form used by Ref.~\cite{Fan2026}, which is closely
related to the Phoenix-based cluster parameterizations discussed by
Refs.~\cite{Gao2012,Han2012}, the projected subhalo component has the shape
\begin{equation}
 T_{\rm sub}(\theta)
 \propto
 \frac{1}
 {\theta_{200}^2+16\theta^2},
 \qquad \theta\leq\theta_{200}.
 \label{eq:Tsub}
\end{equation}
The overall normalization is conventionally expressed through a
substructure boost $b_{\rm sh}$.  Since both the normalization and radial
profile of substructure are subject to substantial systematic uncertainty,
we use fixed benchmark values here and do not infer $b_{\rm sh}$ or fit
the subhalo concentration.  More recent reviews
emphasize the large model dependence of annihilation boosts
\cite{Ando2019}.

Smooth NFW
annihilation is much more central than decay, while the adopted boosted
substructure profile is shallower still. The velocity-dependent Jeans proxy
is an illustration, not a separately calibrated
event-level hypothesis.

\subsection{Event-level comparison of physical radial templates}
\label{subsec:radialtemplates}

To quantify the morphology without fitting the derived annular points, we
perform a second likelihood analysis directly on the photon energies in each
cluster--annulus cell. Our primary comparison uses the conservative
$r_{\rm mask}=0.35^\circ$ source mask. Every 4FGL-DR4 source close enough to
intersect the enlarged $1.4R_{200}$ analysis region is excised from the data,
and the identical circular masks are applied to Monte-Carlo photons drawn
from each PSF-convolved spatial template. The resulting retained template
fractions therefore incorporate the actual source geometry rather than being
approximated by a simple loss of solid angle. For a morphology hypothesis
$H$, the expected line counts are
\begin{equation}
 \mu_{ij}=A\,X_i^{(H)}\,f_{ij}^{(H)}
 \,\epsilon_i(E_{{\rm line},i}),
 \label{eq:radialweights}
\end{equation}
where $A$ is one global line normalization, $f_{ij}^{(H)}$ is the
PSF-convolved fraction of the cluster signal in annulus $j$, and
$\epsilon_i$ is the cluster exposure. For smooth annihilation,
$X_i=J_i$; for decay we use $X_i\propto M_{200,i}/D_{A,i}^{\,2}$; and for
substructure-enhanced annihilation we combine the smooth $J_i$ with the
cluster-dependent boost factors of Ref.~\cite{Fan2026}. An independent
power-law slope is profiled in each cluster--annulus cell. The line response
is the exact P8R3 ULTRACLEAN EDISP1+2+3 kernel at the appropriate cluster
redshift.

We also fit a constant-surface-brightness disk inside $R_{200}$ as an empirical
benchmark for a very shallow projected source. Its uniform intensity and sharp
edge are not generic predictions of subhalos, unresolved sources, or diffuse
cluster emission. It is a reference shape, not an identified emission mechanism.

The substructure normalization requires care: the printed projected
expression in Ref.~\cite{Fan2026} integrates to $b_{\rm sh}J_{\rm NFW}/16$,
whereas the boost-normalized convention integrates to
$b_{\rm sh}J_{\rm NFW}$. We test both conventions rather than assume which
was used in the published numerical implementation. The derivation and its
normalization implications are given in Appendix~\ref{app:subnorm}.

This construction deliberately conditions on the photons retained after
masking and gives each annulus a simple local spectral background. It is
therefore useful for locating line support, but its likelihood gaps should not
be read as the final ranking once the structured sky background and catalog
sources are modeled explicitly.

\subsection{Joint model ranking}

\paragraph{Event-level conditional result.}

For our baseline PSF choice, $r_{68}=0.12^\circ$, the masked
event-level likelihood gives the ordering in Table~\ref{tab:radialtemplates}.
The empirical uniform profile gives the largest likelihood and retains
${\rm TS}=12.55$ relative to the no-line hypothesis. The boost-normalized
substructure profile is the best-performing physical dark-matter template,
with
\begin{equation}
 -2\Delta\ln{\cal L}_{\rm sub}=5.31
 \label{eq:subboostmorph}
\end{equation}
relative to uniform and ${\rm TS}=7.24$ versus no line. NFW decay is worse
than uniform by 8.09. Most strikingly, the smooth NFW-annihilation fit is
driven to the boundary at essentially zero line counts, giving
${\rm TS}=0$ and a likelihood penalty of 12.55 relative to uniform. Thus
excising catalogued source regions does not restore support for a centrally
concentrated line component; the surviving line likelihood is preferentially
associated with shallow, extended morphologies.

\begin{table}[tbp]
\centering
\caption{Primary event-level radial-template comparison after masking every
4FGL-DR4 source with $r_{\rm mask}=0.35^\circ$. The same masks are applied to
the data and to PSF-convolved Monte-Carlo photons from each template. The
baseline PSF is $r_{68}=0.12^\circ$. Because the hypotheses are non-nested,
the final column is a raw likelihood difference and is not mapped directly to
a Gaussian significance.}
\label{tab:radialtemplates}
\begin{ruledtabular}
\begin{tabular}{lrr}
Template & ${\rm TS}$ & $-2\Delta\ln{\cal L}$\\
\hline
uniform surface brightness & 12.55 & 0.00\\
boost-normalized substructure & 7.24 & 5.31\\
substructure, printed & 5.19 & 7.36\\
NFW decay & 4.46 & 8.09\\
smooth NFW annihilation & 0.00 & 12.55\\
\end{tabular}
\end{ruledtabular}
\end{table}

The ordering is stable to the mask radius. With a $0.25^\circ$ mask the
likelihood penalties relative to uniform are 4.38, 7.75, 9.11, and 14.59 for
boost-normalized substructure, the printed substructure prescription, decay,
and smooth annihilation, respectively; the corresponding $0.35^\circ$ values
are 5.31, 7.36, 8.09, and 12.55. The total line information decreases as the
mask becomes more aggressive---the uniform-template TS falls from 23.46 in
the unmasked sample to 14.59 and 12.55 at $0.25^\circ$ and $0.35^\circ$---so
the smaller smooth-annihilation likelihood gap after masking should not be
misread as improved support for that model. At both masked radii its fitted
line normalization is numerically consistent with zero.

The origin of the hierarchy is transparent in the underlying radial
fractions. Before masking, smooth annihilation places approximately 97\%,
92\%, and 90\% of the Virgo, Fornax, and Ophiuchus signal, respectively,
inside $0.2R_{200}$ for the baseline $0.12^\circ$ PSF. Decay is substantially
broader, and boost-normalized substructure is broader still. Applying the
catalog masks alters these fractions non-uniformly---most strongly in the
Virgo center because of M87---but does not make a smooth $\rho^2$ profile
resemble the observed virial-scale distribution. Conversely, the empirical
uniform template places most of its retained weight in the outer independent
annuli, matching the location of the surviving line support.

\subsection{Conditional bootstrap calibration}
\label{subsec:radialcalibration}

Because the principal templates are non-nested and several fits encounter
the non-negative line-amplitude boundary, we calibrate their likelihood gaps
with a conditional parametric bootstrap rather than an asymptotic Gaussian
mapping. Each pseudo-experiment preserves the observed event count in every
cluster--annulus cell, draws energies from the fitted line-plus-background
model under the specified null morphology, and refits all five templates with
the identical $0.35^\circ$ masks, masked template weights, and exact-response
likelihood used for the data. For a specified alternative $H_1$, the statistic
is $T=2[\ln{\cal L}(H_1)-\ln{\cal L}(H_0)]$; ``best'' denotes the maximum
likelihood over all five fitted templates.

Table~\ref{tab:maskedbootstrap} gives the resulting tail probabilities. We use
the finite-simulation estimate $p_{\rm MC}=(k+1)/(N+1)$, where $k$ is the
number of pseudo-experiments with $T\geq T_{\rm obs}$. Under smooth NFW
annihilation, none of 2000 trials reaches either the observed best-alternative
gap of 12.55 or the pairwise substructure gap of 7.24. The corresponding
one-sided 95\% Clopper--Pearson bound is $p<1.50\times10^{-3}$; the quoted
$p_{\rm MC}=5.00\times10^{-4}$ records the Monte-Carlo resolution rather than
a resolved tail measurement. Smooth NFW annihilation therefore shows a strong
conditional template mismatch with the masked radial distribution. Under NFW decay, 5 of 2500 trials reach the
best-alternative gap, giving $p_{\rm MC}=0.0024$. That result is driven by the
very shallow empirical alternative: for the physically more specific
substructure-versus-decay comparison, 72 of 2500 trials reach the observed
gap of 2.78, giving $p_{\rm MC}=0.0292$. Thus standard NFW decay
shows a strong conditional mismatch with the shallower best-fitting morphology at
$p_{\rm MC}=0.0024$, while the conditional substructure-over-decay preference
is only suggestive. The latter does not establish substructure annihilation.

\begin{table*}[tbp]
\centering
\caption{Conditional masked-bootstrap calibration of the non-nested radial
template comparisons. Here $k/N$ is the number of pseudo-experiments reaching
or exceeding the observed statistic and $p_{\rm MC}=(k+1)/(N+1)$. Intervals
for nonzero $k$ are two-sided 95\% Clopper--Pearson intervals for the
underlying tail probability; entries for $k=0$ give the one-sided 95\% upper
bound. The empirical uniform profile is a shallow benchmark rather than a
physical dark-matter model.}
\label{tab:maskedbootstrap}
\begin{ruledtabular}
\begin{tabular}{llrrrr}
Null $H_0$ & Alternative $H_1$ & $T_{\rm obs}$ & $k/N$ & $p_{\rm MC}$ & 95\% finite-MC range\\
\hline
smooth annihilation & best of five & 12.55 & $0/2000$ & $5.00\times10^{-4}$ & $<1.50\times10^{-3}$\\
smooth annihilation & boost substructure & 7.24 & $0/2000$ & $5.00\times10^{-4}$ & $<1.50\times10^{-3}$\\
NFW decay & best of five & 8.09 & $5/2500$ & $2.40\times10^{-3}$ & $[6.50\times10^{-4},4.66\times10^{-3}]$\\
NFW decay & uniform & 8.09 & $4/2500$ & $2.00\times10^{-3}$ & $[4.36\times10^{-4},4.09\times10^{-3}]$\\
NFW decay & boost substructure & 2.78 & $72/2500$ & $2.92\times10^{-2}$ & $[2.26\times10^{-2},3.61\times10^{-2}]$\\
\end{tabular}
\end{ruledtabular}
\end{table*}

These probabilities calibrate the baseline joint comparison conditional on
the observed cell counts, fitted null parameters, fixed energy, mask, and
adopted halo scalings. They are not probabilities that a physical model is
true, nor do they test unknown foreground or detector effects. The systematic
variants check the ranking but have not each received an independent
bootstrap calibration. The finite-Monte-Carlo intervals in the table quantify
simulation counting uncertainty, not the full uncertainty in the null model.

\paragraph{Complete sky-model result.}
The strict-GTI source-plus-diffuse likelihood of
Sec.~\ref{subsec:fullskylike} preserves this qualitative ordering under a much
more complete treatment of the fields. With independent cluster amplitudes,
uniform brightness and substructure give summed TS values of 14.45 and 13.20,
respectively, compared with 10.30 for NFW decay, 1.61 for smooth NFW
annihilation, and 0.70 for central point sources. The uniform--substructure
gap is too small to identify either profile, but the negligible smooth-halo
annihilation support independently confirms that the candidate photons do not
behave like a centrally concentrated annihilation signal. All fits converge,
the diffuse and nearby-source normalizations remain stable, and the broadband
residual maps show no cluster-centered sky-model failure. The complete
likelihood therefore supersedes the numerical event-level ranking while
supporting its central inference: any surviving component is extended and is
incompatible with the simplest smooth-annihilation morphology.

\subsection{Robustness to spatial assumptions}
\label{subsec:radialpsf}

For the primary $0.35^\circ$ masked likelihood, we repeat the fit for PSF
containment radii $r_{68}=0.08^\circ$ and $0.18^\circ$, for $c_{200}$ rescaled by 0.8 and
1.2, and for four $0.05R_{200}$ center displacements. The PSF and concentration
changes have little effect, as summarized in Fig.~\ref{fig:masked_template_systematics}.
Centering has a larger impact on the distinction between the empirical uniform
and boost-substructure templates: the two become nearly degenerate for one
displacement and substructure is mildly preferred for another. Crucially,
however, boost-dominated substructure remains preferred to NFW decay in every
variation, with a pairwise $-2\ln\mathcal{L}$ advantage of 2.21--3.49, while
the smooth-annihilation penalty remains 12.03--12.82 relative to the best
template. Thus the robust statement is the ordering: the data prefer an
extended morphology, smooth annihilation performs poorly, and conditional on
dark matter substructure ranks above decay in every tested systematic
variation. The calibrated strength of the last comparison is only suggestive,
$p_{\rm MC}=0.0292$; the stronger empirical distinction between uniform
emission and substructure is centering-sensitive.

The detailed mask-dependence and systematic-variation plots are in
Appendix~\ref{app:spatialchecks}. In particular, centering sensitivity prevents
us from treating the uniform disk as a measured physical surface-brightness
law, even though it is the best baseline benchmark.

We separately audited the lower bound on every annular background slope,
repeating the extended free-amplitude likelihood for
$\gamma_{\min}=-5,-8,-12,-20,$ and $-40$ while retaining $\gamma_{\max}=1$.
The substructure--decay contrast changes only from 5.954 to 6.075, the
uniform--decay contrast from 16.070 to 16.454, and the rest--observer energy
contrast remains 3.0572. Smooth annihilation retains zero TS and both boost
profiles select the pure-substructure endpoint throughout. The original
boundary flag is traced to the innermost Virgo cell, which contains one photon
at 22.354 GeV and has an unconstrained background-only slope near $-30.3$;
it is a sparse-cell nuisance effect rather than the source of the morphology
ordering. At $\gamma_{\min}=-40$ no morphology or boost fit remains on the
slope boundary.

\subsection{Separating morphology, cluster scaling, and boost inference}
\label{subsec:inferencescope}

Equation~(\ref{eq:radialweights}) ties the three cluster amplitudes through
$X_i^{(H)}$. Its likelihood gaps therefore contain radial and target-to-target
information. Our morphology-only follow-up instead uses
\begin{equation}
 \mu_{ij}=A_i f_{ij}^{(H)}\epsilon_i(E_{{\rm line},i}),
 \qquad A_i\geq0,
 \label{eq:freeclustermorph}
\end{equation}
and profile a separate amplitude for each cluster. We performed this refit
with the same 21 cluster--annulus cells, extended likelihood, physical window
weights, exact energy response, $0.35^\circ$ masks, and five spatial templates.
The free-amplitude TS values are 0.00, 6.49, 8.41, 12.44, and 22.56 for smooth
annihilation, decay, printed substructure, boost-normalized substructure, and
uniform brightness, respectively. Thus smooth annihilation still chooses zero
signal, while uniform exceeds boost substructure by
$2\Delta\ln\mathcal L=10.12$ and boost substructure exceeds decay by 5.95.

\begin{figure}[tbp]
\includegraphics[width=\columnwidth]{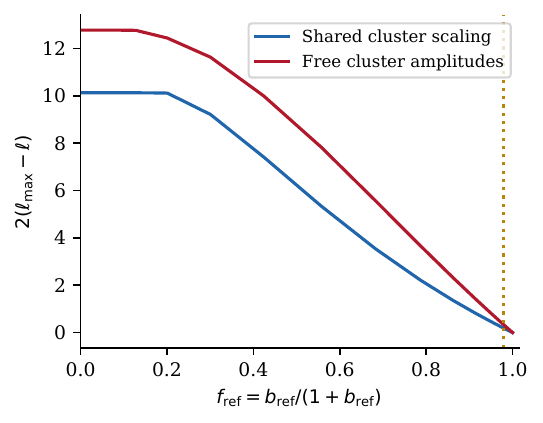}
\caption{Continuous smooth-plus-substructure profile using the same exact
energy response and masked spatial templates as the free-amplitude follow-up.
The horizontal coordinate is $f_{\rm ref}=b_{\rm ref}/(1+b_{\rm ref})$;
$f_{\rm ref}=1$ is evaluated as an exact pure-substructure endpoint, not as a
large finite boost. Both curves reach their minimum there. The vertical line
marks the Virgo-reference benchmark $b_{\rm ref}=47.56$. The ordinate is a
raw likelihood loss from each profile maximum; no confidence threshold is
assigned to it.}
\label{fig:followupboost}
\end{figure}

Plug-in parametric simulations calibrate the two decay-null contrasts for this
frozen free-amplitude protocol. For decay versus the best of five, none of
1000 trials reaches the observed statistic 16.07, giving
$p_{\rm MC}=9.99\times10^{-4}$ and a one-sided 95\% finite-simulation bound
$p<2.99\times10^{-3}$. For substructure versus decay, one of 1000 trials
exceeds 5.95, giving $p_{\rm MC}=0.001998$ and a 95\% binomial interval
$[2.53\times10^{-5},5.56\times10^{-3}]$. These are conditional rather than
global probabilities; the sparse tails are reported at their actual simulation
resolution. Table~\ref{tab:followupfits} and Fig.~\ref{fig:followupboost}
separate this follow-up from the original conditional likelihood.
Table~\ref{tab:followupfits} and Fig.~\ref{fig:followupboost} separate
this follow-up from the original conditional likelihood.

We also fit a continuous smooth-plus-substructure family, profiling the line
normalization and backgrounds at each boost. Both the shared-scaling and
free-amplitude profiles maximize at the pure-substructure endpoint,
$b_{\rm ref}\rightarrow\infty$. The fixed Virgo-reference benchmark
$b_{\rm ref}=47.56$ lies only $2\Delta\ln\mathcal L=0.187$ below the shared
maximum and 0.332 below the free-amplitude maximum. There is therefore no
finite boost measurement: order-tens boosts merely approximate the preferred
limiting shape. Even pure substructure remains below the free-amplitude
uniform template by $2\Delta\ln\mathcal L=9.78$.

\begin{table}[tbp]
\centering
\caption{Extended-likelihood follow-up at fixed source energy 43.2 GeV.
The shared and free columns use the same response, window acceptance and
background treatment; they are not the original conditional-likelihood TS
values in Table~\ref{tab:maskedbootstrap}. TS is measured against the
corresponding background-only fit, not converted into significance.}
\label{tab:followupfits}
\begin{ruledtabular}
\begin{tabular}{lrr}
Template & Shared TS & Free TS\\
\hline
Smooth NFW annihilation & 0.00 & 0.00\\
NFW decay & 6.16 & 6.49\\
Lower substructure benchmark & 7.44 & 8.41\\
Order-tens substructure benchmark & 9.94 & 12.44\\
Uniform brightness & 13.05 & 22.56\\
\end{tabular}
\end{ruledtabular}
\end{table}

\section{Observer-Frame versus Cluster-Frame Energy Coherence}
\label{app:redshift}

Energy coherence provides a complementary test of a cluster origin after
the instrumental and non-dark-matter checks in Sec.~\ref{sec:nondm}. A common
rest-frame energy is expected for either dark-matter emission or a celestial
mechanism producing the same intrinsic feature in different clusters; it
would not, by itself, identify dark matter. This diagnostic therefore probes
a different question from both the stacked line significance and the
physical morphology comparisons. Our primary result uses the event-level P8R3
energy-response kernels and independent line amplitudes for the three
clusters. We retain the earlier Gaussian-response reconstruction only as a
transparent historical cross-check; it is not used for the quoted
source-frame likelihood contrast.
\label{sec:redshift}

A genuine narrow feature produced in an extragalactic source should track the
source redshift. Conversely, an instrumental feature tied to reconstructed
energy should remain fixed in the observer frame. The original stacked
analysis incorporates the cluster redshifts when constructing its line
template \cite{Fan2026}, but this does not test whether the data actually
prefer the redshifted hypothesis. We therefore promote the redshift scaling
to a fitted parameter.

\subsection{Generalized redshift-scaling likelihood}
\label{subsec:alpha}

We parameterize the line energy in cluster $i$ as
\begin{equation}
 E_{{\rm line},i}=\frac{E_0}{(1+z_i)^\alpha},
 \label{eq:alpha}
\end{equation}
where $\alpha=0$ corresponds to a fixed observer-frame energy and $\alpha=1$
to a fixed source-frame energy. For the earlier Gaussian reconstruction,
the conditional likelihood is
\begin{equation}
 \begin{aligned}
 &{\cal L}_z(E_0,\alpha)=\prod_i\max_{f_i\ge0,\gamma_i}\\
 &\quad\times
 \prod_{j\in i}
 \left[(1-f_i)b_i(E_j\mid\gamma_i)
 +f_iG_i(E_j;E_0,\alpha)\right],
 \end{aligned}
 \label{eq:Lalpha}
\end{equation}
where each cluster has an independent line fraction and background slope.
This construction intentionally removes any assumption about relative dark
matter fluxes. The fit is performed over the fixed 20--70 GeV interval, so
changing $\alpha$ does not change the set of photons entering the likelihood.

We use the profile statistic
\begin{equation}
 q(\alpha)=-2\ln\frac{{\cal L}_z[\widehat E_0(\alpha),\alpha]}
 {{\cal L}_z(\widehat E_0,\widehat\alpha)}
 \label{eq:qalpha}
\end{equation}
and the direct source-frame versus observer-frame comparison
\begin{equation}
 \Delta{\rm TS}_z=
 2\ln\frac{{\cal L}_z(\alpha=1)}{{\cal L}_z(\alpha=0)}.
 \label{eq:deltaTSz}
\end{equation}

\subsection{Results}
\label{subsec:redshiftresults}

Our final event-level response calculation gives
\begin{align}
 \alpha=0:&\quad \widehat E_0=43.779~{\rm GeV},\\
 \alpha=1:&\quad \widehat E_0=44.368~{\rm GeV},\\
 \widehat\alpha=1.795:&\quad \widehat E_0=44.806~{\rm GeV}.
\end{align}
The source-frame hypothesis improves on observer-frame coherence by
$2\Delta\ln\mathcal L=3.057$; freeing the exponent improves on $\alpha=1$
by only 0.696. In a leave-one-out fit excluding Ophiuchus, the source-frame
advantage is 0.0038. Thus essentially all redshift leverage comes from
Ophiuchus. A dedicated observer-frame line-null calibration is now complete;
it calibrates this pairwise preference, not a global detection or a measurement
of the free exponent.

\subsection{Production calibration of rest-frame coherence}
\label{subsec:coherencecal}

The production comparison collapses the five masked annuli inside $R_{200}$
into one spectrum per cluster and keeps the reconstructed-energy window fixed
at 21.6--64.8 GeV as $E_0$ and $\alpha$ change. It fits
the extended likelihood specified in the reproducibility supplement, with independent cluster signal amplitudes,
background counts, and power-law slopes. Each cluster's line PDF is evaluated
at true energy $E_0/(1+z_i)^\alpha$ using its tabulated exact energy response.
At each fixed $\alpha$, the search evaluates 25 initial energies between 40
and 46 GeV and refines the two best grid neighborhoods by bounded scalar
optimization. The observer-frame null has $\alpha=0$, fitted
$E_0=43.7788$ GeV, and signal counts $(9.382,6.294,11.811)$ for Virgo,
Fornax, and Ophiuchus. Its background counts are $(53.618,39.706,87.189)$
and slopes $(-3.013,-3.862,-3.177)$; none reaches a slope boundary.

For every null realization, each cluster count is drawn from the Poisson mean
$B_i+\mu_i$. Energies are drawn by inverse-CDF sampling of the fitted
exposure-weighted continuum plus exact-response line mixture. Both
$\alpha=0$ and $\alpha=1$ are refitted, including their energy searches and
cluster nuisance parameters. The signed statistic is
\begin{equation}
 T_z=2\left[\max_{E_0,A_i,B_i,\gamma_i}\ell(\alpha=1)
 -\max_{E_0,A_i,B_i,\gamma_i}\ell(\alpha=0)\right].
 \label{eq:productioncoherencestat}
\end{equation}
It is not truncated at zero. Simulating an observer-frame line, rather than a
no-line sky, answers whether the observed preference for a redshifted line
is unusual under that particular fitted null. It does not calibrate line
discovery, arbitrary detector distortions, or a search over $\alpha$.

\begin{figure}[tbp]
\includegraphics[width=\columnwidth]{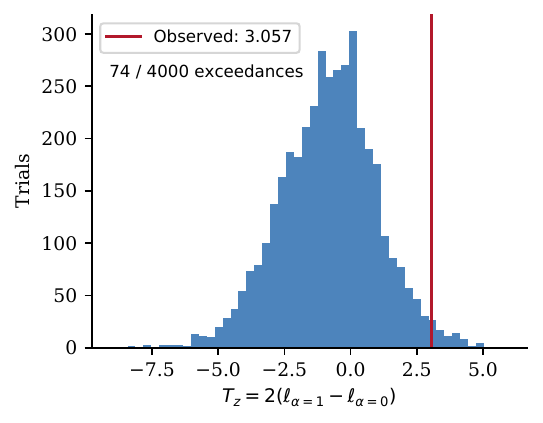}
\caption{Exact-response observer-frame line-null calibration of the signed
rest-minus-observer statistic. The vertical line marks $T_z=3.057$ in the
data; 74 of 4000 production trials lie at or above it. The test profiles the
common energy separately under each hypothesis, with independent cluster
amplitudes. Its modest preference retains the Ophiuchus leverage limitation
and does not include differential energy-scale nuisance parameters.}
\label{fig:coherencecal}
\end{figure}

Four production batches use seeds 943210--943213 and 1000 trials each.
The generator is initialized by the seed pair (batch seed, trial index),
so local trial indices 0--999 in different batches identify distinct draws.
Their exceedance counts are 25, 15, 19, and 15. All 4000 recorded statistics
are finite, the batch protocols have the same cache and observed statistic,
and the duplicated 10-trial smoke-test prefix of batch 943210 is excluded.
Pooling counts, not averaging rounded batch probabilities, gives
\begin{equation}
 \begin{aligned}
 p_{\rm MC}&=\frac{74+1}{4000+1}=0.018745,\\
 p_{95\%,\rm MC}&=[0.01455,0.02317].
 \end{aligned}
 \label{eq:coherencep}
\end{equation}
The interval is the two-sided exact binomial interval for the simulated tail,
not a confidence interval for a cosmological model or an allowance for response
uncertainty. The one-sided Gaussian equivalent is approximately $2.08\sigma$;
we retain the empirical probability as the primary result. In particular,
the 3.057 likelihood gain must not be interpreted through $\sqrt{T_z}$.
The calibrated distribution and observed statistic are shown in
Fig.~\ref{fig:coherencecal}.

The interpretation of this number is limited by the sample. Virgo and
Fornax have nearly identical redshifts, so Ophiuchus at $z=0.028$ supplies
almost the entire lever arm, and removing it eliminates the source-frame
preference altogether. The present redshift result is therefore a consistency
check on cluster association, not independent evidence for a celestial line,
and it does not measure the free exponent. Event-type and detector-coordinate
partitions are tested separately in Appendix~\ref{app:instrumental}; their
probabilities are not combined with this one.

\subsection{What would make redshift a useful cross-check?}
\label{subsec:redshiftrequirements}

The relevant quantity is relative centroid precision across redshifts, not
whether the cosmological shift exceeds the width of one detected photon.
For intuition, let $y_i=\ln E_{{\rm line},i}$ be independent measured
centroids with Gaussian errors $\sigma_{y_i}$, and
$x_i=\ln(1+z_i)$. The model is $y_i=\ln E_0-\alpha x_i$. Profiling the
common intercept gives the weighted-regression information
\begin{equation}
 I_{\alpha}=\sum_i w_i(x_i-\bar x_w)^2,
 \quad w_i=\sigma_{y_i}^{-2},\quad
 \bar x_w=\frac{\sum_i w_ix_i}{\sum_i w_i},
 \label{eq:redshiftinformation}
\end{equation}
and $\sigma_\alpha\simeq I_\alpha^{-1/2}$ in this approximation. This is a
design relation, not a reanalysis or a forecast from the present few-photon
data. It shows that extra exposure at almost identical redshifts supplies
less leverage than precise centroids in multiple well-separated redshift
bins. For Virgo and Ophiuchus, the $\alpha=1$ energy separation is only
about 1 GeV near this line; additional signal-bearing clusters must therefore
provide both redshift spread and useful centroid precision. A single
high-redshift target cannot establish a reproducible redshift trend.

The present production fit already replaces the exploratory Gaussian kernels
with tabulated cluster-specific P8R3 responses and retains a fixed photon
window and independent cluster amplitudes. A more decisive test would also
profile differential calibration nuisances constrained by matched control photons.
A common energy-scale shift is largely absorbed into $E_0$; relative shifts
correlated with sky position or event selection are the dangerous degeneracy.
Simulation calibration under both $\alpha=0$ and $\alpha=1$ should reproduce
the full fitting and selection protocol, including non-negative amplitudes.
Leave-one-cluster-out and event-partition fits would then test stability.
Redshift permutation is a supplementary diagnostic only if the permutation
scheme respects the target-dependent exposure, response, and selection; an
unrestricted shuffle need not define an exchangeable null. The energy rule,
target list, and comparison statistic should be frozen before applying the
test to new photons or additional clusters.

\section{Conditions on a Dark-Matter Interpretation}
\label{sec:dm}
\label{sec:dmimplications}

The spectral and spatial results constrain different layers of a dark-matter
hypothesis, and it is worth separating them before combining them. Particle
microphysics fixes the line energy, the branching fractions, the continuum
yield, and the velocity dependence. The halo and subhalo populations translate
those ingredients into a cluster morphology and into relative fluxes across
targets. A viable interpretation must satisfy both layers at once, and the
purpose of this section is to show that the requirements are not independent:
the measures that relieve one of them generically aggravate another.

\subsection{The spectral requirement is unusually easy to satisfy at this mass}
\label{subsec:spectralreq}

A line at $43.2$ GeV implies $2m_\chi\simeq86.4$ GeV for annihilation, or
$m_\phi\simeq86.4$ GeV for two-body decay. Both lie below the on-shell
$\gamma Z$ threshold, and $ZZ$, $W^+W^-$, $\gamma h$, and $t\bar t$ are
likewise closed. This is a genuinely favorable accident. A gauge-invariant
field-strength operator that would predict a family of electroweak lines and
an unavoidable tree-level continuum at higher masses can produce an
essentially isolated $\gamma\gamma$ feature here, with the residual continuum
arising only from off-shell bosons and electroweak radiation
\cite{Rajaraman2013,WeinerYavin2012,WeinerYavin2013}.

The restrictive case is the opposite one, in which the line is a loop closure
of an open tree-level channel. For a single charged Standard-Model fermion
saturating the loop, the line-to-tree ratio is calculable and lies six to seven
orders of magnitude below unity at this mass, which no plausible reading of the
same-target Virgo continuum nondetection can accommodate
\cite{Abazajian2012,Asano2013,Jackson2013}. This is what disfavors minimal
Higgs and $Z$ portals, the minimal supersymmetric standard model (MSSM)
pseudoscalar resonance, a Higgs-mixed secluded
mediator, and---once the absolute rate and the LEP chargino bound are included
as well---the canonical low-mass MSSM neutralino
\cite{ALEPH2004,Belanger2013,ChalonsMcCabe2013,DeLaTorreLuque2024}.

What survives spectrally is the class in which photons are a leading
interaction rather than a radiative afterthought: Rayleigh and field-strength
operators, heavy charged mediators that are closed as real final states,
photophilic axion-like or singlet pseudoscalar resonances, nearly degenerate
secluded mediators with ${\rm Br}(\varphi\to\gamma\gamma)\simeq1$, and
field-strength decay
\cite{Rajaraman2013,LeeParkPark2012,ChalonsMcCabe2013,Ibarra2012,BuckleyHooper2012,Yang2020,Feng2016Line}.
We emphasize that this is a statement about spectral architecture only. It is
not a global preference, and the classification is qualitative because our
continuum comparison is not a new joint Virgo likelihood.

\subsection{The morphological requirement is severe, and no tested template meets it}
\label{subsec:morphreq}

The measured support lies predominantly between $0.2$ and $1.0R_{200}$, and its
best empirical description is approximately uniform surface brightness. Uniform
brightness in projection gives a cumulative fraction
$F(<xR_{200})=x^2$, hence $F(<0.3)=0.09$. The tested physical templates give,
for the adopted Virgo halo, $F(<0.3)=0.988$ for smooth $s$-wave annihilation,
$0.990$ for the $p$-wave Jeans proxy, $0.536$ for NFW decay, $0.329$ for the
benchmark substructure profile with $b_{\rm sh}=47.56$, and $0.315$ in the pure
substructure limit $b_{\rm sh}\to\infty$.

Two conclusions follow. First, the gap between the shallow family and smooth
annihilation is not marginal: it is a factor of order ten in the fraction of
flux predicted inside $0.3R_{200}$, and it is what produces summed
${\rm TS}=1.61$ for smooth NFW annihilation against $14.45$ for uniform
brightness in the complete sky model of Sec.~\ref{sec:fullsky}. Second, and
less comfortably, \emph{even the pure-substructure limit is more centrally
concentrated than the best-fitting empirical template}. The data do not
distinguish them---uniform exceeds substructure by only
$2\Delta\ln\mathcal{L}=1.25$---but no dark-matter spatial model we test
actually reaches the measured flatness. A constant surface brightness truncated
near the virial radius is not a prediction of any annihilation or decay
profile; it is, however, the generic signature of an aperture-scale
background residual. We regard this as the single strongest argument against a
dark-matter reading of the excess, and it is stronger than the
template-ranking statement alone.

If the shallow profile is nevertheless attributed to substructure, the
requirement is quantitative and stiff. A boost of $b_{\rm sh}=47.56$
corresponds to a subhalo luminosity fraction $f_{\rm sh}=0.979$: nearly the
entire annihilation luminosity inside $R_{200}$ must be supplied by subhalos.
This is not a normalization correction to a smooth halo but a qualitatively
different luminosity budget. Its \emph{radial} behavior is well motivated---
tidal stripping makes the surviving subhalo population much less centrally
concentrated than $\rho^2$, and simulated cluster subhalos are anti-biased
toward the outskirts \cite{Gao2012,Han2012}. Its \emph{normalization} is the
demanding part: modern calibrated extrapolations give cluster boosts of order
tens, with factors of a few of model dependence
\cite{SanchezConde2014,Bartels2015,Ando2019,Moline2017}. A boost of order
$50$ is therefore demanding but defensible; a boost of order $10^3$ is not.

\subsection{Velocity dependence is squeezed from both sides}
\label{subsec:velreq}

Write the low-velocity rate as $\langle\sigma v\rangle\propto v^{\,n}$. The
morphology requirement fixes the sign of $n$ almost by itself.

For $n>0$ the emissivity retains the factor $\rho^2$, so the projected profile
is barely distinguishable from $s$-wave: the Jeans proxy gives
$F(<0.3)=0.990$ against $0.988$. Worse, positive velocity dependence suppresses
the dynamically cold subhalo population relative to the hot host and therefore
works directly against the substructure-dominated budget that the profile
demands. Velocity-suppressed annihilation is thus disfavored twice over, and
$p$-wave models cannot be rescued by appealing to the outer halo.

For $n<0$---Sommerfeld enhancement, or an $s$-channel resonance sitting just
below threshold---the sign is favorable: the enhancement weights cold
substructure over the hot host and so supplies part of the required flattening
dynamically, relaxing the structural boost needed from the subhalo mass
function. This is the one direction in which velocity dependence helps, and it
should be said plainly, because the draft literature on this candidate has
tended to treat velocity dependence as uniformly unhelpful
\cite{Lacroix2022,Piccirillo2022}.

The cost appears elsewhere. A Sommerfeld enhancement at $m_\chi\simeq43$ GeV
requires a mediator light compared with $\alpha m_\chi$, i.e.\ well below the
GeV scale, and that mediator must then decay dominantly to photons for the
spectrum to remain line-rich---a strong additional requirement on top of the
spectral conditions of Sec.~\ref{subsec:spectralreq}. More importantly, the
enhancement does not discriminate between cluster subhalos and dwarf
spheroidals, whose internal velocity dispersions are comparable. It therefore
does not relieve the counterpart tension discussed next; it relocates it.

\subsection{Why the dwarf and Galactic constraints are the binding ones,
and what relieves them}
\label{subsec:counterpartreq}

It is worth being precise about which counterpart limit actually binds, because
the natural expectation---dwarf spheroidals---is not the strongest one for a
line. Dwarf line searches are weak because dwarf $J$-factors are small; the
strongest model-independent line limits at tens of GeV come from the Galactic
halo, where the dark-matter column is large and the LAT line analysis is most
sensitive \cite{FermiLAT2015,Huang2012Consistency}. Dwarfs are the sharper test
of the \emph{continuum} accompanying the line, and of velocity dependence,
rather than of the line itself.

The mechanism that relieves both is the same one the morphology already
demands. If the cluster line is supplied by substructure with boost
$b_{\rm sh}$, then the particle-level cross section required to reproduce the
observed cluster flux is smaller by roughly $b_{\rm sh}$ than in a
smooth-halo interpretation, and the predicted Galactic-halo and dwarf line
fluxes---whose own boosts are of order unity to a few, because those systems
offer far less dynamic range below the host mass---fall by the same factor.
The profile requirement and counterpart consistency are therefore
\emph{aligned}, not opposed: substructure domination is simultaneously what the
morphology asks for and what buys roughly one to one and a half orders of
magnitude of relief against the halo line limits.

The difficulty is that the two requirements want different amounts of it. The
first dedicated model-building study of this candidate estimated a required
particle-level line cross section of about
$5\times10^{-28}\,\cms$ under an assumed cluster boost of $10^3$
\cite{Feng2016Line}. Rescaling to the boost regime that the ${\Lambda}$CDM
subhalo literature actually supports, $b_{\rm sh}\sim30$--$50$, raises that
requirement by a factor of roughly twenty, to
\begin{equation}
 \langle\sigma v\rangle_{\gamma\gamma}
 \;\sim\;
 1\times10^{-26}\,\cms ,
 \label{eq:requiredsigmav}
\end{equation}
that is, a two-photon cross section at the scale of the full thermal relic
annihilation rate. Since $\gamma\gamma$ would then have to carry a branching
fraction of order unity, or the total rate would have to exceed the thermal
value, this is the point at which the interpretation becomes genuinely
difficult, and it is where the Galactic-halo line limits should be confronted
directly.

This produces the central conditional statement of this work. A dark-matter
explanation of the 43 GeV cluster feature is squeezed between two requirements
that pull on the same parameter:

\begin{enumerate}
 \item the profile requires substructure to dominate the cluster annihilation
   luminosity, hence $b_{\rm sh}\gtrsim30$;
 \item the absolute line normalization requires $b_{\rm sh}$ to be as
   \emph{large} as possible, because the needed
   $\langle\sigma v\rangle_{\gamma\gamma}$ scales as $b_{\rm sh}^{-1}$ and
   must remain below the Galactic-halo line limit.
\end{enumerate}

Requirement (i) is a lower bound that ${\Lambda}$CDM can supply. Requirement
(ii) pushes toward boosts of order $10^3$, which the same simulation
literature does not support. The model-building task, stated as sharply as the
present data allow, is to find a construction that is
(a) photon dominated in the sense of Sec.~\ref{subsec:spectralreq},
(b) velocity independent or mildly low-velocity enhanced,
(c) compatible with a cluster substructure boost of order tens rather than
thousands, and
(d) still below the Galactic-halo line limit at $m_\chi\simeq43$ GeV with
$\langle\sigma v\rangle_{\gamma\gamma}$ of order $10^{-26}\,\cms$.

We have not constructed such a model, and we regard the fourth condition as the
one most likely to fail. A photophilic resonant singlet or a Rayleigh operator
with an order-unity photon branching fraction satisfies (a) and can satisfy
(b) and (c); whether any such construction satisfies (d) is a calculation that
requires the Galactic line limit to be evaluated for the same operator and the
same halo assumptions, and it is the natural next step for a dedicated
model-building study rather than for this measurement paper.

\subsection{Two consistency tests this work does not perform}
\label{subsec:openconsistency}

Two further requirements follow from the above, and neither is settled here.
First, a cluster line implies a definite pattern of relative luminosities
across a cluster sample---and, under substructure domination, one that follows
total subhalo luminosity and cluster mass rather than the conventional central
$J$-factor ordering. Testing it requires a common event-level likelihood over
the full published target list, not a comparison of separately reported
per-cluster results, and the required virial-scale matched filter differs from
the aperture used in a smooth-annihilation search. Second, any model with an
appreciable charged-particle branch predicts synchrotron and inverse-Compton
counterparts whose normalization depends on the electron yield, the
magnetic-field structure, and transport; converting a radio nondetection into a
bound on that yield requires calibrated injection-and-recovery tests for smooth
cluster-scale emission, because interferometric spatial filtering removes much
of such a signal \cite{JeltemaProfumo2011,Marchegiani2025Radio}. Both are worth
doing. Neither changes the conclusion drawn from the profile, which is what the
present data measure.

\section{Discussion and Conclusions}
\label{sec:discussion}

The reported 43 GeV feature presents two physical questions: what could
produce its narrow spectrum, and why does its support extend well beyond
the cluster centers? Our reconstruction recovers the spectral excess in
the same public LAT observations used in the original report. The annular
analysis then shows that the feature is supported predominantly between
$0.2$ and $1.0R_{200}$, including after catalogued-source masking. Explaining
both properties is more restrictive than explaining a spectral excess in
a stacked aperture. Before assigning either property to a celestial
source, however, instrumental and sky-background explanations must be
assessed.

\subsection{The complete sky model as the decisive internal check}

The analyses in this paper have two distinct levels. The event-level
reconstruction, masks, and annuli establish which photons generate the
candidate and how that support is distributed under simple local backgrounds.
They do not determine whether a structured Galactic foreground or the fitted
wings and spectra of catalog sources can account for the same photons. The
strict-GTI spatial--spectral likelihood of
Sec.~\ref{subsec:fullskylike} addresses that question directly and must carry
greater weight in the final interpretation.

The complete likelihood retains a modest line-like component when the three
clusters are combined. Uniform brightness gives the largest summed value,
${\rm TS}=14.45$, followed closely by substructure at 13.20 and NFW decay at
10.30. Smooth NFW annihilation and a central point source give only 1.61 and
0.70. Virgo provides the largest uniform-template contribution
(${\rm TS}=7.35$); Ophiuchus, despite its low Galactic latitude, contributes
${\rm TS}=3.12$, and Fornax contributes 3.98 while individually preferring
substructure at 5.66. The fitted Galactic and isotropic normalizations remain
away from their boundaries and are stable across morphologies, as are the
well-constrained nearby catalog sources. The broadband residual maps show no
coherent cluster-centered structure and no conspicuous Ophiuchus-specific
foreground failure. The source-plus-diffuse result therefore becomes the main
quantitative evidence for persistence of the candidate, while the event-level
annuli remain the more transparent diagnostic of which photons and radial
ranges supply it.

For the uniform template, the fixed-energy three-amplitude boundary
distribution gives a local significance of $3.23\sigma$. The direct
energy-and-morphology calibration gives
$p_{\rm global}=3.60\times10^{-3}$, or approximately $2.7\sigma$, with the
maximum at 44.5 GeV for uniform brightness. This establishes that the feature
remains modest after accounting for the defined sky-model search. A direct
full-profile evaluation at that maximum gives $T_{\max}=19.938$, only 0.133
above the cached statistic used to rank the simulations. The result is
conditional on the preselected three-cluster sample and is distinct from the
earlier event-level discovery statistic and its calibration.

The event-level ${\rm TS}=30.62$ and complete-sky-model ${\rm TS}=14.45$
are therefore not two estimates of the same significance. The aperture-only
fit models the local energy distribution but does not use spatial information
to assign photons among Galactic diffuse emission, isotropic emission,
catalogued sources, and the candidate. The complete sky model performs that
assignment while profiling the relevant nuisance components. It is consequently
the authoritative significance calculation in this work; the larger
event-level statistic is retained as a reproduction and photon-level
diagnostic, not as the detection significance.

\subsection{What the controls and emission tests establish}

The event partitions, temporal holdout, and detector-coordinate tests reveal
no clear instrumental explanation for the feature. In particular, the
calibrated detector-coordinate test finds no significant combined amplitude
or centroid inconsistency ($p_{\rm MC}=0.454$). These checks constrain
selection-dependent artifacts, but cannot eliminate an energy-response
effect shared by the event classes. The matched-sky analysis supplies a
separate comparison with background fields; its sensitivity to individual
fields and the reuse of fields in stacks limit the precision of its tail
probabilities. Neither those probabilities nor the temporal and event-class
checks can be combined as independent detection significances. A common-mode
response effect, a statistical fluctuation, or an inadequately modeled
source or diffuse background remains a possible explanation.

The conventional celestial alternatives face a different difficulty:
producing a sufficiently narrow spectrum. Physical pion-decay calculations
convolve the proton population with broad hadronic production and decay
kinematics. Power-law, cutoff, and broken-power-law proton distributions
change the broad-band continuum but leave a line likelihood improvement
of ${\rm TS}_{\rm line}=21.49$--26.38 near 43.2 GeV. These fits do not
exclude cosmic rays as a contributor to cluster gamma-ray emission; they
show that the tested hadronic spectra do not replace the localized residual.

Inverse-Compton emission on the CMB, infrared, and optical radiation fields
is also broad, even for monoenergetic electrons. A narrow feature can arise
in the deep Klein--Nishina regime, where a photon can receive most of an
electron's energy. This possibility prevents a narrow gamma-ray feature
from being an unambiguous dark-matter signature on spectral grounds alone.
For the extended cluster interpretation, however, the required EUV or
soft-X-ray photon field is problematic. At the two narrowness benchmarks
studied here, making the power in the line interval comparable to the
unavoidable total CMB-IC power requires target energy densities roughly
$10^5$--$10^6$ times our generous virial-scale X-ray benchmark. At the
benchmark density, the same electrons instead radiate overwhelmingly into
a broad MeV component. Maintaining an extremely narrow electron population
throughout the emitting region adds a further difficulty. Compact cold-wind
sources can evade the assumption of a dilute cluster-scale radiation field,
but then require a population capable of reproducing both the common energy
and the extended masked profile. No such population is established here.

\subsection{Implications of the spatial and energy distributions}

The morphology disfavors a centrally concentrated origin under the tested
spatial models. Smooth NFW annihilation shows a strong conditional template
mismatch, and standard NFW decay also fits less
well than the best shallow alternative. The strength of a specific
substructure-versus-decay comparison depends on the treatment of the
relative cluster amplitudes: its calibrated probability is 0.0292 for the
adopted weights and approximately 0.002 with free amplitudes, with finite
simulation uncertainty in both cases. These probabilities test specified
spatial hypotheses conditional on the candidate; they do not establish
the significance or celestial origin of the line.

Uniform surface brightness remains the best empirical template, and the
continuous smooth-plus-subhalo fit reaches the pure-substructure boundary, so
the data favor a shallow distribution without measuring a finite subhalo boost.
Section~\ref{sec:dm} sets out what that distribution would require of a
dark-matter model.

Energy coherence tests cluster association independently of these spatial
interpretations. The common source-frame line improves the likelihood over
a common observer-frame line by $T_z=3.057$, with a fitted observer-frame
line-null calibration of $p_{\rm MC}=0.0187$. Almost all leverage comes
from Ophiuchus; removing it reduces the likelihood advantage to 0.0038.
The present result therefore provides modest support for source-frame
alignment, rather than a measurement of cosmological redshifting. Several
signal-bearing clusters with different redshifts and controlled relative
energy calibration are needed to distinguish a reproducible redshift
pattern from a target-specific centroid shift. Such a pattern would support
a cluster origin but would not distinguish dark matter from another
cluster-associated emission mechanism.

\subsection{Conclusions}
\label{sec:conclusions}

The public event sample contains the reported 43 GeV excess, and several
orthogonal instrumental checks reveal no simple partition-specific pathology.
The tested hadronic and inverse-Compton mechanisms fail to explain its narrow
spectrum under ordinary cluster conditions. Instrumental common modes,
structured foregrounds, unresolved sources, and a statistical fluctuation
nevertheless remain viable, because neither reproduction nor an event-level
masking analysis establishes a celestial source.

The complete sky-model search quantifies that limitation. Its fixed-energy
uniform result has a local significance of $3.23\sigma$, while 17 of 5000
background-only simulations produce an equal or larger maximum somewhere in
the 20--70 GeV, five-morphology scan. This gives a conditional global
significance of approximately $2.7\sigma$
($p_{\rm global}=3.60\times10^{-3}$; 95\% Monte Carlo interval
$(1.98$--$5.44)\times10^{-3}$). The feature is therefore evidence worth
explaining, not a standalone detection. The complete profiled fit at the
observed maximum agrees with the accelerated statistic to 0.133 TS. Because
the three clusters were selected from the original 13-cluster sample, this
probability is not an unconditional significance for the historical search,
and any genuinely data-dependent target-selection correction would reduce it.
An unconditional calibration would have to reproduce the historical
target-selection rule inside every background realization. As an
order-of-magnitude illustration, 10--20 effective selection opportunities
would reduce the nominal evidence to approximately $1.5$--$1.8\sigma$. We do
not apply the full factor of $\binom{13}{3}$ because the targets were not
searched with equal freedom, the candidate triplets are strongly correlated,
and the ranking also used external halo information.

The central measurement is spatial. In a strict-GTI likelihood containing the
Galactic and isotropic diffuse emission and all 4FGL-DR4 sources, the
surviving line-like component is broad: uniform surface brightness and
substructure-enhanced annihilation retain summed TS values of 14.45 and 13.20,
NFW decay 10.30, while smooth NFW annihilation gives 1.61 and a central point
source 0.70. The independent event-level annular analysis agrees, placing the
masked support predominantly between $0.2$ and $1.0R_{200}$. The data thus
select a broad family rather than a unique flat profile, but they separate that
family decisively from the centrally concentrated profile that smooth-halo
annihilation predicts.

These template rows are non-nested likelihood rankings and are not converted
into Gaussian significances; their role is to identify the morphology favored
conditional on the candidate.

This is the result that constrains interpretation. A line at tens of GeV points
most naturally to two-body annihilation in the smooth halo, and that is the one
morphology the data disfavor. Substructure-enhanced annihilation moves in the
required direction and is the closest tested dark-matter template, but the
continuous boost fit runs to the pure-substructure boundary without measuring a
finite enhancement, and even that limiting profile fits worse than the uniform
benchmark. Taken together, the modest conditional significance, the
uncalibrated three-cluster preselection, and the sensitivity of the original
aperture statistic to spatially structured backgrounds point to a chance
fluctuation amplified by analysis and target-selection effects as the most
likely origin of the feature. Residual Galactic diffuse emission or
unresolved-source mismodeling, particularly in the complicated Ophiuchus
field, provides a plausible additional contribution. The instrumental tests
make a simple detector artifact less likely, while the narrow spectrum
disfavors ordinary cluster emission. If some part of the excess is nevertheless
celestial, its broad virial-scale morphology remains essentially incompatible
with smooth-halo dark-matter annihilation.

The immediate priority is to establish whether the feature recurs with energy,
target selection, and spatial tests fixed before new photons are examined. The
source-plus-diffuse likelihood and its global calibration address the principal
internal sky-model and search-trials questions for the selected three-cluster
sample. Independent gamma-ray observations can test recurrence and
spectral width, and a larger cluster sample can test redshift coherence and
relative luminosities within a single event-level fit. If the feature is
celestial, those measurements, together with Galactic and dwarf constraints,
determine whether any common dark-matter model can accommodate it. The central
requirement is a single explanation for the narrow spectrum, the extended
spatial distribution, and the presence or absence of associated signals in
other bands and other systems.

\paragraph*{Data and code availability.}
The Fermi-LAT observations analyzed here are public. Analysis code,
machine-readable fit products, Monte Carlo records, run manifests, and the
reproducibility supplement are available in the associated Zenodo record,
\href{https://doi.org/10.5281/zenodo.22740677}{doi:10.5281/zenodo.22740677}.

\begin{acknowledgments}
This work was supported by the U.S.\ Department of Energy, Office of Science,
Office of High Energy Physics, under Award Number DE-SC0010107, and by a
\mbox{ChatGPT for Academic Researchers} grant from OpenAI. ChatGPT and Codex were
used as research-assistance tools for code development, numerical auditing,
workflow documentation, and editorial revision; all scientific judgments,
validation decisions, and responsibility for the manuscript remain with the
author. The author executed the analyses from the public LAT event data,
inspected the fit diagnostics and calibration outputs, and verified the
reported numerical values against the archived machine-readable results. I am grateful to Tesla Jeltema for suggestions and feedback.
\end{acknowledgments}

\clearpage
\appendix

\section{Independent Reproduction of the 43 GeV Excess}
\label{sec:replication}

We first ask whether the public photon sample independently reproduces the
reported feature before introducing any new physical test. Following the
baseline prescription of Ref.~\cite{Fan2026}, for each trial source-frame
energy $E_0$ we fit the merged top-three data over
\begin{equation}
 E\in[0.5E_0,1.5E_0]
 \label{eq:baselinewindow}
\end{equation}
with a power-law background and the exact redshifted P8R3 EDISP template of
Sec.~\ref{subsec:response}.

The event-count comparison provides a stringent first check. After the strict
quality filter of Eq.~(\ref{eq:strictgti}) we find
\begin{equation}
 \boxed{N_{40-46}=44},
 \label{eq:4046counts}
\end{equation}
identical to the published top-three ULTRACLEAN EDISP$(1+2+3)$ count. The
same analysis with the looser $\texttt{DATA\_QUAL}>0$ convention also gives
44 events in this interval.

\begin{figure}[tbp]
\includegraphics[width=\columnwidth]{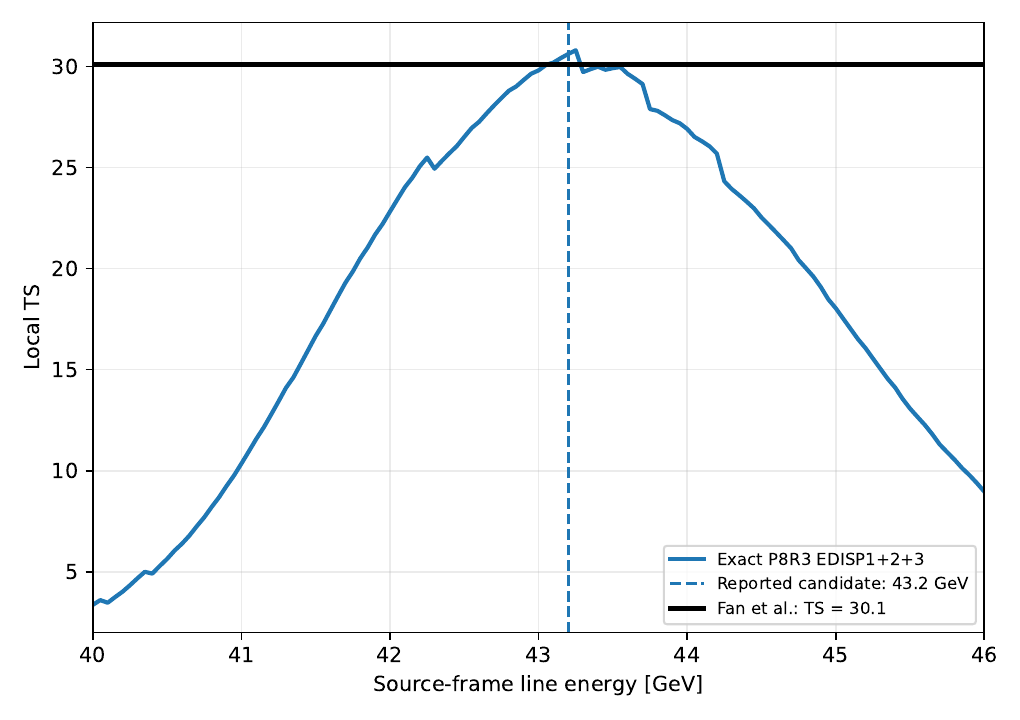}
\caption{Independent top-three line scan using the strict spacecraft-quality
selection and the exact P8R3\_ULTRACLEAN\_V3 EDISP1+2+3 response. The vertical
dashed line marks 43.2 GeV. The thick black horizontal line marks the
published baseline ${\rm TS}=30.1$. We obtain ${\rm TS}=30.62$ at 43.2 GeV
and a maximum of 30.81 at 43.25 GeV.}
\label{fig:replication_scan}
\end{figure}

\begin{figure}[tbp]
\includegraphics[width=\columnwidth]{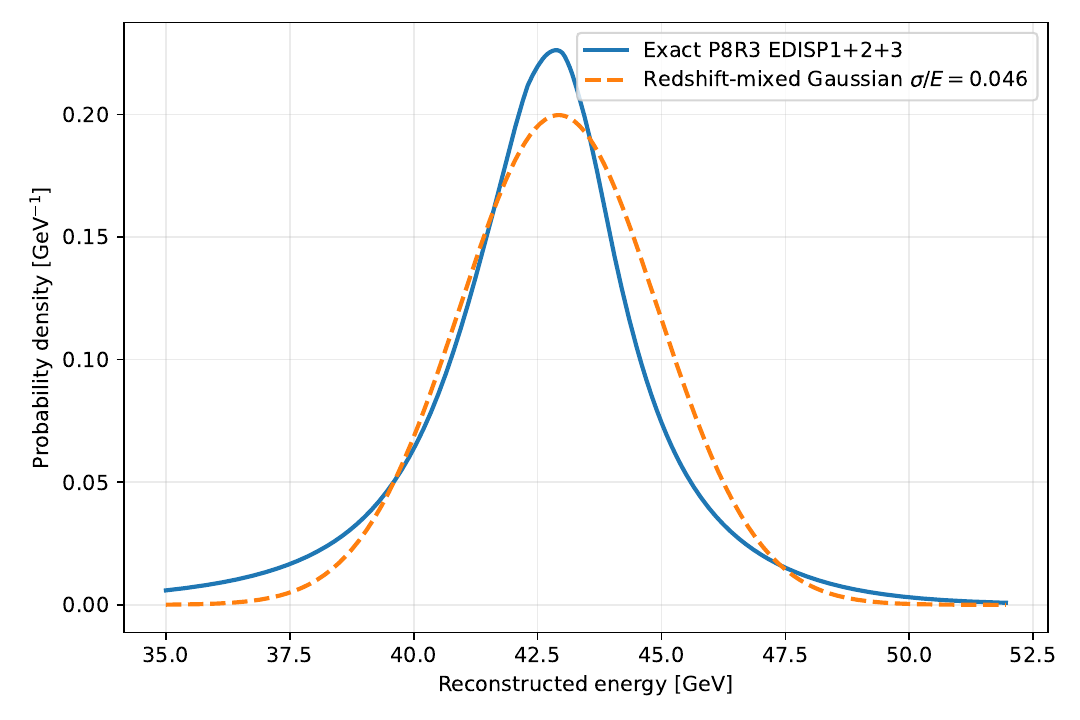}
\caption{Exposure- and redshift-weighted stacked response for a 43.2 GeV
source-frame line. The solid curve uses the exact P8R3 EDISP1+2+3 IRFs and
mission-long incidence-angle distribution. The dashed curve is the properly
redshift-mixed Gaussian approximation with $\sigma_E/E=0.046$. The exact
response has a sharper core and asymmetric non-Gaussian tails.}
\label{fig:exact_edisp}
\end{figure}

Using the exact CALDB line response, the likelihood at the reported energy is
\begin{equation}
 \boxed{{\rm TS}(43.2~{\rm GeV})=30.62},
 \label{eq:repTS432}
\end{equation}
compared with ${\rm TS}\simeq30.1$ in Ref.~\cite{Fan2026}. The scan reaches
\begin{equation}
 \widehat E_0=43.25~{\rm GeV},\qquad
 {\rm TS}_{\max}=30.81.
 \label{eq:repbest}
\end{equation}
Thus the independent implementation agrees with the published baseline to
within $\Delta{\rm TS}\simeq0.5$ at 43.2 GeV (Fig.~\ref{fig:exact_edisp}) while reproducing the event count
exactly. Figure~\ref{fig:replication_scan} shows the local scan.

As a useful internal cross-check, our earlier Gaussian-response analysis,
performed before the FT2 filtering was available, gave
${\rm TS}(43.2~{\rm GeV})=30.37$ and a maximum of 30.42 at 43.25 GeV. The
small change after implementing the exact response demonstrates why the
first-pass reconstruction was already accurate in the narrow energy window,
while the exact calculation removes the remaining instrumental-response
approximation.

The agreement in both photon counts and the exact-response likelihood
establishes that the public data and our independent response implementation
recover the same spectral candidate. We therefore use this event sample for
the additional consistency tests below.

\subsection{Response-validation summary}
\label{subsec:exactstatus}

The response calculations provide several independent internal checks. The
strict spacecraft-quality cut changes the raw 40--46 GeV count from 45 to 44,
exactly matching Ref.~\cite{Fan2026}. Folding the broad continua through the
mission-long exposure reduces the preliminary broad-band line TS from
approximately 27--28 to 21--23 and brings all four published continuum
families to within about one TS unit of the reported values. Finally,
replacing the Gaussian line template by the exact EDISP1+2+3 kernel gives
${\rm TS}(43.2~{\rm GeV})=30.62$, compared with 30.1 in the published
narrow-window analysis. The final hadronic calculation then shows that this
response-level agreement is not specific to phenomenological backgrounds:
physical proton PL, cutoff, and broken-power-law spectra still leave
${\rm TS}_{\rm line}=21.49$, 26.17, and 26.38, respectively, with
$\widehat E_{\rm line}\simeq43.21$ GeV. Together these checks isolate the
remaining small likelihood differences to implementation-level details
rather than event selection, a qualitatively different instrumental
response, or ordinary broad hadronic curvature.

\section{Instrumental, Temporal, and Bootstrap Validation}
\label{app:instrumental}
\label{sec:systematics}

This appendix retains only the diagnostics that materially determine the
claim hierarchy. Complete selections, response grids, optimizer settings,
control-field inventories, seeds, hashes, and run manifests are given in the
separate reproducibility supplement.

\subsection{Event partitions and detector coordinates}
\label{subsec:eventpartitions}

\begin{figure*}[tbp]
\centering
\includegraphics[width=0.96\textwidth]{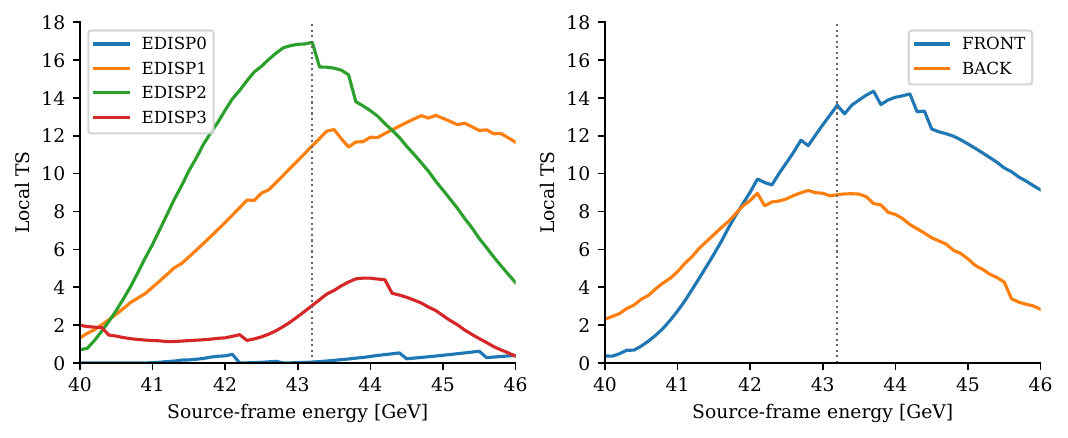}
\caption{Partition-specific exact-response scans for the four EDISP
quartiles (left) and FRONT/BACK conversion types (right). The line support is
distributed across multiple better-resolved partitions rather than isolated
in a single event class.}
\label{fig:splitfullscan}
\end{figure*}

The strict-GTI exact-response analysis reproduces the published aggregate
partitions: EDISP0+1+2+3, EDISP1+2+3, EDISP2+3, FRONT, and BACK give peak
${\rm TS}=22.83,30.24,18.71,14.35,$ and 9.11, with exactly the published
40--46 GeV counts in every subset. Disjoint EDISP1 and EDISP2 samples carry
substantial support (${\rm TS}=13.08$ and 16.93), EDISP3 is weaker, and
EDISP0 is null; the feature is therefore not confined to one reconstruction
quality or conversion type. A 1000-permutation omnibus test over incidence
angle, instrument azimuth, rocking angle, and epoch gives calibrated
probabilities 0.257, 0.119, and 0.454 for the maximum amplitude, centroid,
and total contrasts (Fig.~\ref{fig:splitfullscan}). These are response checks, not independent detections,
and they do not exclude a common-mode instrumental effect.

\subsection{Matched skies and temporal holdout}
\label{subsec:blankcompleted}
\label{subsec:heldout}

\begin{figure*}[tbp]
\centering
\includegraphics[width=0.92\textwidth]{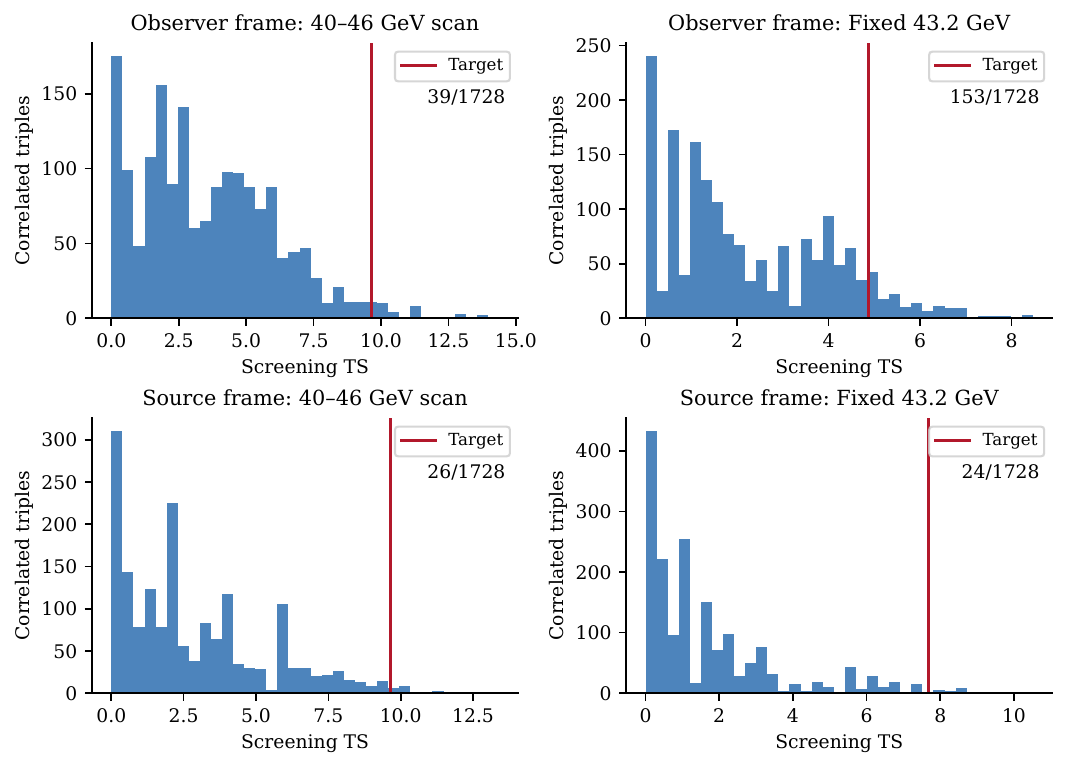}
\caption{Matched-control statistics and target values for observer- and
source-frame energy rules. Reused fields make the histogram entries
correlated; the figure is a finite-control screening diagnostic.}
\label{fig:blanktails}
\end{figure*}

The frozen matched-sky screen compares the target with all 1728 triples
formed from 36 non-overlapping fields matched in aperture, latitude, exposure,
and catalog environment. Source-frame scan and fixed-energy ranks are 0.0156
and 0.0145, but all corresponding exceedances involve one Ophiuchus-matched
field; leave-one-field-out and resampling audits therefore show that the
nominal tail is limited by the finite control set
(Table~\ref{tab:blanktails} and Fig.~\ref{fig:blanktails}). The historically specified
42.7 GeV observer-frame energy provides a quasi-independent temporal check:
post-2015 photons give ${\rm TS}_{\rm hold}=11.02$ with the strict-GTI exact
response and maximize near 43.25 GeV when scanned. Neither diagnostic is
combined with the discovery statistic.

\begin{table}[tbp]
\centering
\caption{Matched-sky screening. Add-one ranks reuse the same 36 fields and
must not be assigned the precision of 1728 independent skies.}
\label{tab:blanktails}
\begin{ruledtabular}
\begin{tabular}{lrrr}
Test & Target TS & $k/1728$ & Add-one rank\\
\hline
Observer scan & 9.655 & 39 & 0.02313\\
Observer, 43.2 GeV & 4.867 & 153 & 0.08907\\
Source scan & 9.652 & 26 & 0.01562\\
Source, 43.2 GeV & 7.685 & 24 & 0.01446\\
\end{tabular}
\end{ruledtabular}
\end{table}

\subsection{Calibration scope}
\label{subsec:trials}

\begin{figure*}[tbp]
\centering
\includegraphics[width=0.95\textwidth]{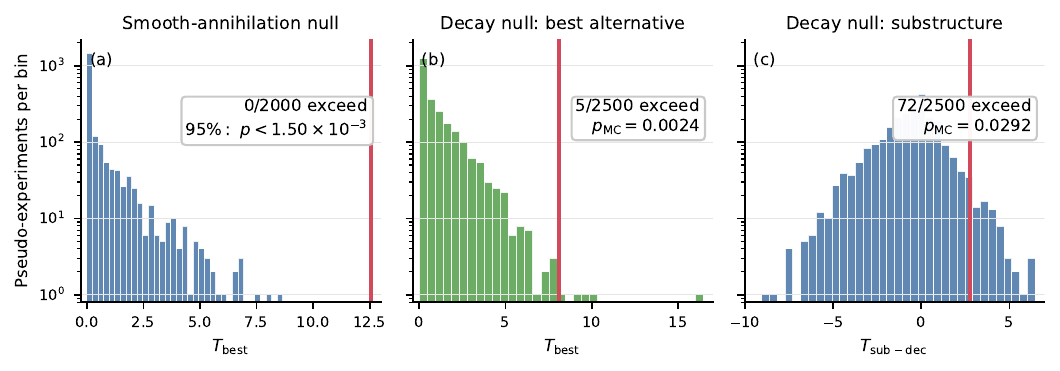}
\caption{Conditional masked-bootstrap distributions for the primary
non-nested morphology comparisons. The pairwise substructure--decay panel,
rather than the best-of-five panel, calibrates the physical choice between
the two shallow dark-matter templates.}
\label{fig:masked_bootstrap_distributions}
\end{figure*}

The non-nested morphology likelihood gaps are calibrated with the frozen
$0.35^\circ$-masked conditional bootstrap. Smooth NFW annihilation produces
no exceedance in 2000 trials; under NFW decay, 5 of 2500 trials reach the
best-of-five gap and 72 reach the physically specific substructure--decay
gap, corresponding to $p_{\rm MC}=0.0024$ and 0.0292. The free-cluster-
amplitude follow-up uses a distinct extended-likelihood ensemble and is not
pooled with these trials. All probabilities are conditional on the plug-in
null, frozen masks, line energy, cell counts or Poisson means, and adopted
halo templates; their bookkeeping is moved to the reproducibility supplement. The simulated
distributions are shown in Fig.~\ref{fig:masked_bootstrap_distributions}.
The sky-model global calibration of Sec.~\ref{subsec:globalcalibration} uses a
separate Poisson count-cube ensemble and calibrates the maximum over energy and
all five morphologies. It is not pooled with the conditional annular-template
ensembles summarized here.

\section{Technical Response and Spatial Checks}
\label{app:technical}

The exact line response is constructed separately for EDISP1--3 by weighting
\label{subsec:response}
the P8R3 energy-dispersion kernels with the ROI livetime and effective area,
then combining clusters at the redshifted true line energy. At 43.2 GeV the
stacked response has a central 68\% reconstructed-energy interval of
40.42--44.50 GeV. The detailed response grids and interpolation conventions
are archived in the reproducibility supplement.

\begin{figure*}[tbp]
\centering
\includegraphics[width=0.325\textwidth]{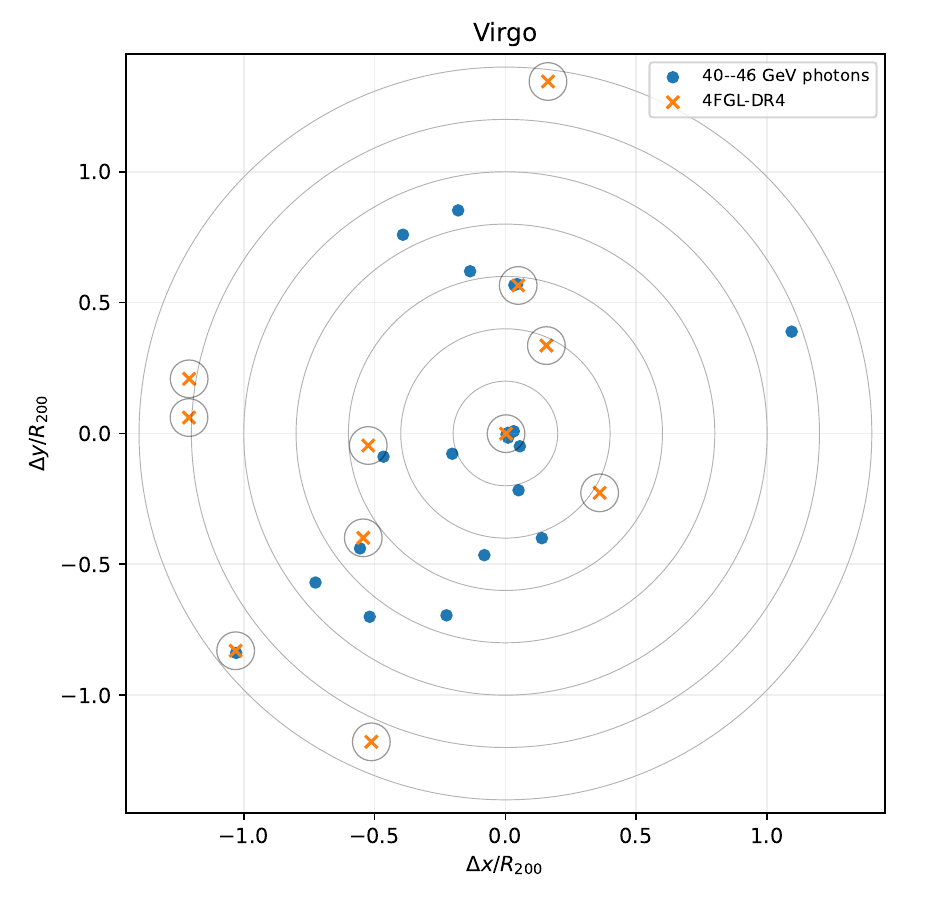}
\hfill
\includegraphics[width=0.325\textwidth]{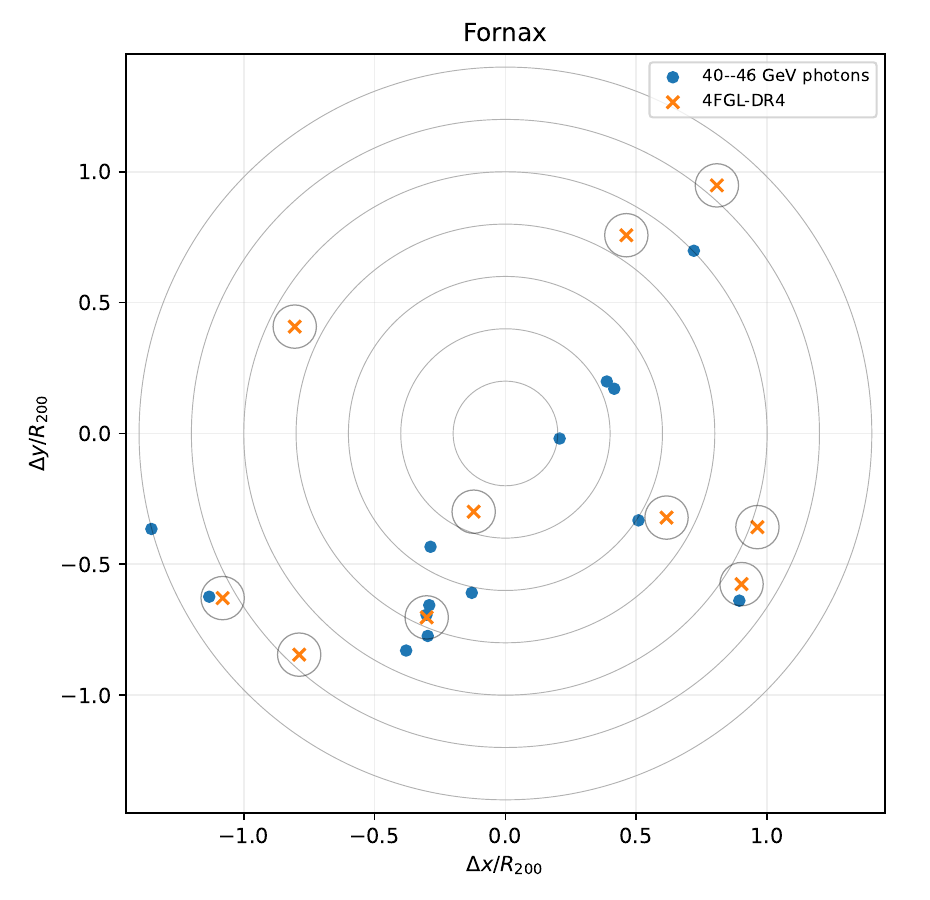}
\hfill
\includegraphics[width=0.325\textwidth]{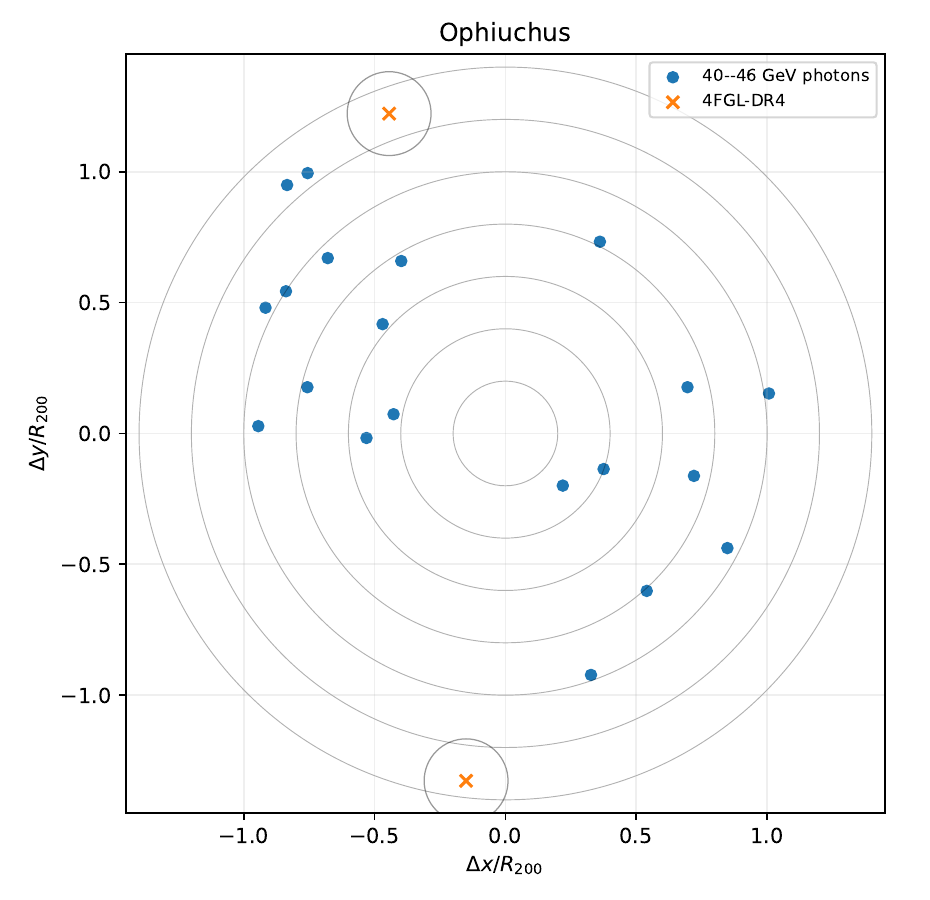}
\caption{Line-band photons, 4FGL-DR4 sources, and representative
$0.25^\circ$ masks within $1.4R_{200}$. The absence of catalog coincidences
in Ophiuchus does not establish that its low-latitude field is source-free.}
\label{fig:psmaps}
\end{figure*}

The 4FGL-DR4 cross-match finds 11 catalogued sources within $1.4R_{200}$ of
\label{app:sourcechecks}
Virgo, 10 around Fornax, and two around Ophiuchus. Of 55 photons between
40 and 46 GeV, 16 lie within $0.35^\circ$ of a catalogued source, but none of
the 20 Ophiuchus photons does. Masking removes the central Virgo contribution
and weakens one intermediate annulus without erasing the virial-scale pattern
(Fig.~\ref{fig:psmaps}).
This establishes robustness to catalogued sources only; unresolved sources
and structured diffuse emission, especially at the low Galactic latitude of
Ophiuchus, require the separate spatial--spectral likelihood.

The printed substructure profile of Ref.~\cite{Fan2026} integrates to
\label{app:subnorm}
$b_{\rm sh}J_{\rm NFW}/16$ in the small-angle limit, whereas the conventional
boost-normalized form contains an additional factor of 16. We therefore test
both the literal printed form and the boost-normalized form. This ambiguity
changes the normalization assigned to the subhalo component but not its
projected radial shape. In the absence of a machine-readable normalization
definition for the original implementation, we do not identify either
convention as uniquely reproducing that implementation; all comparisons state
which convention is used.

\subsection{Condensed spatial-systematics summary}
\label{app:spatialchecks}

Across $r_{68}=0.08^\circ$--$0.18^\circ$, concentration rescalings of
0.8--1.2, four $0.05R_{200}$ center shifts, and mask radii of $0.25^\circ$
and $0.35^\circ$, smooth annihilation remains the worst-fitting template and
boost-dominated substructure remains above NFW decay. Centering materially
affects only the stronger distinction between the empirical uniform benchmark
and substructure. Widening the annular background-slope floor from $-5$ to
$-40$ changes the free-amplitude substructure--decay gap from 5.954 to 6.075
and leaves the ordering and pure-substructure endpoint unchanged. These
variations test stability of the conditional ranking; they are not separate
calibrated exclusions.

\begin{figure*}[tbp]
\centering
\includegraphics[width=0.93\textwidth]{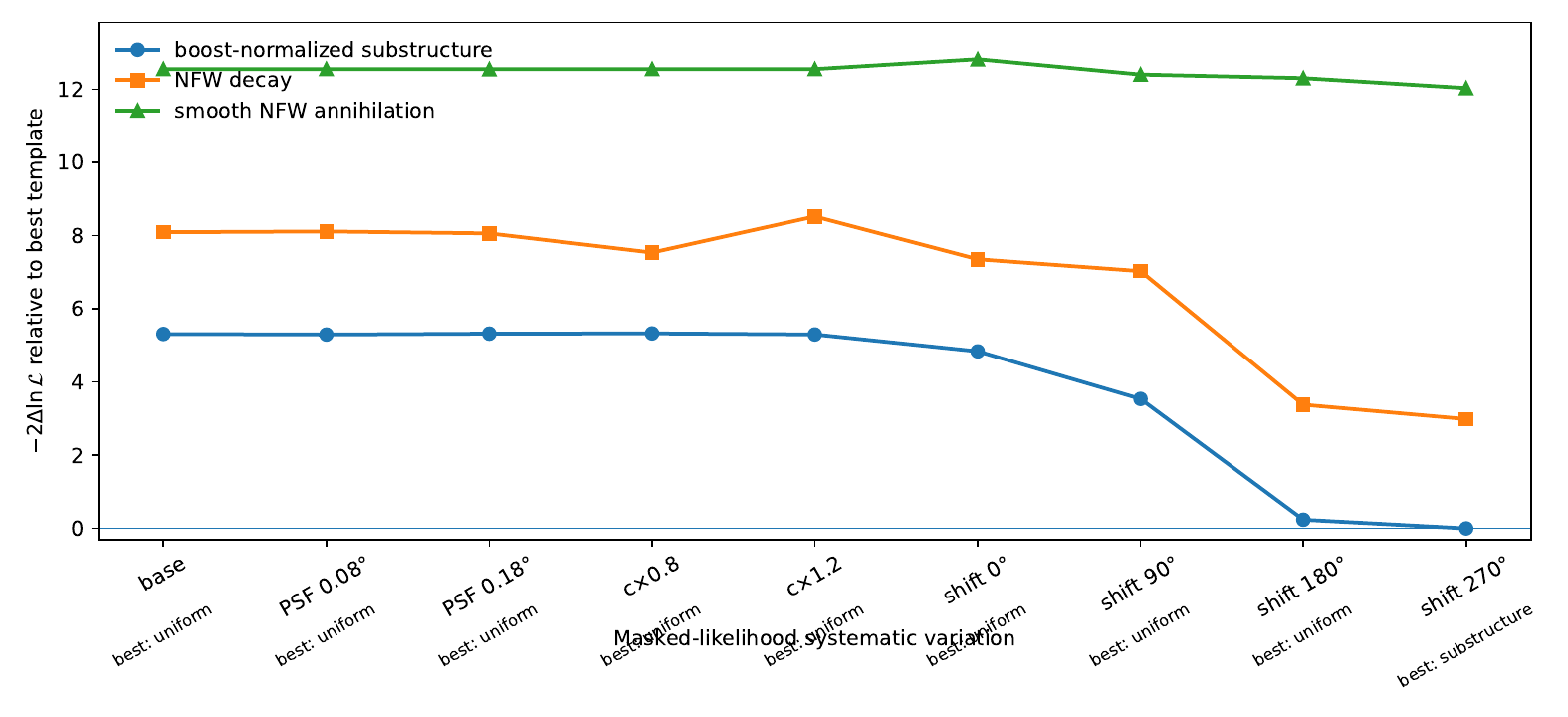}
\caption{Primary masked morphology ranking under PSF, concentration, and
centering variations. Smooth NFW annihilation retains a strong conditional
template mismatch; substructure remains above NFW decay, while its separation
from uniform brightness is centering-sensitive.}
\label{fig:masked_template_systematics}
\end{figure*}

\clearpage

\section{Phenomenological Continuum Fits}
\label{sec:continuum}

This appendix gives the phenomenological continuum fits supporting the
spectral analysis. The physical hadronic and inverse-Compton alternatives
are treated in Sec.~\ref{sec:nondm}. The final broad-band
continuum analysis uses the strict GTI-selected 2.5--450 GeV top-three sample,
containing 6390 photons and 44 photons between 40 and 46 GeV, and folds every
continuum through the energy-dependent ROI-averaged LAT exposure.

\subsection{Phenomenological continuum models}
\label{subsec:phenomenological}

We compare five continuum shapes:
\begin{align}
 {\rm PL}:&\quad F(E)\propto E^\gamma,\\
 {\rm PLEC}:&\quad F(E)\propto E^\gamma e^{-E/E_c},\\
 {\rm LogP}:&\quad F(E)\propto
 \exp[-a\ln(E/E_p)-b\ln^2(E/E_p)],\\
 {\rm BPL}:&\quad F(E)\propto
 \begin{cases}
 (E/E_b)^{\gamma_1}, & E<E_b,\\
 (E/E_b)^{\gamma_2}, & E\ge E_b,
 \end{cases}\\
 {\rm SBPL}:&\quad F(E)\propto
 x^{\gamma_1}\left(1+x^{1/s}\right)^{s(\gamma_2-\gamma_1)},
 \quad x=E/E_b.
 \label{eq:continuummodels}
\end{align}
For these broad-band fits the predicted count spectrum is the source spectrum
multiplied by the solid-angle-weighted ROI exposure. The line response in this
stage uses the Gaussian EDISP approximation; the narrow-window analysis above
shows that replacing it by the exact kernel changes the local TS only at the
sub-unit level.

The exposure correction materially improves agreement with the published
broad-band analysis. We obtain
\begin{align}
 {\rm PL}:&\quad {\rm TS}_{\rm line}=21.12,\\
 {\rm PLEC}:&\quad {\rm TS}_{\rm line}=21.41,\\
 {\rm LogP}:&\quad {\rm TS}_{\rm line}=23.16,\\
 {\rm BPL}:&\quad {\rm TS}_{\rm line}=22.44,\\
 {\rm SBPL}:&\quad {\rm TS}_{\rm line}=22.47.
 \label{eq:finalbroadTS}
\end{align}
For PL, PLEC, LogP, and BPL backgrounds, Ref.~\cite{Fan2026} reports 22.1,
22.4, 24.4, and 23.6, respectively. Our independent values are therefore
lower by only 0.97--1.24 TS units across all four model families. The fitted
line energy is correspondingly stable at 42.83--42.85 GeV.

\begin{table*}[tbp]
\centering
\caption{Strict-GTI, exposure-corrected 2.5--450 GeV continuum fits. The final
column gives the corresponding published top-three value where available.
The likelihood is a normalized spectral-shape likelihood, so absolute
$-2\ln{\cal L}$ values need not coincide with those of Ref.~\cite{Fan2026};
the directly comparable quantity is the likelihood improvement within each
continuum family.}
\label{tab:broadbandmodels}
\begin{ruledtabular}
\begin{tabular}{lrrrr}
Continuum & ${\rm TS}_{\rm line}$ & $\widehat E_0$ [GeV] & $\Delta{\rm AIC}$ & Fan et al. ${\rm TS}_{\rm line}$\\
\hline
PL    & 21.12 & 42.831 & 1.52 & 22.1\\
PLEC  & 21.41 & 42.837 & 1.80 & 22.4\\
LogP  & 23.16 & 42.837 & 0.00 & 24.4\\
BPL   & 22.44 & 42.847 & 0.40 & 23.6\\
SBPL  & 22.47 & 42.847 & 2.45 & ---\\
\end{tabular}
\end{ruledtabular}
\end{table*}

\begin{figure*}[tbp]
\centering
\includegraphics[width=0.88\textwidth]{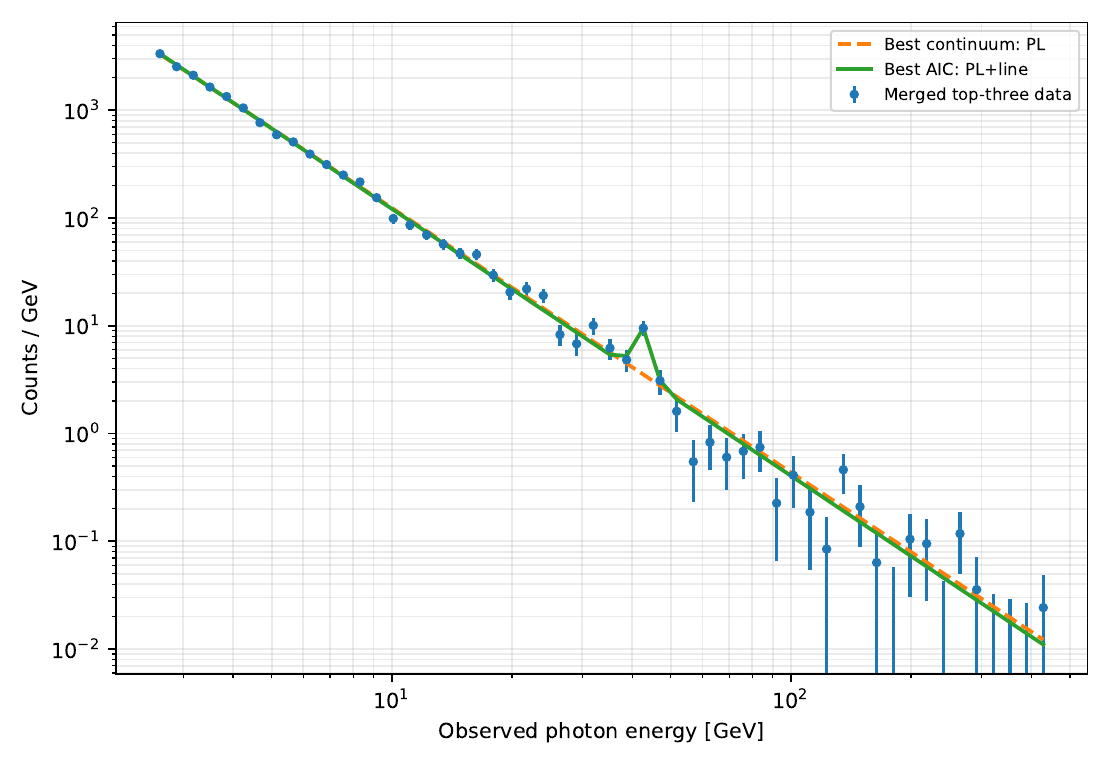}
\caption{Observed 2.5--450 GeV merged count spectrum, shown as a descriptive
diagnostic. The statistical results quoted in the text and Table~\ref{tab:broadbandmodels}
use the strict good-time selection and the full energy-dependent ROI exposure;
the curves in this raw count-spectrum visualization are retained only to show
the localized nature of the 43 GeV excess.}
\label{fig:broadband_spectrum}
\end{figure*}

\begin{figure}[tbp]
\includegraphics[width=\columnwidth]{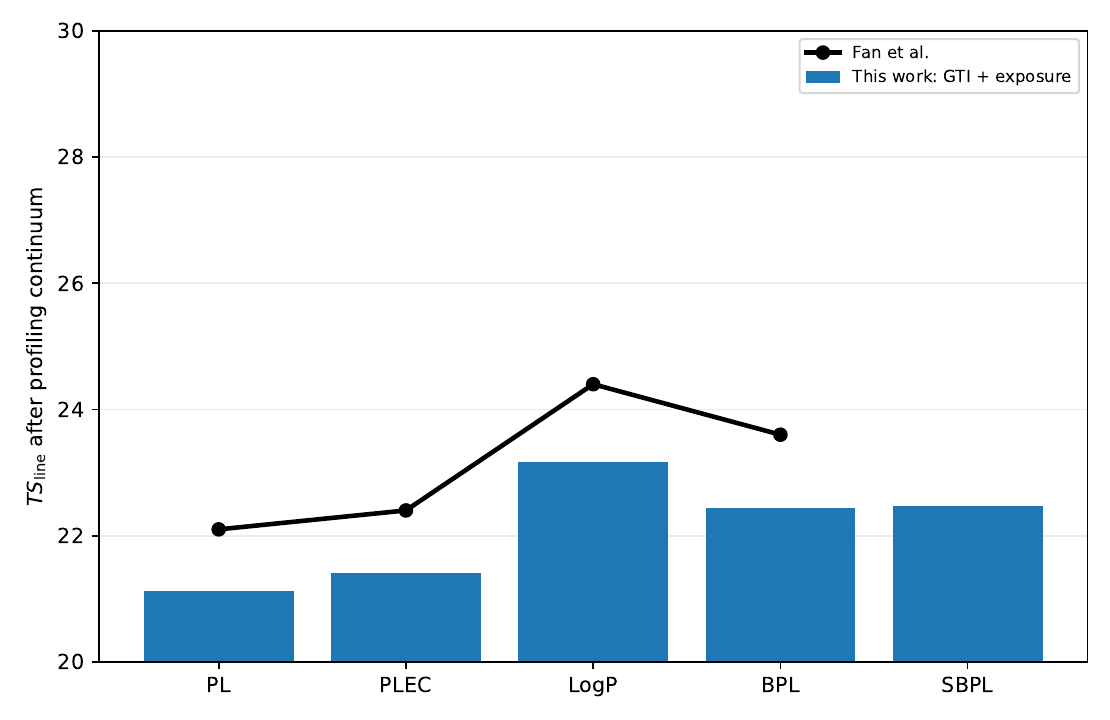}
\caption{Broad-band line likelihood improvement after profiling each
continuum. Bars show our strict-GTI, exposure-corrected analysis. Black points
and connecting line show the published values of Ref.~\cite{Fan2026} for PL,
PLEC, LogP, and BPL. The agreement is at the level of approximately one TS
unit for every published continuum family. The ordinate is truncated and does
not begin at zero.}
\label{fig:continuum_TS}
\end{figure}

The broken-power-law behavior remains particularly informative. Without a
line, the BPL attempts to place its break at
\begin{equation}
 E_b=41.95~{\rm GeV},
 \label{eq:bplnoline}
\end{equation}
with slopes $\gamma_1=-2.39$ and $\gamma_2=-2.68$. Once the line is included,
the break migrates to $E_b=22.60$ GeV while a separate line remains at
42.85 GeV with ${\rm TS}_{\rm line}=22.44$. The SBPL behaves almost
identically: its no-line fit places $E_b=41.87$ GeV and drives the smoothness
to the imposed minimum $s=0.02$, while the line fit moves the break to
22.46 GeV and retains ${\rm TS}_{\rm line}=22.47$. Thus even an artificially
sharp continuum edge does not economically replace the localized feature.
Figure~\ref{fig:broadband_spectrum} shows the corresponding count spectrum and
Fig.~\ref{fig:continuum_TS} the line likelihood improvement after profiling
each continuum family.

\bibliographystyle{apsrev4-2}
\bibliography{references}

@article{Fan2026,
  author        = {Fan, Yi-Zhong and Shen, Zhao-Qiang and Liang, Yun-Feng and Li, Xiang and Duan, Kai-Kai and Xia, Zi-Qing and Huang, Xiao-Yuan and Feng, Lei and Yuan, Qiang},
  title         = {Evidence for a $\sim 43$ GeV $\gamma$-ray line signal in a stacking analysis of the Virgo, Fornax, and Ophiuchus Galaxy clusters},
  journal       = {Phys. Rev. Lett.},
  volume        = {137},
  pages         = {071003},
  year          = {2026},
  doi           = {10.1103/lq5r-sjp7},
  eprint        = {2407.11737},
  archivePrefix = {arXiv},
  primaryClass  = {astro-ph.HE}
}

@article{SelfLiang1987,
  author  = {Self, Steven G. and Liang, Kung-Yee},
  title   = {Asymptotic Properties of Maximum Likelihood Estimators and Likelihood Ratio Tests under Nonstandard Conditions},
  journal = {J. Am. Stat. Assoc.},
  volume  = {82},
  number  = {398},
  pages   = {605--610},
  year    = {1987},
  doi     = {10.1080/01621459.1987.10478472}
}

@article{Liang2016,
  author        = {Liang, Yun-Feng and Shen, Zhao-Qiang and Li, Xiang and Fan, Yi-Zhong and Huang, Xiaoyuan and Lei, Shi-Jun and Feng, Lei and Liang, En-Wei and Chang, Jin},
  title         = {Search for a gamma-ray line feature from a group of nearby galaxy clusters with Fermi LAT Pass 8 data},
  journal       = {Phys. Rev. D},
  volume        = {93},
  pages         = {103525},
  year          = {2016},
  doi           = {10.1103/PhysRevD.93.103525},
  eprint        = {1602.06527},
  archivePrefix = {arXiv},
  primaryClass  = {astro-ph.HE}
}

@article{Ackermann2012,
  author        = {Ackermann, M. and others},
  collaboration = {Fermi-LAT Collaboration},
  title         = {The Fermi Large Area Telescope on Orbit: Event Classification, Instrument Response Functions, and Calibration},
  journal       = {Astrophys. J. Suppl.},
  volume        = {203},
  pages         = {4},
  year          = {2012},
  doi           = {10.1088/0067-0049/203/1/4},
  eprint        = {1206.1896},
  archivePrefix = {arXiv},
  primaryClass  = {astro-ph.IM}
}

@article{FermiLAT2015,
  author        = {Ackermann, M. and others},
  collaboration = {Fermi-LAT Collaboration},
  title         = {Updated search for spectral lines from Galactic dark matter interactions with Pass 8 data from the Fermi Large Area Telescope},
  journal       = {Phys. Rev. D},
  volume        = {91},
  pages         = {122002},
  year          = {2015},
  doi           = {10.1103/PhysRevD.91.122002},
  eprint        = {1506.00013},
  archivePrefix = {arXiv},
  primaryClass  = {astro-ph.HE}
}

@article{Finkbeiner2013,
    author = "Finkbeiner, Douglas P. and Su, Meng and Weniger, Christoph",
    title = "{Is the 130 GeV Line Real? A Search for Systematics in the Fermi-LAT Data}",
    eprint = "1209.4562",
    archivePrefix = "arXiv",
    primaryClass = "astro-ph.HE",
    doi = "10.1088/1475-7516/2013/01/029",
    journal = "JCAP",
    volume = "01",
    pages = "029",
    year = "2013"
}

@article{Lacroix2022,
  author        = {Lacroix, Thomas and Facchinetti, Gaetan and Perez-Romero, Judit and Stref, Martin and Lavalle, Julien and Maurin, David and Sanchez-Conde, Miguel A.},
  title         = {Classification of gamma-ray targets for velocity-dependent and subhalo-boosted dark-matter annihilation},
  journal       = {JCAP},
  volume        = {10},
  pages         = {021},
  year          = {2022},
  doi           = {10.1088/1475-7516/2022/10/021},
  eprint        = {2203.16440},
  archivePrefix = {arXiv},
  primaryClass  = {astro-ph.HE}
}

@inproceedings{Zabalza2015,
  author        = {Zabalza, Victor},
  title         = {naima: a Python package for inference of relativistic particle energy distributions from observed nonthermal spectra},
  booktitle     = {34th International Cosmic Ray Conference (ICRC2015)},
  pages         = {922},
  year          = {2015},
  eprint        = {1509.03319},
  archivePrefix = {arXiv},
  primaryClass  = {astro-ph.IM}
}

@article{Kafexhiu2014,
  author        = {Kafexhiu, Ervin and Aharonian, Felix and Taylor, Andrew M. and Vila, Gabriela S.},
  title         = {Parametrization of gamma-ray production cross sections for pp interactions in a broad proton energy range from the kinematic threshold to PeV energies},
  journal       = {Phys. Rev. D},
  volume        = {90},
  pages         = {123014},
  year          = {2014},
  doi           = {10.1103/PhysRevD.90.123014},
  eprint        = {1406.7369},
  archivePrefix = {arXiv},
  primaryClass  = {astro-ph.HE}
}

@article{Navarro1997,
  author        = {Navarro, Julio F. and Frenk, Carlos S. and White, Simon D. M.},
  title         = {A Universal Density Profile from Hierarchical Clustering},
  journal       = {Astrophys. J.},
  volume        = {490},
  pages         = {493--508},
  year          = {1997},
  doi           = {10.1086/304888},
  eprint        = {astro-ph/9611107},
  archivePrefix = {arXiv}
}

@article{Gao2012,
  author        = {Gao, L. and Navarro, J. F. and Frenk, C. S. and Jenkins, A. and Springel, V. and White, S. D. M.},
  title         = {The Phoenix Project: the dark side of rich Galaxy clusters},
  journal       = {Mon. Not. Roy. Astron. Soc.},
  volume        = {425},
  pages         = {2169--2186},
  year          = {2012},
  doi           = {10.1111/j.1365-2966.2012.21564.x}
}

@article{Ando2019,
  author        = {Ando, Shin'ichiro and Ishiyama, Tomoaki and Hiroshima, Nagisa},
  title         = {Halo Substructure Boosts to the Signatures of Dark Matter Annihilation},
  journal       = {Galaxies},
  volume        = {7},
  number        = {3},
  pages         = {68},
  year          = {2019},
  doi           = {10.3390/galaxies7030068},
  eprint        = {1903.11427},
  archivePrefix = {arXiv},
  primaryClass  = {astro-ph.CO}
}

@article{Abazajian2012,
  author        = {Abazajian, Kevork N. and Agrawal, Prateek and Chacko, Zackaria and Kilic, Can},
  title         = {Lower Limits on the Strengths of Gamma Ray Lines from WIMP Dark Matter Annihilation},
  journal       = {Phys. Rev. D},
  volume        = {85},
  pages         = {123543},
  year          = {2012},
  doi           = {10.1103/PhysRevD.85.123543},
  eprint        = {1111.2835},
  archivePrefix = {arXiv},
  primaryClass  = {hep-ph}
}

@article{Asano2013,
  author        = {Asano, Masaki and Bringmann, Torsten and Sigl, Gunter and Vollmann, Martin},
  title         = {The 130 GeV gamma-ray line and generic dark matter model building constraints from continuum gamma rays, radio and antiproton data},
  journal       = {Phys. Rev. D},
  volume        = {87},
  pages         = {103509},
  year          = {2013},
  doi           = {10.1103/PhysRevD.87.103509},
  eprint        = {1211.6739},
  archivePrefix = {arXiv},
  primaryClass  = {hep-ph}
}

@article{Rajaraman2013,
  author        = {Rajaraman, Arvind and Tait, Tim M. P. and Wijangco, Alexander M.},
  title         = {Effective Theories of Gamma-ray Lines from Dark Matter Annihilation},
  journal       = {Phys. Dark Univ.},
  volume        = {2},
  pages         = {17--21},
  year          = {2013},
  doi           = {10.1016/j.dark.2012.12.001},
  eprint        = {1211.7061},
  archivePrefix = {arXiv},
  primaryClass  = {hep-ph}
}

@article{WeinerYavin2012,
  author        = {Weiner, Neal and Yavin, Itay},
  title         = {How Dark Are Majorana WIMPs? Signals from Magnetic Inelastic Dark Matter and Rayleigh Dark Matter},
  journal       = {Phys. Rev. D},
  volume        = {86},
  pages         = {075021},
  year          = {2012},
  doi           = {10.1103/PhysRevD.86.075021},
  eprint        = {1206.2910},
  archivePrefix = {arXiv},
  primaryClass  = {hep-ph}
}

@article{WeinerYavin2013,
  author        = {Weiner, Neal and Yavin, Itay},
  title         = {UV Completions of Magnetic Inelastic and Rayleigh Dark Matter for the Fermi Line(s)},
  journal       = {Phys. Rev. D},
  volume        = {87},
  pages         = {023523},
  year          = {2013},
  doi           = {10.1103/PhysRevD.87.023523},
  eprint        = {1209.1093},
  archivePrefix = {arXiv},
  primaryClass  = {hep-ph}
}

@article{Jackson2013,
  author        = {Jackson, C. B. and Servant, Geraldine and Shaughnessy, Gabe and Tait, Tim M. P. and Taoso, Marco},
  title         = {Gamma-ray Lines and One-Loop Continuum from s-channel Dark Matter Annihilations},
  journal       = {JCAP},
  volume        = {07},
  pages         = {021},
  year          = {2013},
  doi           = {10.1088/1475-7516/2013/07/021},
  eprint        = {1302.1802},
  archivePrefix = {arXiv},
  primaryClass  = {hep-ph}
}

@article{BergstromUllio1997,
  author        = {Bergstr{\"o}m, Lars and Ullio, Piero},
  title         = {Full One-Loop Calculation of Neutralino Annihilation into Two Photons},
  journal       = {Nucl. Phys. B},
  volume        = {504},
  pages         = {27--44},
  year          = {1997},
  doi           = {10.1016/S0550-3213(97)00530-0},
  eprint        = {hep-ph/9706232},
  archivePrefix = {arXiv}
}

@article{Bringmann2008,
  author        = {Bringmann, Torsten and Bergstr{\"o}m, Lars and Edsj{\"o}, Joakim},
  title         = {New Gamma-Ray Contributions to Supersymmetric Dark Matter Annihilation},
  journal       = {JHEP},
  volume        = {01},
  pages         = {049},
  year          = {2008},
  doi           = {10.1088/1126-6708/2008/01/049},
  eprint        = {0710.3169},
  archivePrefix = {arXiv},
  primaryClass  = {hep-ph}
}

@article{ChalonsMcCabe2013,
  author        = {Chalons, Guillaume and Dolan, Matthew J. and McCabe, Christopher},
  title         = {Neutralino Dark Matter and the Fermi Gamma-Ray Lines},
  journal       = {JCAP},
  volume        = {02},
  pages         = {016},
  year          = {2013},
  doi           = {10.1088/1475-7516/2013/02/016},
  eprint        = {1211.5154},
  archivePrefix = {arXiv},
  primaryClass  = {hep-ph}
}

@article{ALEPH2004,
  author        = {Heister, A. and others},
  collaboration = {ALEPH},
  title         = {Absolute Mass Lower Limit for the Lightest Neutralino of the MSSM from $e^+e^-$ Data at $\sqrt{s}$ up to 209 GeV},
  journal       = {Phys. Lett. B},
  volume        = {583},
  pages         = {247--263},
  year          = {2004},
  doi           = {10.1016/j.physletb.2003.12.066}
}

@article{Belanger2013,
  author        = {B{\'e}langer, Genevi{\`e}ve and Drieu La Rochelle, Guillaume and Dumont, B{\'e}ranger and Godbole, Rohini M. and Kraml, Sabine and Kulkarni, Suchita},
  title         = {LHC Constraints on Light Neutralino Dark Matter in the MSSM},
  journal       = {Phys. Lett. B},
  volume        = {726},
  pages         = {773--780},
  year          = {2013},
  doi           = {10.1016/j.physletb.2013.09.059},
  eprint        = {1308.3735},
  archivePrefix = {arXiv},
  primaryClass  = {hep-ph}
}

@article{Feng2016Line,
  author        = {Feng, Lei and Liang, Yun-Feng and Dong, Tie-Kuang and Fan, Yi-Zhong},
  title         = {Interpretations of the Possible 42.7 GeV Gamma-Ray Line},
  journal       = {Phys. Rev. D},
  volume        = {94},
  pages         = {043535},
  year          = {2016},
  doi           = {10.1103/PhysRevD.94.043535},
  eprint        = {1608.04056},
  archivePrefix = {arXiv},
  primaryClass  = {hep-ph}
}

@article{BuckleyHooper2012,
  author        = {Buckley, Matthew R. and Hooper, Dan},
  title         = {Implications of a 130 GeV Gamma-Ray Line for Dark Matter},
  journal       = {Phys. Rev. D},
  volume        = {86},
  pages         = {043524},
  year          = {2012},
  doi           = {10.1103/PhysRevD.86.043524},
  eprint        = {1205.6811},
  archivePrefix = {arXiv},
  primaryClass  = {hep-ph}
}

@article{LeeParkPark2012,
  author        = {Lee, Hyun Min and Park, Myeonghun and Park, Wan-Il},
  title         = {Fermi Gamma Ray Line at 130 GeV from Axion-Mediated Dark Matter},
  journal       = {Phys. Rev. D},
  volume        = {86},
  pages         = {103502},
  year          = {2012},
  doi           = {10.1103/PhysRevD.86.103502},
  eprint        = {1205.4675},
  archivePrefix = {arXiv},
  primaryClass  = {hep-ph}
}

@article{Yang2020,
  author        = {Yang, Kwei-Chou},
  title         = {A Potentially Detectable Gamma-Ray Line in the Fermi Galactic Center Excess -- In Light of One-Step Cascade Annihilations of Secluded (Vector) Dark Matter via the Higgs Portal},
  journal       = {JHEP},
  volume        = {07},
  pages         = {148},
  year          = {2020},
  doi           = {10.1007/JHEP07(2020)148},
  eprint        = {2001.04946},
  archivePrefix = {arXiv},
  primaryClass  = {hep-ph}
}

@article{DeLaTorreLuque2024,
  author        = {De La Torre Luque, Pedro and Smirnov, Juri and Linden, Tim},
  title         = {Gamma-Ray Lines in 15 Years of Fermi-LAT Data: New Constraints on Higgs Portal Dark Matter},
  journal       = {Phys. Rev. D},
  volume        = {109},
  pages         = {L041301},
  year          = {2024},
  doi           = {10.1103/PhysRevD.109.L041301},
  eprint        = {2309.03281},
  archivePrefix = {arXiv},
  primaryClass  = {hep-ph}
}

@article{Ibarra2012,
  author        = {Ibarra, Alejandro and L{\'o}pez Gehler, Sergio and Pato, Miguel},
  title         = {Dark Matter Constraints from Box-Shaped Gamma-Ray Features},
  journal       = {JCAP},
  volume        = {07},
  pages         = {043},
  year          = {2012},
  doi           = {10.1088/1475-7516/2012/07/043},
  eprint        = {1205.0007},
  archivePrefix = {arXiv},
  primaryClass  = {hep-ph}
}

@article{BrunettiJones2014,
  author        = {Brunetti, Gianfranco and Jones, Thomas W.},
  title         = {Cosmic Rays in Galaxy Clusters and Their Nonthermal Emission},
  journal       = {Int. J. Mod. Phys. D},
  volume        = {23},
  number        = {04},
  pages         = {1430007},
  year          = {2014},
  doi           = {10.1142/S0218271814300079},
  eprint        = {1401.7519},
  archivePrefix = {arXiv},
  primaryClass  = {astro-ph.CO}
}

@article{Ballet2023DR4,
  author        = {Ballet, Jean and Bruel, Philippe and Burnett, T. H. and Lott, Benoit and others},
  collaboration = {Fermi-LAT},
  title         = {Fermi Large Area Telescope Fourth Source Catalog Data Release 4 (4FGL-DR4)},
  journal       = {arXiv e-prints},
  pages         = {arXiv:2307.12546},
  year          = {2023},
  eprint        = {2307.12546},
  archivePrefix = {arXiv},
  primaryClass  = {astro-ph.HE}
}

@article{Han2012,
  author        = {Han, Jiaxin and Frenk, Carlos S. and Eke, Vincent R. and Gao, Liang and White, Simon D. M. and Boyarsky, Alexey and Malyshev, Dmitry and Ruchayskiy, Oleg},
  title         = {Constraining extended gamma-ray emission from galaxy clusters},
  journal       = {Mon. Not. Roy. Astron. Soc.},
  volume        = {427},
  pages         = {1651--1665},
  year          = {2012},
  doi           = {10.1111/j.1365-2966.2012.22080.x}
}

@article{AckermannEnergyScale,
  author        = {Ackermann, M. and others},
  collaboration = {Fermi-LAT Collaboration},
  title         = {In-flight measurement of the absolute energy scale of the Fermi Large Area Telescope},
  journal       = {Astropart. Phys.},
  volume        = {35},
  pages         = {346--353},
  year          = {2012},
  doi           = {10.1016/j.astropartphys.2011.10.007},
  eprint        = {1108.0201},
  archivePrefix = {arXiv},
  primaryClass  = {astro-ph.IM}
}

@article{Weniger2012,
  author        = {Weniger, Christoph},
  title         = {A Tentative Gamma-Ray Line from Dark Matter Annihilation at the Fermi Large Area Telescope},
  journal       = {JCAP},
  volume        = {08},
  pages         = {007},
  year          = {2012},
  doi           = {10.1088/1475-7516/2012/08/007},
  eprint        = {1204.2797},
  archivePrefix = {arXiv},
  primaryClass  = {hep-ph}
}

@article{SuFinkbeiner2012,
  author        = {Su, Meng and Finkbeiner, Douglas P.},
  title         = {Strong Evidence for Gamma-ray Line Emission from the Inner Galaxy},
  journal       = {arXiv e-prints},
  year          = {2012},
  eprint        = {1206.1616},
  archivePrefix = {arXiv},
  primaryClass  = {astro-ph.HE}
}

@article{FermiLAT2013Lines,
  author        = {Ackermann, M. and others},
  collaboration = {Fermi-LAT Collaboration},
  title         = {Search for Gamma-ray Spectral Lines with the Fermi Large Area Telescope and Dark Matter Implications},
  journal       = {Phys. Rev. D},
  volume        = {88},
  pages         = {082002},
  year          = {2013},
  doi           = {10.1103/PhysRevD.88.082002},
  eprint        = {1305.5597},
  archivePrefix = {arXiv},
  primaryClass  = {astro-ph.HE}
}

@article{Aharonian2012ColdWind,
  author        = {Aharonian, Felix and Khangulyan, Dmitry and Malyshev, Denys},
  title         = {Cold ultrarelativistic pulsar winds as potential sources of galactic gamma-ray lines above 100 GeV},
  journal       = {Astron. Astrophys.},
  volume        = {547},
  pages         = {A114},
  year          = {2012},
  doi           = {10.1051/0004-6361/201220092},
  eprint        = {1207.0458},
  archivePrefix = {arXiv},
  primaryClass  = {astro-ph.HE}
}

@article{BlumenthalGould1970,
  author        = {Blumenthal, George R. and Gould, Robert J.},
  title         = {Bremsstrahlung, Synchrotron Radiation, and Compton Scattering of High-Energy Electrons Traversing Dilute Gases},
  journal       = {Rev. Mod. Phys.},
  volume        = {42},
  pages         = {237--270},
  year          = {1970},
  doi           = {10.1103/RevModPhys.42.237}
}

@article{AckermannClusterCR2014,
  author        = {Ackermann, M. and others},
  collaboration = {Fermi-LAT Collaboration},
  title         = {Search for Cosmic-Ray-Induced Gamma-Ray Emission in Galaxy Clusters},
  journal       = {Astrophys. J.},
  volume        = {787},
  pages         = {18},
  year          = {2014},
  doi           = {10.1088/0004-637X/787/1/18},
  eprint        = {1308.5654},
  archivePrefix = {arXiv},
  primaryClass  = {astro-ph.HE}
}

@article{SanchezConde2014,
  author        = {S{\'a}nchez-Conde, Miguel A. and Prada, Francisco},
  title         = {The flattening of the concentration--mass relation towards low halo masses and its implications for the annihilation signal boost},
  journal       = {Mon. Not. Roy. Astron. Soc.},
  volume        = {442},
  pages         = {2271--2277},
  year          = {2014},
  doi           = {10.1093/mnras/stu1014},
  eprint        = {1312.1729},
  archivePrefix = {arXiv},
  primaryClass  = {astro-ph.CO}
}

@article{Bartels2015,
  author        = {Bartels, Richard H. and Ando, Shin'ichiro},
  title         = {Boosting the annihilation boost: Tidal effects on dark matter subhalos and consistent luminosity modeling},
  journal       = {Phys. Rev. D},
  volume        = {92},
  number        = {12},
  pages         = {123508},
  year          = {2015},
  doi           = {10.1103/PhysRevD.92.123508},
  eprint        = {1507.08656},
  archivePrefix = {arXiv},
  primaryClass  = {astro-ph.CO}
}

@article{Moline2017,
  author        = {Molin{\'e}, {\'A}ngeles and S{\'a}nchez-Conde, Miguel A. and Palomares-Ruiz, Sergio and Prada, Francisco},
  title         = {Characterization of subhalo structural properties and implications for dark matter annihilation signals},
  journal       = {Mon. Not. Roy. Astron. Soc.},
  volume        = {466},
  pages         = {4974--4990},
  year          = {2017},
  doi           = {10.1093/mnras/stx026}
}

@article{Piccirillo2022,
  author        = {Piccirillo, Erin and Blanchette, Keagan and Bozorgnia, Nassim and Strigari, Louis E. and Frenk, Carlos S. and Grand, Robert J. J. and Marinacci, Federico},
  title         = {Velocity-dependent annihilation radiation from dark matter subhalos in cosmological simulations},
  journal       = {J. Cosmol. Astropart. Phys.},
  volume        = {08},
  pages         = {058},
  year          = {2022},
  doi           = {10.1088/1475-7516/2022/08/058},
  eprint        = {2203.08853},
  archivePrefix = {arXiv},
  primaryClass  = {astro-ph.GA}
}

@article{BergstromSnellman1988,
  author = {Bergstr{\"o}m, Lars and Snellman, H{\aa}kan},
  title = {Observable monochromatic photons from cosmic photino annihilation},
  journal = {Phys. Rev. D},
  volume = {37},
  pages = {3737},
  year = {1988},
  doi = {10.1103/PhysRevD.37.3737},
}

@article{Pullen2007,
  author = {Pullen, Anthony R. and Chary, Ranga-Ram and Kamionkowski, Marc},
  title = {Search with EGRET for a Gamma Ray Line from the Galactic Center},
  journal = {Phys. Rev. D},
  volume = {76},
  pages = {063006},
  year = {2007},
  doi = {10.1103/PhysRevD.76.063006},
  eprint = {astro-ph/0610295},
  archivePrefix = {arXiv},
}

@article{FermiLAT2010Lines,
  author = {Abdo, A. A. and others},
  title = {Fermi LAT Search for Photon Lines from 30 to 200 GeV and Dark Matter Implications},
  journal = {Phys. Rev. Lett.},
  volume = {104},
  pages = {091302},
  year = {2010},
  doi = {10.1103/PhysRevLett.104.091302},
  eprint = {1001.4836},
  archivePrefix = {arXiv},
}

@article{DAMPE2022Lines,
  author = {Alemanno, Francesca and others},
  title = {Search for gamma-ray spectral lines with the DArk Matter Particle Explorer},
  journal = {Science Bulletin},
  volume = {67},
  pages = {679--684},
  year = {2022},
  doi = {10.1016/j.scib.2021.12.015},
  eprint = {2112.08860},
  archivePrefix = {arXiv},
}

@article{HESS2018Lines,
  author = {Abdallah, H. and others},
  title = {Search for $\gamma$-ray line signals from dark matter annihilations in the inner Galactic halo from ten years of observations with H.E.S.S.},
  journal = {Phys. Rev. Lett.},
  volume = {120},
  pages = {201101},
  year = {2018},
  doi = {10.1103/PhysRevLett.120.201101},
  eprint = {1805.05741},
  archivePrefix = {arXiv},
}

@article{MAGIC2023Lines,
  author = {Abe, H. and others},
  title = {Search for Gamma-Ray Spectral Lines from Dark Matter Annihilation up to 100 TeV toward the Galactic Center with MAGIC},
  journal = {Phys. Rev. Lett.},
  volume = {130},
  pages = {061002},
  year = {2023},
  doi = {10.1103/PhysRevLett.130.061002},
}

@article{HESS2026Lines,
  author = {Aharonian, F. and others},
  title = {Search for gamma-ray spectral lines from dark matter annihilation with the H.E.S.S. Inner Galaxy Survey},
  journal = {Phys. Rev. Lett.},
  volume = {137},
  pages = {091002},
  year = {2026},
  doi = {10.1103/d8tj-55kc},
  eprint = {2608.07234},
  archivePrefix = {arXiv},
}

@article{Huang2012Consistency,
  author = {Huang, Xiaoyuan and Yuan, Qiang and Yin, Peng-Fei and Bi, Xiao-Jun and Chen, Xuelei},
  title = {Constraints on the dark matter annihilation scenario of Fermi 130 GeV $\gamma$-ray line emission by continuous gamma-rays, Milky Way halo, galaxy clusters and dwarf galaxies observations},
  journal = {JCAP},
  volume = {11},
  pages = {048},
  year = {2012},
  note = {Erratum: JCAP 05 (2013) E02},
  doi = {10.1088/1475-7516/2012/11/048},
  eprint = {1208.0267},
  archivePrefix = {arXiv}
}

@article{Marchegiani2025Radio,
  author = {Marchegiani, P. and Vacca, V. and Govoni, F. and Murgia, M. and Loi, F.},
  title = {Constraints on dark matter annihilation and turbulent reacceleration set by high-frequency observations of the radio halo in the Coma cluster},
  journal = {Mon. Not. Roy. Astron. Soc.},
  volume = {542},
  pages = {2901--2909},
  year = {2025},
  doi = {10.1093/mnras/staf1426},
  eprint = {2508.19354},
  archivePrefix = {arXiv},
}

@article{Speckhard2016,
  author = {Speckhard, Eric G. and Ng, Kenny C. Y. and Beacom, John F. and Laha, Ranjan},
  title = {Dark Matter Velocity Spectroscopy},
  journal = {Phys. Rev. Lett.},
  volume = {116},
  pages = {031301},
  year = {2016},
  doi = {10.1103/PhysRevLett.116.031301},
  eprint = {1507.04744},
  archivePrefix = {arXiv},
  primaryClass = {astro-ph.CO}
}

@article{Thompson1993EGRET,
  author = {Thompson, D. J. and others},
  title = {Calibration of the Energetic Gamma-Ray Experiment Telescope ({EGRET}) for the Compton Gamma-Ray Observatory},
  journal = {Astrophys. J. Suppl.},
  volume = {86},
  pages = {629--656},
  year = {1993},
  url = {https://adsabs.harvard.edu/pdf/1993ApJS...86..629T}
}

@article{JeltemaKehayiasProfumo2009,
  author = {Jeltema, Tesla E. and Kehayias, John and Profumo, Stefano},
  title = {Gamma Rays from Clusters and Groups of Galaxies: Cosmic Rays versus Dark Matter},
  journal = {Phys. Rev. D},
  volume = {80},
  pages = {023005},
  year = {2009},
  doi = {10.1103/PhysRevD.80.023005},
  eprint = {0812.0597},
  archivePrefix = {arXiv},
  primaryClass = {astro-ph}
}

@article{JeltemaProfumo2011,
  author = {Jeltema, Tesla E. and Profumo, Stefano},
  title = {Implications of {Fermi} Observations for Hadronic Models of Radio Halos in Clusters of Galaxies},
  journal = {Astrophys. J.},
  volume = {728},
  pages = {53},
  year = {2011},
  doi = {10.1088/0004-637X/728/1/53},
  eprint = {1006.1648},
  archivePrefix = {arXiv},
  primaryClass = {astro-ph.HE}
}

\end{document}